\documentclass{emulateapj}
\slugcomment{Accepted to AJ}

\shortauthors{Matthews et al.}
\shorttitle{HI Imaging of AGB Stars}

\begin{document}
\newcommand{\ang}{\rm \AA}
\newcommand{\msun}{M$_\odot$}
\newcommand{\lsun}{L$_\odot$}
\newcommand{\days}{$d$}
\newcommand{\degree}{$^\circ$}
\newcommand{\ud}{{\rm d}}
\newcommand{\as}[2]{$#1''\,\hspace{-1.7mm}.\hspace{.0mm}#2$}
\newcommand{\am}[2]{$#1'\,\hspace{-1.7mm}.\hspace{.0mm}#2$}
\newcommand{\ad}[2]{$#1^{\circ}\,\hspace{-1.7mm}.\hspace{.0mm}#2$}
\newcommand{\lsim}{~\rlap{$<$}{\lower 1.0ex\hbox{$\sim$}}}
\newcommand{\gsim}{~\rlap{$>$}{\lower 1.0ex\hbox{$\sim$}}}
\newcommand{\HA}{H$\alpha$}
\newcommand{\HII}{\mbox{H\,{\sc ii}}}
\newcommand{\kms}{\mbox{km s$^{-1}$}}
\newcommand{\HI}{\mbox{H\,{\sc i}}}
\newcommand{\KI}{\mbox{K\,{\sc i}}}
\newcommand{\nan}{Nan\c{c}ay}
\newcommand{\galex}{{\it GALEX}}
\newcommand{\jks}{Jy~km~s$^{-1}$}

\title{An \HI\ Imaging Survey of Asymptotic Giant Branch Stars}

\author{L. D. Matthews\altaffilmark{1},  T. Le~Bertre\altaffilmark{2}, 
E. G\'erard\altaffilmark{3}, 
M. C. Johnson\altaffilmark{4}}

\altaffiltext{1}{MIT Haystack Observatory, Off Route 40, Westford, MA
  01886 USA; lmatthew@haystack.mit.edu}
\altaffiltext{2}{LERMA, UMR 8112, Observatoire de Paris, 61 av.
de l'Observatoire, F-75014 Paris, France}
\altaffiltext{3}{GEPI, UMR 8111, Observatoire de Paris, 5 Place J.
Janssen, F-92195 Meudon Cedex, France}
\altaffiltext{4}{University of Texas at Austin, Department of
  Astronomy, 2515 Speedway, Stop C1400, Austin, TX 78712 USA}

\begin{abstract}
We present an imaging study of
a sample of eight
asymptotic giant branch (AGB) stars in the \HI\ 21-cm line.
Using observations from the Very Large Array, we have unambiguously detected
\HI\ emission associated with the extended circumstellar envelopes of six
of the targets. The detected \HI\ masses range from $M_{\rm
  HI}\approx0.015-0.055M_{\odot}$. 
The \HI\ morphologies and kinematics are diverse, but in all cases
appear to be significantly influenced by  
the interaction between the circumstellar envelope and the surrounding 
medium. Four stars (RX~Lep, Y~UMa,
Y~CVn, and V1942~Sgr) are surrounded by detached \HI\ shells ranging from
0.36 to 0.76~pc across. We interpret these shells as resulting from
material entrained in a stellar outflow being abruptly slowed at a
termination shock where it meets the local medium. RX~Lep and TX~Psc, two stars with
moderately high space velocities ($V_{\rm space}>56$~\kms), 
exhibit extended gaseous wakes ($\sim$0.3 and 0.6~pc in the plane of
the sky),
trailing their motion through space. The other detected star, R~Peg, displays a
peculiar ``horseshoe-shaped'' \HI\ morphology with emission extended on scales up to
$\sim$1.7~pc; in this case, the circumstellar debris may have been
distorted by transverse flows in the local interstellar medium. 
We briefly discuss our new results in the context of the
entire sample of evolved stars that has been imaged in \HI\ to date.

\end{abstract}


\section{Introduction}
Stars on the asymptotic giant branch (AGB) undergo substantial mass-loss
through cool, low-velocity winds
($V_{\rm out}\sim$10~\kms). These winds are one of the dominant
means through which chemically enriched material 
is recycled back into the Galaxy to sustain
new generations of star and planet formation (e.g., Schr\"oder \&
Sedlmayr 2001; Cristallo et al. 2009; Leitner \& Kravtsov 2011).
The distribution of circumstellar debris on the
largest scales may also be important for the subsequent evolution of novae,
planetary nebulae
and some Type~Ia supernovae (e.g., Wang et al. 2004; Mohamed
\& Podsiadlowski 2007; Wareing et al. 2007a; Chiotellis et al. 2012;
Roy et al. 2012). 
In addition, the outflows and space motions of these mass-losing stars
play a role in shaping the structure of the interstellar medium (ISM)
on parsec and sub-parsec scales (Villaver et al. 2002; 
Wareing et al. 2007b; Begum et al. 2010). 

It has long been recognized that the interstellar
environment of evolved stars can have a profound impact on the observed
mass, 
size, shape, and chemical composition of their circumstellar envelopes
(hereafter CSEs; 
e.g., Smith 1976; Isaacman 1979;  
Young et al. 1993b; Villaver
et al. 2002, 2012; Wareing et
al. 2006; 2007a, b, c; Wareing 2012; see also the review by Stencel 2009). 
Nonetheless, 
the interfaces through which AGB stars interact with their environments have
remained poorly studied observationally. 
One problem is that CSEs
can be enormously extended ($\gsim$1~pc), and
their chemical compositions change as a
function of distance from the star as densities drop, and 
molecules become dissociated by
the interstellar radiation field or other radiation sources. 
The result is that some of the most frequently used observational 
tracers of CSEs
(e.g., CO; SiO, H$_{2}$O, and OH masers) 
do not probe the outermost CSE or its interaction zone
with the ISM. In some cases, more extended material can be traced via
imaging observations in the far-infrared (FIR; e.g., 
Young et al. 1993a, b; Zijlstra \& Weinberger
2002; Ueta et
al. 2006, 2010; Cox et al. 2012a) or in the
far-ultraviolet (FUV; Martin et al. 2007; Sahai \& Chronopoulos 2010),
but such data do not supply any direct kinematic information.

Recent studies have demonstrated that the \HI\ 21-cm line of neutral
hydrogen is a powerful tool for the study
of extended CSEs
and their interface with their larger-scale environments 
(e.g., G\'erard \& Le~Bertre 2006; 
Matthews \& Reid 2007; Libert
et al. 2007, 2008, 2010a, b; Matthews et
al. 2008, 2011a, b, 2012; G\'erard et al. 2011a, b).  
Not only is hydrogen the most plentiful constituent of CSEs, 
but \HI\ line measurements provide valuable kinematic information
and can be directly translated into measurements of atomic hydrogen
mass. 
Furthermore,  because \HI\ is not destroyed by the interstellar
radiation field, \HI\ measurements are particularly well-suited to probing the
outermost reaches of CSEs.

While the number of evolved stars detected in the \HI\ line has grown
steadily over the past decade thanks to ongoing surveys with the
Nan\c{c}ay Radio Telescope (NRT; 
Le~Bertre \& G\'erard 2001, 2004; G\'erard \& Le Bertre 2003; 
G\'erard \& Le~Bertre 2006; Gardan et al. 2006; Libert
et al. 2008, 2010a; G\'erard et al. 2011a, b), the number of stars
that has been imaged in \HI\ with aperture synthesis arrays 
has remained small (Bowers \& Knapp
1987, 1988;
Matthews \& Reid 2007; Matthews et al. 2008, 2011b, 2012; Le~Bertre et
al. 2012).  Such imaging observations offer a crucial
complement to single-dish measurements, since the high angular
resolution of interferometers provides detailed information on the
structure and spatially resolved kinematics of the CSE and enables detailed comparisons
with observations of the CSE at other wavelengths. In addition,
interferometers act as effective filters against large-scale Galactic
contamination along the line-of-sight, allowing stellar signals to be
more readily disentangled from background radiation.

In this paper we present new \HI\ imaging observations of eight mass-losing
AGB stars, effectively doubling the existing sample. We describe
in detail
the results for each individual star, followed by a discussion of
emerging trends from the combined sample of
evolved stars that has been imaged in \HI\ to date.

\section{Sample Selection\protect\label{sample}}
The targets for the present study of circumstellar \HI\
comprise a set of eight Galactic
AGB stars spanning a range in properties, including effective
temperature, mass-loss rate, variability class, 
and chemistry. Some basic parameters
of the sample are summarized in Table~1. None of the stars in our
current program is a known or suspected
binary, although two binary AGB stars, R~Aqr and Mira, have been observed in \HI\ as
part of our earlier studies (Matthews \& Reid 2007; Matthews et al. 2008).

%
\begin{deluxetable*}{lllccccccccc}
\tabletypesize{\tiny}
\tablewidth{0pc}
\tablenum{1}
\tablecaption{Properties of the Target Stars}
\tablehead{
\colhead{Name} & \colhead{$\alpha$(J2000.0)} &
\colhead{$\delta$(J2000.0)} & \colhead{$l$} & \colhead{$b$}
 & \colhead{$V_{\star,\rm LSR}$} &  \colhead{$d$}  &
\colhead{$T_{\rm eff}$} & \colhead{Spec.} & \colhead{Var.} &
\colhead{$\dot{M}$} & \colhead{$V_{\rm out}$} 
  \\
\colhead{}     & \colhead{}     & \colhead{} & \colhead{(deg)} & \colhead{(deg)} &
\colhead{(\kms)} & \colhead{(pc)} &\colhead{(K)}  & \colhead{Type} & \colhead{Class}
&\colhead{($M_{\odot}$ yr$^{-1}$)} & \colhead{(\kms)}  \\
  \colhead{(1)} & \colhead{(2)} & \colhead{(3)} &
\colhead{(4)} & \colhead{(5)} & \colhead{(6)} & \colhead{(7)}
& \colhead{(8)} & \colhead{(9)} & \colhead{(10)} & \colhead{(11)} & \colhead{(12)}
}

\startdata

\tableline
\multicolumn{12}{c}{Oxygen-rich stars} \\
\tableline

IK Tau & 03 53 28.87& +11 24 21.7 & 177.95& $-$31.41& 34.6 & 245 &
2000-3000 & M6e-M10e & Mira & $4.2\times10^{-6}$ & 22.0 \\ 

RX Lep & 05 11 22.87& $-$11 50 56.72 & 212.54 & $-$27.51 & 28.9 & 149 &
3300 & M7 & SRb & $2.0\times10^{-7}$ & 4.1 \\ 

Y UMa & 12 40 21.28 & +55 50 47.61 & 126.16 & +61.21 & 19.0 & 385 &3100 
& M7II-III & SRb & 2.6$\times10^{-7}$ & 5.4\\ 

R Peg & 23 06 39.17 & +10 32 36.10 &85.41& $-$44.56& 24.0 & 400 &2300-2900 &
M7e & Mira & $5.3\times10^{-7}$ & 4.3 \\ 

\tableline
\multicolumn{12}{c}{Carbon-rich stars}\\
\tableline

Y CVn & 12 45 07.82 & +45 26 24.92& 126.44& +71.64& 21.1 & 272& 2730 &
C5.4J & SRb & $1.7\times10^{-7}$ & 7.8 \\ 

V1942 Sgr & 19 19 09.59& $-$15 54 30.03& 21.61& $-$13.19 & $-33.7$ & 535 &
2960 & C & Lb & $(6.1,1.0)\times10^{-7}$ & 17.5,5.0 \\ 

AFGL~3099 & 23 28 17.50& +10 54 36.9 & 92.23& $-$46.94& 46.6 & 1500&
1800-2000 & C & Mira & $1.3\times10^{-5}$ & 10.1 \\ 

TX Psc & 23 46 23.51 & +03 29 12.52& 93.28 & $-$55.59 & 12.2 & 275 &
3115 & C5II & Lb & $3.1\times10^{-7}$ & 10.7 \\

\enddata

\tablecomments{Units of right ascension are hours, minutes, and
seconds. Units of declination are degrees, arcminutes, and
arcseconds. Explanation of columns: (1) star name; (2) \& (3) right
ascension and declination (J2000.0); (4) \& (5) Galactic coordinates;
(6) systemic velocity relative to the Local Standard of
Rest (LSR); (7) adopted distance in parsecs; (8) stellar effective
temperature in kelvins; (9)
spectral type; (10) variability class; 
(11) mass-loss rate in solar masses per year; two values are quoted in
cases of multi-component molecular line profiles (12) outflow velocity
in \kms\ as determined from CO observations. Coordinates
and spectral classifications were taken from
SIMBAD (http://simbad.harvard.edu).  References and additional
descriptions for the 
other parameters are provided throughout \S\ref{iktau}-\S\ref{txpsc}.}

\end{deluxetable*}


Six of our
targets (RX~Lep, Y~UMa, R~Peg, Y~CVn, V1942~Sgr, and TX~Psc)
were selected from the growing sample of AGB stars
that has been detected in \HI\ using
single-dish observations with the NRT (see above). As we were
interested in imaging examples of both carbon-rich and oxygen-rich stars, we
selected three from each chemistry type, all of which are
relatively nearby and exhibited relatively strong peak \HI\ signals
based on the previous NRT measurements ($\gsim$0.1~Jy).  In the case of Y~UMa, R~Peg,
Y~CVn and V1942~Sgr, an additional motivation for selection was that 
line-of-sight confusion from Galactic emission near the stellar
systemic velocity is low. Imposing this selection criterion
results in the selection of
stars at high Galactic latitudes that are more likely to be
Population~II stars. In contrast, RX~Lep and TX~Psc exhibit moderate and
moderately strong line-of-sight confusion,
respectively, but were included in our sample owing to
evidence for elongated \HI\ morphologies based on
single-dish measurements (G\'erard \& Le~Bertre 2006).
Finally, we
downloaded previously unpublished data for two stars (IK~Tau and
AFGL~3099) from the Very
Large Array (VLA) archive (see Table~2). Both of these stars were
previously targeted in single-dish \HI\ surveys but neither was detected
(Zuckerman et al. 1980; G\'erard \& Le~Bertre 2006).

%
\begin{deluxetable*}{llccccccccc}
\tabletypesize{\tiny}
\tablewidth{0pc}
\tablenum{2}
\tablecaption{Summary of VLA Observations}
\tablehead{
\colhead{Star} & \colhead{Obs. date} &
\colhead{No.} & \colhead{Mode} & \colhead{No.} 
& \colhead{Bandwidth}&  \colhead{$V_{1}$,$V_{2}$}  & \colhead{$V_{\rm cent}$}   &
\colhead{$\Delta\nu$} & \colhead{$\Delta v$} & \colhead{$t$} \\
\colhead{}     &  \colhead{} & \colhead{antennas} & \colhead{} &
\colhead{pol.} & \colhead{(MHz)} & \colhead{(\kms)}  & \colhead{(\kms)} & \colhead{(kHz)}
& \colhead{(\kms)} & \colhead{(hours)}   \\
\colhead{(1)} & \colhead{(2)} & \colhead{(3)} &
\colhead{(4)} & \colhead{(5)} & \colhead{(6)} & \colhead{(7)}
& \colhead{(8)} & \colhead{(9)} & \colhead{(10)} & \colhead{(11)} 
}

\startdata

IK Tau$^{*}$ & 1989-Dec-18    & 27 & 2AC& 2 & 0.39& $-4.6$,72.6   & 34.0 & 12.2
&  2.6 & 3.97\\

RX Lep & 2009-Dec-30/31 & 22 & 1A & 1 & 1.5 & $-135.3$,193.2 & 29.0 & 3.05
& 1.29 & 3.85\\ 
       & 2010-Jan-1/2   & 22 & 1A & 1 & 1.5 & $-135.3$,193.2 & 29.0 & 3.05
& 1.29 & 2.75\\ 

Y UMa & 2007-Apr-27 & 26 & 2AC & 2 & 0.78 & $-62.8$,100.8 & 19.0 &
                        3.05 & 0.64 & 3.35\\
      & 2007-Apr-28 & 26 & 2AC & 2 & 0.78 & $-62.8$,100.8 & 19.0 &
                        3.05 & 0.64 & 3.45\\

R Peg & 2007-Apr-8 & 26 & 2AC & 2 & 0.78 & $-81.8$,81.8 & 0.0 & 3.05 &
0.64 & 3.40\\
      & 2007-May-4 & 26 & 2AC & 2 & 0.78 & $-81.8$,81.8 & 0.0 & 3.05 &
0.64 & 3.42\\

Y CVn & 2010-Jan-1/2 & 22 & 1A & 1 & 1.5 & $-143.1$,183.5 & 21.1 &
3.05 & 1.29 & 5.10\\
      & 2010-Jan-2/3 & 21 & 1A & 1 & 1.5 & $-143.1$,183.5 & 21.1 & 
3.05 & 1.29 & 0.90\\ 

V1942 Sgr & 2009-Dec-5   & 23 & 1A & 1 & 1.5 & $-196.2$,132.2 & $-32.0$ & 3.05 &
1.29 & 2.93\\
          & 2009-Dec-7   & 22 & 1A & 1 & 1.5 & $-196.2$,132.2 & $-32.0$ & 3.05 &
1.29 & 2.88\\

AFGL~3099$^{*}$ & 1989-Dec-17   & 27 & 2AC& 2 & 0.19& 26.7,65.3     & 46.0    & 6.10 &
1.29 & 4.23 \\

TX Psc  & 2009-Dec-5/6 & 23 & 1A & 1 & 1.5 & $-151.3$,177.3 &13.0    & 3.05 &
1.29 & 0.80\\
        & 2009-Dec-7/8 & 22 & 1A & 1 & 1.5 & $-151.3$,177.3 &13.0    &3.05  & 
1.29 & 0.82\\
        & 2010-Jan-4/5 & 21 & 1A & 1 & 1.5 & $-151.3$,177.3 &13.0    &3.05  & 
1.29 & 4.00\\

\enddata

\tablecomments{Explanation of columns: (1) star name; (2) observing
  date; (3) number of available antennas; (4) VLA correlator mode; 
(5) number of circular polarizations;
(6) total bandwidth; (7) minimum and maximum 
LSR velocity covered by the observing band; (8)
  LSR velocity at band center; (9) channel width;
(10) effective velocity resolution of the data presented in this work;
  all data with the exceptions of Y~UMa and R~Peg were Hanning smoothed
  prior to analysis;
(11) total on-source integration time.}
\tablenotetext{*}{Archival data (see \S\ref{archival}).}

\end{deluxetable*}

%
\begin{deluxetable*}{lccccl}
\tabletypesize{\tiny}
\tablewidth{0pc}
\tablenum{3}
\tablecaption{Calibration Sources}
\tablehead{
\colhead{Source} & \colhead{$\alpha$(J2000.0)} &
\colhead{$\delta$(J2000.0)} & \colhead{Flux Density (Jy)} &
\colhead{$\nu_{0}$ (MHz)} & \colhead{Date}
}

\startdata

3C48$^{a}$  & 01 37 41.2994 & +33 09 35.132 & 15.8851$^{*}$ & 1419.5 &2007-Apr-8\\
            &               &               & 15.8675$^{*}$ & 1420.5 &2007-Apr-8\\
            &               &               & 15.8847$^{*}$ & 1419.5 &2007-May-4\\
            &               &               & 15.8671$^{*}$ & 1420.5 &2007-May-4\\
            &               &               & 15.8905$^{*}$ & 1418.9 & 2009-Dec-6\\
            &               &               & 15.8691$^{*}$ & 1421.3 & 2009-Dec-6\\
            &               &               & 15.8905$^{*}$ & 1418.9 & 2009-Dec-8\\
            &               &               & 15.8691$^{*}$ & 1421.3 & 2009-Dec-8\\
           &               &               & 15.8885$^{*}$ & 1419.1 & 2009-Dec-31\\
           &               &               & 15.8701$^{*}$ & 1421.5 & 2009-Dec-31\\
           &               &               & 15.8885$^{*}$ & 1419.1 & 2010-Jan-2\\
           &               &               & 15.8701$^{*}$ & 1421.5 & 2010-Jan-2\\     
            &               &               & 15.8868$^{*}$ & 1419.3 &2010-Jan-5\\
            &               &               & 15.8696$^{*}$ & 1421.2 & 2010-Jan-5\\

0022+002$^{b}$ & 00 22 25.42 & +00 14 56.16  & 2.92$\pm$0.06 & 1420.2&2009-Dec-6\\
               &             &              & 3.02$\pm$0.07 & 1420.2&2009-Dec-8\\
               &             &              & 2.73$\pm$0.05 & 1420.2&2010-Jan-2\\

0409+122$^{c}$ & 04 09 22.00 & +12 17 39.84 & 0.814$\pm$0.002&1420.7 & 1989-Dec-18\\

0459+024$^{d}$ & 04 59 52.05 & +02 29 31.17 & 1.72$\pm$0.04 & 1420.1 & 2009-Dec-31\\
               &             &             & 1.79$\pm$0.03 & 1420.1 & 2010-Jan-2\\

1227+365$^{e}$ & 12 27 58.72 & +36 35 11.82 & 1.97$\pm$0.04 & 1420.4 & 2010-Jan-2\\
               &             &             & 2.07$\pm$0.06 & 1420.4 & 2010-Jan-3\\

1252+565$^{f}$ &12 52 26.38 & +56 34 19.41 & 2.20$\pm$0.02  & 1420.3 & 2007-Apr-27\\
               &            &             & 2.22$\pm$0.01  & 1420.3 & 2007-Apr-28\\

3C286$^{g}$ & 13 31 08.2879 & +30 30 32.958 & 16.0394$^{*}$ & 1420.3&1989-Dec-18\\
            &               &              & 14.7317$^{*}$ & 1419.3&2007-Apr-27\\ 
            &               &              & 14.7218$^{*}$ & 1421.3&2007-Apr-27\\
            &               &              & 14.7317$^{*}$ & 1419.3&2007-Apr-28\\ 
            &               &              & 14.7218$^{*}$ & 1421.3&2007-Apr-28\\
            &               &               & 14.7305$^{*}$ & 1419.5 &2010-Jan-2\\
            &               &               & 14.7193$^{*}$ & 1421.8 &2010-Jan-2\\
            &               &               & 14.7305$^{*}$ & 1419.5 &2010-Jan-3\\
            &               &               & 14.7193$^{*}$ & 1421.8 &2010-Jan-3\\

1411+522$^{h}$ & 14 11 20.64 & +52 12 09.14 & 21.63$\pm$0.08 & 1419.3 &2007-Apr-27\\
               &             &             & 21.72$\pm$0.06 & 1421.3 &2007-Apr-27\\
               &             &             & 21.84$\pm$0.05 & 1419.3 &2007-Apr-28\\ 
               &             &             & 21.78$\pm$0.05 & 1421.3 &2007-Apr-28\\
               &             &             & 21.83$\pm$0.20 & 1420.4 &2010-Jan-2\\
               &             &             & 22.01$\pm$0.73 & 1420.4 &2010-Jan-3\\

1911-201$^{i}$ & 19 11 09.65 & $-$20 06 55.10 & 2.17$\pm$0.04 & 1420.5& 2009-Dec-5\\
               &             &                & 2.24$\pm$0.08 & 1420.5& 2009-Dec-7\\

2253+161$^{j}$  & 22 53 57.74 & +16 08 53.56 & 14.71$\pm$0.04  & 1419.5 &2007-Apr-8\\
                &             &             & 14.69$\pm$0.04  & 1421.5 &2007-Apr-8\\
                &             &             & 14.46$\pm$0.05  & 1419.5 &2007-May-4\\
                &             &             & 14.51$\pm$0.04  & 1421.5 &2007-May-4\\
                &             &             & 14.13$\pm$0.10 & 1418.9&2009-Dec-7\\
               &             &              & 13.42$\pm$0.17 & 1419.5&2010-Jan-4\\
               &             &              & 13.31$\pm$0.14 & 1421.8&2010-Jan-4\\

2255+132$^{k}$ & 22 55 03.73 & +13 13 33.01 & 2.65$\pm$0.01  &1420.5 &2007-Apr-8\\
               &             &             & 2.66$\pm$0.02 & 1420.5&2007-May-4\\

2330+110$^{l}$ & 23 30 40.85 & +11 00 18.70    & 1.117$\pm$0.003&1420.0&1989-Dec-17\\

\enddata

\tablecomments{Units of right ascension are hours, minutes, and
seconds, and units of declination are degrees, arcminutes, and
arcseconds. $\nu_{0}$ is the frequency at which the flux density in
the fourth column was computed.
For the archival data (\S\ref{archival}), the phase
center coordinates and phase calibrator names have been translated from
B1950.0 to J2000.0 coordinates.}
\tablenotetext{*}{Adopted flux density, computed according to the VLA
Calibration Manual (Perley \& Taylor 2003).}

\tablenotetext{a}{Primary flux calibrator and bandpass calibrator for
  RX~Lep, R~Peg, V1942~Sgr, and TX~Psc}
\tablenotetext{b}{Secondary gain calibrator for TX~Psc}
\tablenotetext{c}{Secondary gain calibrator for IK~Tau}
\tablenotetext{d}{Secondary gain calibrator for RX~Lep}
\tablenotetext{e}{Secondary gain calibrator for Y~CVn}
\tablenotetext{f}{Secondary gain calibrator for Y~UMa}
\tablenotetext{g}{Primary flux calibrator for IK~Tau, Y~UMa, Y~CVn,
  and AFGL~3099; bandpass
  calibrator for Y~UMa and Y~CVn}
\tablenotetext{h}{Bandpass calibrator for  Y~UMa and Y~CVn}
\tablenotetext{i}{Secondary gain calibrator for V1942~Sgr}
\tablenotetext{j}{Bandpass calibrator for R~Peg and TX~Psc}
\tablenotetext{k}{Secondary gain calibrator for R~Peg}
\tablenotetext{l}{Secondary gain calibrator for AFGL~3099}

\end{deluxetable*}


\section{Observations and Data Reduction\protect\label{observations}}
The data that we present here are a combination of new and archival
\HI\ 21-cm line observations obtained with the VLA of
the National Radio Astronomy Observatory (NRAO).\footnote{The National
Radio Astronomy Observatory is operated by Associated Universities,
Inc., under cooperative agreement with the National Science
Foundation.} All of the data were obtained 
using the most compact (D)
configuration (0.035-1.0~km baselines) of the VLA, providing
sensitivity to emission on scales of up to $\sim$15$'$. The
primary beam of the VLA at our observing
frequency is $\sim 31'$.  The data were reduced and imaged
using the Astronomical Image Processing System
(AIPS). Further details  are
described below.


%
\begin{deluxetable*}{lcccccccc}
\tabletypesize{\tiny}
\tablewidth{0pc}
\tablenum{4}
\tablecaption{Deconvolved Image Characteristics}
\tablehead{
\colhead{Source} & \colhead{Type} & \colhead{{$\cal R$}} & \colhead{Taper} &
\colhead{$\theta_{\rm FWHM}$} & \colhead{PA} &
\colhead{rms} & \colhead{Continuum velocities} & \colhead{Clean Boxes?}\\
\colhead{}  & \colhead{}   & \colhead{}  & \colhead{(k$\lambda$,k$\lambda$)}
& \colhead{($''\times ''$)} & \colhead{(degrees)} & \colhead{(mJy
beam$^{-1}$)} & \colhead{(\kms)} & \colhead{}\\
\colhead{(1)} & \colhead{(2)} & \colhead{(3)} &
\colhead{(4)} & \colhead{(5)} &
\colhead{(6)} & \colhead{(7)} & \colhead{(8)} & \colhead{(9)}
}

\startdata
IK~Tau & line & +5 & ... & \as{69}{2}$\times$\as{51}{7} & 32.1
& 1.4 & 57.2 to 72.6 & no \\

IK~Tau & cont. & +1 & ... & \as{69}{2}$\times$\as{51}{7} & 32.1
& 0.61 &  57.2 to 72.6 & yes \\

RX Lep & line & +5 & ... & \as{76}{2}$\times$\as{47}{9} & 10.6 & 2.0 & 
$-37.3$ to $-121.0$; 34.8 to 168.7 & yes \\

RX Lep & cont. & +1 & ... & \as{67}{5}$\times$\as{43}{8}& 10.7
& 0.4 & $-37.3$ to $-121.0$; 34.8 to 168.7 & yes \\

Y UMa$^{a}$ & line & +1 & ... & \as{61}{2}$\times$\as{52}{1} &51.2 & 1.7 &
$-41.5$ to $-15.1$; 0.96 to 12.5; 24.2 to 88.6 & yes\\ 

Y UMa$^{a}$ & line & +5 & ... & \as{72}{1}$\times$\as{58}{1} &54.9 & 1.7 &
$-41.5$ to $-15.1$; 0.96 to 12.5; 24.2 to 88.6 & yes\\ 

Y UMa & cont. & +1 & ... & \as{61}{7}$\times$\as{52}{4} &51.4 & 0.18 &
$-41.5$ to $-15.1$; 0.96 to 12.5; 24.2 to 88.6 & yes\\ 

R Peg$^{a}$ & line & +5 & ... & \as{62}{9}$\times$\as{56}{9} & 60.9 & 1.7 &
$-68.9$ to $-32.8$; 35.4 to 57.3  & no\\

R Peg & cont. & +1 & ... & \as{56}{9}$\times$\as{67}{2} & 67.2 & 0.24
& $-68.9$ to $-32.8$; 35.4 to 57.3 & yes\\

Y~CVn & line & +1 & ... & \as{53}{6}$\times$\as{44}{4} & 30.1 & 2.2 &
$-125.1$ to $-25.9$; 57.8 to 100.3; 124.8 to 150.5 & yes\\

Y~CVn & line & +5 & ... & \as{60}{9}$\times$\as{48}{6} & 32.8 & 2.1 &
$-125.1$ to $-25.9$; 57.8 to 100.3; 124.8 to 150.5 & yes\\

Y~CVn & line & +5 & 2,2 & \as{106}{9}$\times$\as{83}{5} & 38.6 & 2.6 &
$-125.1$ to $-25.9$; 57.8 to 100.3; 124.8 to 150.5 & yes\\

Y~CVn & cont. & +1 & ... & \as{53}{6}$\times$\as{44}{4} & 30.0 & 0.48
& $-125.1$ to $-25.9$; 57.8 to 100.3; 124.8 to 150.5 & yes\\

V1942 Sgr & line & +5 & ... & \as{89}{5}$\times$\as{49}{4} & 12.8 &
2.7 & $-175.6$ to $-50.6$ & yes\\

V1942 Sgr & cont. & +1 & ... & \as{79}{6}$\times$\as{45}{7} & 13.8 &
0.48 & $-175.6$ to $-50.6$ & yes\\

AFGL~3099 & line & +5 & ... & \as{63}{6}$\times$\as{51}{9} & 15.6& 1.7
& 26.7 to 39.5; 52.4 to 65.3 & no \\

AFGL~3099 & cont. & +1 & ... & \as{56}{4}$\times$\as{47}{9} & 13.8& 0.45
& 26.7 to 39.5; 52.4 to 65.3 & yes \\

TX~Psc & line & +5  & ...  & \as{71}{6}$\times$\as{48}{8} & 5.0 &2.0 &
$-130.6$ to $-25.0$; 75.5 to 150.2 & no \\

TX~Psc & line & +5  & 2.2,2.2  & \as{104}{2}$\times$\as{78}{6} & 4.2 &2.2 &
$-130.6$ to $-25.0$; 75.5 to 150.2 & no \\

TX~Psc & cont. & +1 & ... & \as{63}{6}$\times$\as{45}{6} & 6.9 & 0.29 &
  $-130.6$ to $-25.0$; 75.5 to 150.2 & yes\\

\enddata

\tablenotetext{a}{Data were not Hanning smoothed for the present analysis.}

\tablecomments{Explanation of columns: (1) target name; (2) indication
  of whether the image contains line or continuum emission; the
  continuum images comprise a single spectral channel representing an
  average of the line-free portions of the band; (3) AIPS robust
parameter used in image deconvolution; $\cal R$=+5 is comparable to
natural weighting; (4) Gaussian taper applied in $u$ and
$v$ directions, expressed as
distance to 30\% point of Gaussian in units of kilolambda;
(5) FWHM dimensions of
synthesized beam; (6) position angle of synthesized beam (measured
east from north); (7) rms
noise per channel (1$\sigma$); (8) LSR velocities of spectral channels used for continuum
subtraction (line data) or that were averaged to compute a continuum image;
(9) indication of whether or not clean boxes were used during
image deconvolution. }

\end{deluxetable*}


\subsection{New VLA Observations\protect\label{newdata}}
Our new VLA observations for six stars 
were obtained throughout the transition period during
which the original VLA antennas were being 
retrofitted as part of the Expanded Very Large Array (EVLA)
upgrade. Throughout this period, the array contained fewer than the nominal
27 antennas (Table~2). 
As described below, 
the use of this ``hybrid'' array necessitated special care during the
calibration  process.

For the observations of R~Peg and Y~UMa,
the VLA correlator was used in dual polarization (2AC) 
mode with a 0.78~MHz bandpass, yielding 256 spectral
channels with 3.05~kHz spacing, providing an effective velocity
resolution of $\sim$0.64~\kms.  
For V1942~Sgr, RX~Lep, TX~Psc, and Y~CVn, the VLA correlator was used
in 1A mode with a 1.5~MHz bandwidth and 512 spectral channels. This mode
provided a single right circular (RR) polarization and a 3.05~kHz
channel spacing. 

The band center velocities
adopted for each star are summarized in Table~2. 
In some cases, the band center was offset from
the systemic velocity of the star 
to shift strong Galactic emission away from the band edges. 
Total integration times spent on each target
are also summarized in Table~2, along with additional details related to the
observations. 

The observations of the target stars were interspersed with observations
of a complex gain calibrator approximately every 20-25
minutes. Either 
3C48 (0137+331) or 3C286 (1331+305) was used as an absolute flux calibrator, and
in all cases, an
additional strong point source  was observed as a bandpass 
calibrator (see Table~3). To 
insure that the absolute flux scale and bandpass calibration were not
corrupted by Galactic emission in the band, the
flux and bandpass calibrators were each observed twice, with
the band shifted by sufficient positive and negative frequency offsets
so as to shift Galactic \HI\ emission out of the band. 
\HI\ survey spectra from Kalberla et al. (2005) were used to determine
the
required frequency offsets, and the adopted center frequencies
are summarized in Table~3. Each of the gain calibrators was
also observed once at each of these offset frequencies to
insure accurate bootstrapping of the absolute 
flux scale. We estimate that the resulting flux density scale for each
target star has
an uncertainty of
$\lsim$10\% (see also \S\ref{imaging}). 

\subsubsection{Calibration of the R~Peg and Y~UMa 
Observations\protect\label{rpegandyuma}}
Traditionally, the calibration of VLA data sets employs a
``channel 0'' data set formed by taking a vector average of the inner
75\% of the observing band. However, the
mismatch in bandpass response between the VLA and EVLA
antennas
causes closure errors in the channel~0 data computed this
way.\footnote{http://www.vla.nrao.edu/astro/guides/evlareturn/postproc/}
Furthermore, the hardware used to convert the digital signals from the
EVLA antennas into analog signals for the VLA correlator caused
aliased power in the bottom 0.5~MHz of the baseband (i.e., at low-numbered
channels).\footnote{http://www.vla.nrao.edu/astro/guides/evlareturn/aliasing/}
This aliasing is of consequence because of the narrow bandwidth used for
our observations. However, it affects EVLA-EVLA baselines only, as it does not
correlate on ELVA-VLA baselines.

To mitigate the above effects,
we used the following modified approach to the gain calibration for
the R~Peg and Y~UMa data sets.
After applying the latest available corrections to the
antenna positions and performing an initial excision of corrupted data,
we  computed and applied an initial bandpass calibration to our
spectral line data to remove closure errors on
VLA-EVLA baselines. EVLA-EVLA baselines were flagged during this step
and unflagged afterward. After applying the bandpass solutions,
we then computed
a new frequency-averaged (channel~0) data set (using the inner 75\% of
the band) for use in calibrating
the frequency-independent complex
gains.  
After flagging the EVLA-EVLA baselines on the channel 0 data, we solved 
for the frequency-independent portion of the complex
gain and transferred these solutions to the full spectral line data
set.  Lastly, we applied time-dependent frequency
shifts to the data to compensate for changes caused by
the Earth's motion during the course of the observations.

\subsubsection{Calibration of the RX~Lep, Y~CVn, V1942~Sgr, and TX~Psc Observations}
During the observations of RX~Lep, Y~CVn, V1942~Sgr, and TX~Psc, there
were too few non-EVLA antennas left in the array to utilize the
calibration strategy described in \S\ref{rpegandyuma}.  Instead, we 
increased our observing bandwidth by a factor of two
(at the cost of limiting ourselves to a single polarization product)
to provide a suitable amount of bandwidth uncorrupted by aliasing. We
then employed the following calibration approach.

After applying the latest available antenna position corrections 
and performing an initial excision of corrupted data,
we  computed and applied a bandpass calibration to the
spectral line data to remove closure errors on
VLA-EVLA baselines. All baselines were used in the calculation, but 
the bandpass was normalized using only channels
163 through 448, thus excluding the portion of the band affected by aliasing.
We next computed
a frequency-averaged channel~0 data set for use in calibrating
the frequency-independent complex gains,  using a vector average of channels 163
through 448.
Following gain calibration, we
applied time-dependent frequency
shifts to the data to compensate for changes caused by
the Earth's motion during the course of the observations. Lastly, 
we applied Hanning smoothing in velocity and
discarded every other channel, yielding a 256 channel data set with an effective
velocity resolution of $\sim$1.3~\kms.

Issues affecting two stars merit special comment.
The data for RX~Lep were heavily contaminated by narrow-band radio frequency
interference (RFI), necessitating considerable channel-dependent
flagging. This resulted in a loss of continuum sensitivity, although 
fortunately, little RFI was present near the
systemic velocity of the star.
In the case of V1942~Sgr,
a portion of the data taken on 2009 December 7 between
23:07:25~UT and 23:25:46~UT were affected by a bug in the VLA archive
software that resulted in the frequencies assigned to each of the
spectral channels being mirror reversed relative to their respective
channel numbers. The affected data were corrected using the AIPS task
{\small\sc FLOPM}.

\subsubsection{Continuum Subtraction}
Prior to imaging the line data for all of the newly observed 
target stars, we performed a continuum
subtraction using a linear fit to the 
real and imaginary
components of the visibilities using the AIPS task {\small\sc UVLIN}. 
The portions of the band that were
determined to be line-free channels and were used for these
fits are summarized in Column~7 of Table~4. 
Although the spectral shape of the aliased portion of the
continuum as measured toward our continuum calibrators was 
better approximated by a higher order polynomial,
there was generally insufficient continuum signal in our target fields
to adequately constrain higher order fits. 

\subsection{Archival VLA Data\protect\label{archival}}
\HI\ data for two stars (AFGL~3099 and IK~Tau) 
were retrieved from the VLA archive. These data were originally taken
as part of program AK237  
in late 1989 (Principal Investigator: G. Knapp) and to our knowledge are
previously unpublished.

The data for IK~Tau comprise 32 frequency channels with a channel spacing of
12.2~kHz (2.6~\kms) after on-line Hanning smoothing, resulting in
a  total bandwidth of 0.39~MHz
($\sim$82~\kms) in dual circular polarizations. The
AFGL~3099 data comprise 32 frequency channels with a channel spacing of
6.1~kHz (1.3~\kms) after on-line Hanning smoothing, providing 
a  total bandwidth of 0.195~MHz
($\sim$41~\kms), also with dual circular polarization.
Additional details are provided in Table~2. 

We processed both of the archival data sets in a similar manner. After
flagging corrupted data, we used a channel~0 data set
to calibrate the antenna-based, frequency independent complex 
gains. The absolute flux level was established using observations of
3C48 (0137+331). We adopted the VLA-determined flux density values for
3C48 from the year 1990, as calculated by the AIPS task {\small\sc SETJY}
(see Table~3).

Because the 3C48 observations
suffered from \HI\ absorption across much
of the band, we estimate that 
the uncertainty in the absolute flux scale for these data
may be as high as 
$\sim$20\%. This \HI\ absorption also rendered 3C48 unsuitable for
use as a bandpass calibrator, hence no bandpass calibration was applied to 
the archival data. This had minimal impact on the data quality
given the narrow bandwidths and the fact that the channels at the 
band edges had already been discarded by the VLA on-line system.

To create a continuum-subtracted spectral data set for each of the archival
sources, we produced a clean
image of the continuum using an average of the line-free channels (see
Table~4) and then
subtracted the clean components from the line data in the visibility
domain using the AIPS task {\small\sc UVSUB}.

\subsection{Imaging the VLA Data\protect\label{imaging}}
\subsubsection{Continuum Data\protect\label{contimaging}}
We produced an image of the 21-cm continuum
emission in the field of each target star using the line-free portion
of the
band (see Table~4). No continuum emission was detected from any of the
stars. 
For the six newly observed target fields (\S\ref{newdata}), 
we compared the measured flux densities of several continuum
sources in each field with values measured from NRAO VLA Sky Survey
(NVSS) 
images (Condon
et al. 1998). In all cases we found agreement to within better than
10\%. For each target, we also searched for \HI\ absorption against the
brightest continuum sources in the field surrounding each star, but in no case
did we detect
statistically significant absorption.

\subsubsection{Line Data}
We imaged the VLA \HI\ line data (both new and archival) 
using the standard AIPS CLEAN deconvolution 
algorithm. For each target, we produced data cubes using various 
weighting  schemes. The characteristics of those used for the present
analysis are summarized in Table~4. 


\section{Results\protect\label{results}}
In \S\ref{iktau}-\ref{txpsc}
we present the results of our most recent \HI\ imaging study of
AGB stars. In six cases
we have unambiguously detected \HI\ emission
associated with the circumstellar envelopes of the target
stars. For the other two targets we derive upper limits on the \HI\
content of
their CSEs. We discuss the VLA data for each star and their implications for
understanding the mass loss history of each star
in the following subsections. In \S5
we discuss emerging trends from the entire sample of AGB and related
stars that has been imaged in \HI\ to date.

\begin{figure*}
\vspace{-1.5cm}
\centering
\scalebox{0.8}{\rotatebox{-90}{\includegraphics{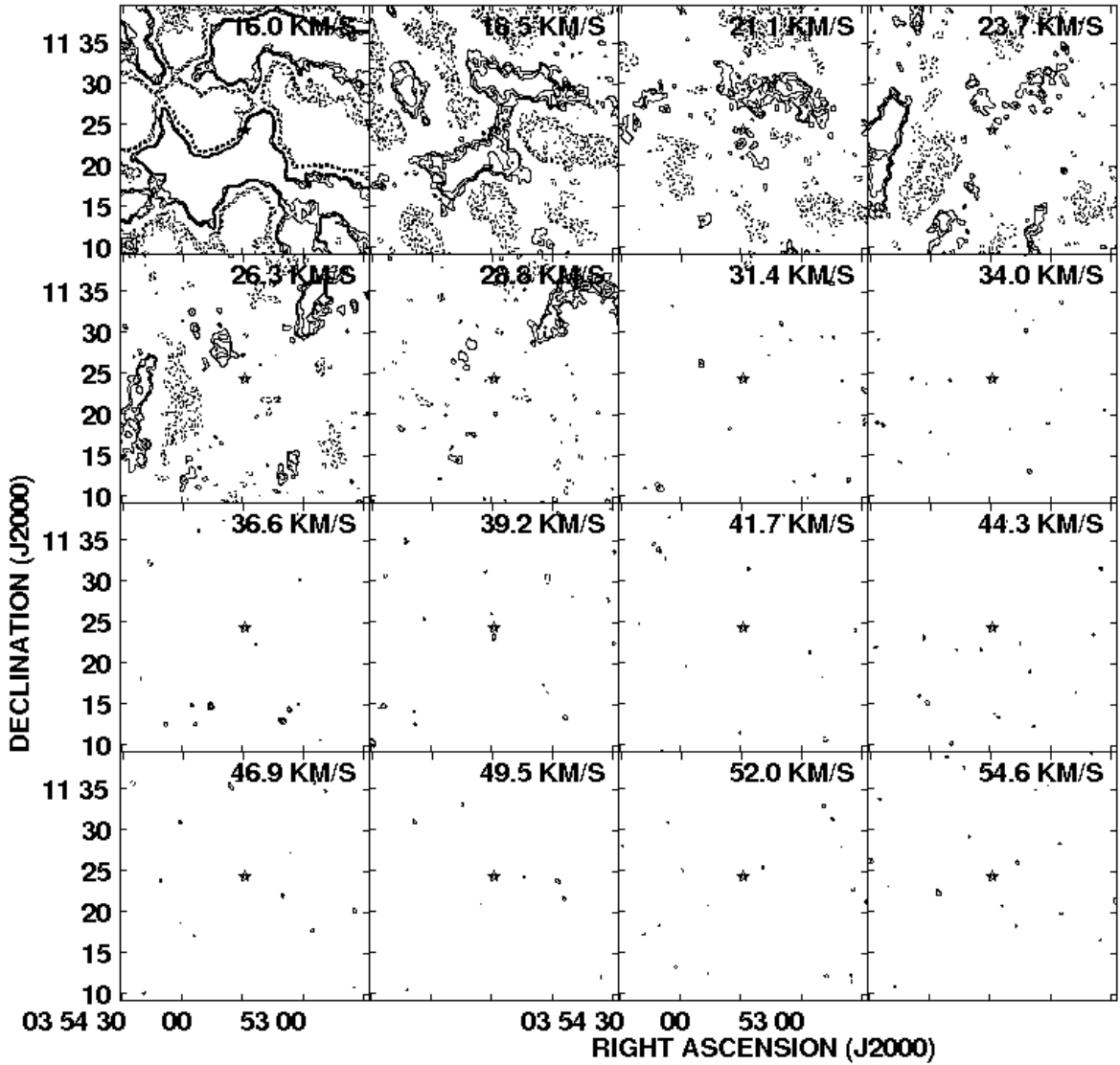}}}
\caption{\HI\ channel maps bracketing the systemic velocity of IK~Tau,
  derived from naturally weighted VLA data. The spatial resolution is $\sim69''\times52''$.
A star symbol indicates the stellar position. 
Contour levels are 
$(-8.4,-6,-4.2,-3,3...8.4)\times$1.4~mJy beam$^{-1}$. The lowest
contour is $\sim3\sigma$ and negative contours are shown as dashed lines.  The
field-of-view shown is comparable to the VLA primary beam. 
  }
\label{fig:iktaucmaps}
\end{figure*}

\begin{figure}
\centering
\scalebox{0.5}{\rotatebox{0}{\includegraphics{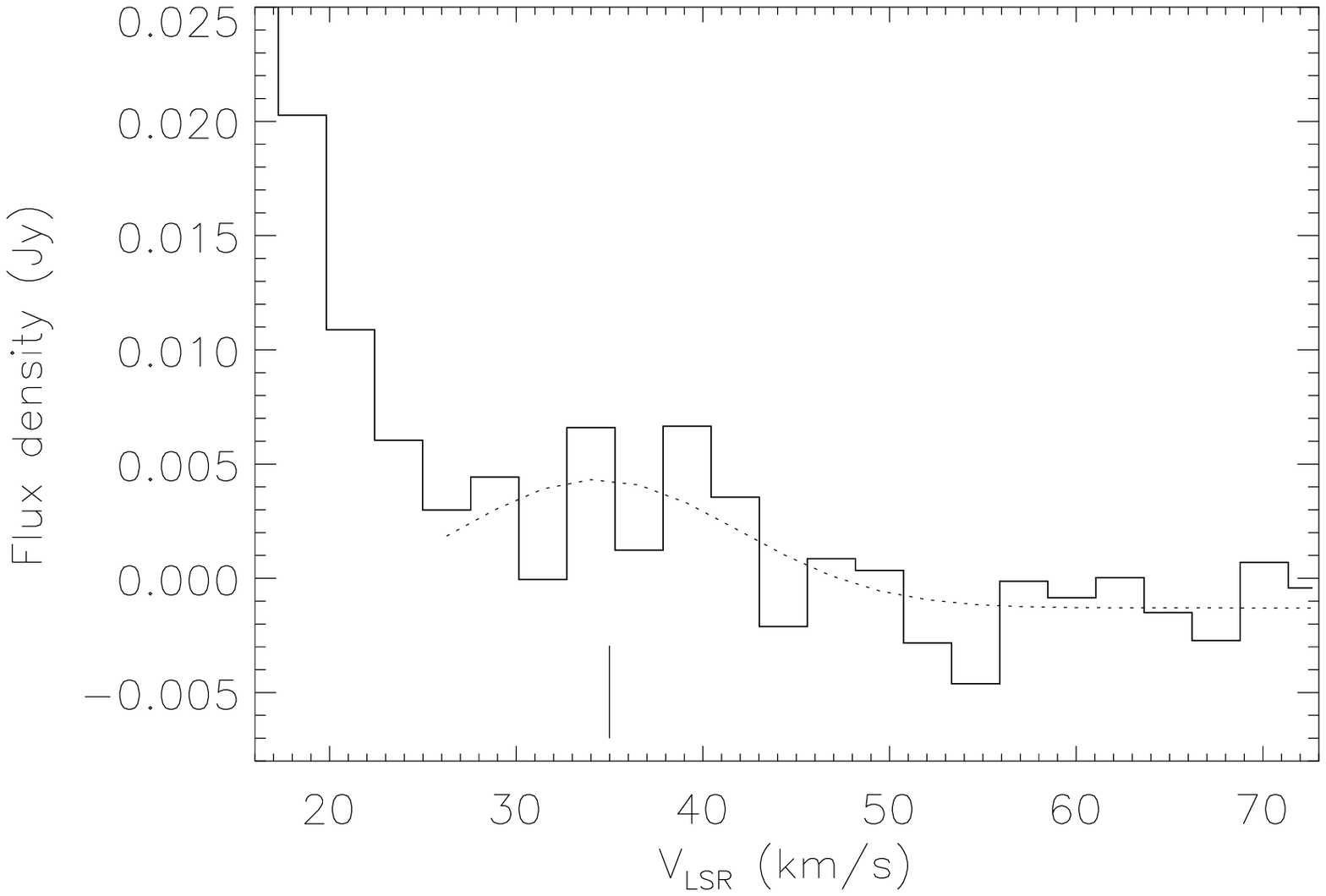}}}
\caption{Spatially integrated \HI\ spectrum of IK~Tau
  derived from naturally weighted VLA data. 
The flux density in each channel was derived by integrating within a
  circular aperture with radius $85''$ (see \S\ref{iktauvla}). 
The vertical bar indicates the
  systemic velocity of the star derived from CO 
observations. Spectral channels with $V_{\rm
  LSR}\lsim26$~\kms\ are contaminated by Galactic emission along the
  line-of-sight. The dotted line is a Gaussian fit to the data over
  the velocity range
  $26.3 \le V_{\rm LSR} \le 72.4$~\kms; we consider this  
  feature to be of marginal significance (see \S\ref{iktauvla}).}
\label{fig:iktauspec}
\end{figure}

\subsection{IK Tauri (IK Tau)\protect\label{iktau}}
IK Tauri (IK~Tau=NML~Tau) is an extremely red, large-amplitude Mira
variable with an oxygen-rich chemistry and a period of 462 days
(Le~Bertre 1993).  
The large amplitude of its variability corresponds to
variations in its effective temperature from
$\sim$2000~K to 3000~K (Dyck et al. 1974; G\'erard \& Le~Bertre
2006). Following Le~Bertre \& Winters (1998), we adopt a distance to the
star of 245~pc based on the period-luminosity relation of Feast (1996).

IK~Tau has been extensively studied at multiple
wavelengths and is known to have a rich molecular chemistry (e.g.,
Milam et al. 2007; Decin et al. 2010; Claussen et al. 2011). Based on
CO(1-0) observations, Knapp \& Morris (1985) derived a stellar systemic
velocity of $V_{\star, \rm
  LSR}=34.6\pm0.4$~\kms, a mass-loss rate ${\dot
  M}=4.2\times10^{-6}~M_{\odot}$~yr$^{-1}$ (scaled to our adopted
distance), and an outflow velocity $V_{\rm out}=22.0\pm0.6$~\kms. 

\subsubsection{VLA Results for IK~Tau\protect\label{iktauvla}}
\HI\ channel maps bracketing the systemic velocity of IK~Tau  are
presented in Figure~\ref{fig:iktaucmaps}. The velocity range shown
roughly corresponds to the 
velocities over which CO emission has been previously detected  in the
CSE.
We find no \HI\
emission features with $\ge3\sigma$ significance coincident with the stellar
position in any of the spectral channels with $V_{\rm LSR}\ge23.7$~\kms.
For velocities 
$V_{\rm LSR}\le28.7$~\kms, the data suffer from confusion from line-of-sight
Galactic emission that is poorly spatially sampled by the VLA, 
producing a ``mottled'' appearance in the channel images. There is no
evidence that any of the emission seen at these velocities is related
to the CSE of IK~Tau.

In Figure~\ref{fig:iktauspec} we present a spatially integrated \HI\
spectrum derived by summing each plane of 
the spectral line data cube within a circular aperture of
radius $85''$, centered on the star. This aperture was chosen to match
the extent of the FIR-emitting shell detected by Cox et al. (2012a; see
\S\ref{iktaudisc}). Consistent with Figure~\ref{fig:iktaucmaps}, 
the resulting \HI\ spectrum is contaminated by Galactic
emission at velocities $V_{\rm LSR}\lsim24$~\kms. However, near the stellar systemic
velocity, we see marginal evidence 
for a weak, broad spectral feature that is 
distinct from the Galactic emission component at lower velocities. 
Using the spectral
data spanning $26.3 \le V_{\rm LSR} \le 72.4$~\kms, 
we performed a Gaussian fit to this feature. The best-fit Gaussian
is overplotted on Figure~\ref{fig:iktauspec}; it has a peak amplitude of
5.6$\pm$1.5~mJy, a velocity centroid of 34.4$\pm$2.3~\kms, a
FWHM velocity width of 17.6$\pm$2.9~\kms, and a DC offset term of
$-0.0013\pm$0.0008~Jy. The peak amplitude of the Gaussian
is roughly 3.5 times the RMS noise in the line-free portion of
the spectrum, and the velocity centroid agrees with 
the stellar systemic velocity derived from CO observations. 
Despite this, 
we consider this feature to be of marginal significance given the 
combination of several factors. The blend with the Galactic emission
on the blue edge of the profile makes our Gaussian fit parameters
more uncertain than the formal errors would indicate, 
and the emission giving rise to this possible spectral
feature cannot be
unambiguously associated with the CSE of IK~Tau in any of the individual
channel images. Finally,  our data cube contains too few line-free
channels to  permit a robust characterization of 
the intrinsic instrumental band shape. 
Deeper \HI\ imaging observations with higher spectral resolution and a
wider band could help to overcome these issues.

For the present analysis
we use the integral of the Gaussian fit in
Figure~\ref{fig:iktauspec}
to provide a $\sim3\sigma$ upper limit to the integrated \HI\ flux density from the
CSE of IK~Tau within a
85$''$ (0.10~pc) radius of
$\int S_{\rm HI}~d\nu<$0.10~Jy~\kms. This translates to an 
\HI\ mass of $M_{\rm
  HI}\le 1.4\times10^{-3}M_{\odot}$ at our adopted distance.

\subsubsection{Discussion of IK~Tau Results\protect\label{iktaudisc}}
IK~Tau was previously targeted in single-dish \HI\ surveys by
Zuckerman et al. (1980) and G\'erard \& Le~Bertre (2006) but was not
detected.
IK~Tau is a challenging target for single-dish studies owing to 
the strong line-of-sight contamination across the velocity range $V_{\rm
  LSR}\approx$$-40$ to 40~\kms, which encompasses the stellar systemic
velocity (see 
G\'erard \& Le~Bertre 2006). 

With the VLA, we almost entirely filter out the contaminating emission for
$V_{\rm LSR}\ge$31.4~\kms, allowing us to place more
robust constraints on the total amount of atomic hydrogen in
the CSE. However, we note some important caveats to the \HI\ mass upper
limit that we derived in the previous section. First, the
rather coarse spectral resolution employed for the IK~Tau observations
($\sim$2.6~\kms) may have diminished our sensitivity to emission components
with narrow velocity widths ($\Delta V\sim1$~\kms). 
Such components may be present even when the global
(spatially integrated) \HI\ line width is several times
larger. A second caveat is that 
the gas temperature in CSEs is expected to depend
strongly on the mass-loss rate (Jura et al. 1988; Sahai 1990), with
cooler CSE
temperatures expected for stars with higher values of ${\dot M}$. Thus a
significant fraction of the stellar wind may be molecular (Glassgold
\& Huggins 1983) and most of the remaining fraction of the atomic
hydrogen in the CSE of IK~Tau may be simply too cold to detect in emission.
This argument
may also be relevant to the case of AFGL~3099
(\S\ref{crl3099}), another star that remains undetected in \HI\
emission despite its high mass loss rate.

Recent {\it Herschel} imaging at 70$\mu$m and 160$\mu$m by Cox
et al. (2012a) revealed 
that IK~Tau is surrounded by extended infrared emission with radius
$\sim85''$ that appears to be consistent with a bow shock and related
structures.  The total mass
of the CSE estimated from these FIR measurements is $\sim 0.037M_{\odot}$ (assuming a
gas-to-dust ratio of 200; Cox et al. 2012b), roughly 20 times higher
than the upper limit we
derive from our \HI\ measurements (after scaling by a factor of
1.4 to account for the mass of helium).

\subsection{RX Leporis (RX Lep)\protect\label{rxlep}}
RX~Leporis (RX~Lep) is an oxygen-rich AGB star that was
reclassified by Samus et al. (2004) from an irregular variable to a semi-regular variable of
type SRb. We adopt here a distance of 149~pc based on the {\it
  Hipparcos} parallax of 6.71$\pm$0.44~mas 
(van~Leeuwen 2007). 

Libert et al. (2008) presented CO(2-1) observations for RX~Lep from
which they derived a stellar systemic velocity  of $V_{\star,\rm
  LSR}=28.9\pm$0.1~\kms, a mass-loss rate ${\dot
  M}\approx2.0\times10^{-7}~M_{\odot}$~yr$^{-1}$ (scaled to our adopted
distance), and an outflow velocity
for the wind $V_{\rm out}=4.2\pm0.1$~\kms. The proper motions measured by {\it Hipparcos} are
31.76$\pm$0.58~mas yr$^{-1}$ in right ascension and 56.93$\pm$0.50~mas
yr$^{-1}$ in declination, implying a peculiar space motion of
$V_{\rm space}\approx$56.6~\kms\ along a position angle of \ad{24}{6}
(see Table~5).

RX~Lep has an effective temperature of $\sim$3300~K 
(Dumm \& Schild 1998),
and using the bolometric correction from 
Le~Bertre et al. (2001), its luminosity
is 5300~$L_{\odot}$ at our adopted distance (Libert et al. 2008). 
The initial mass of RX~Lep is believed to
be in the range 2.5-4.0~$M_{\odot}$ (Mennessier et al. 2001). As discussed further below,
RX~Lep was previously detected in \HI\ by Libert et al. (2008) using
single-dish observations from the NRT. 

Technetium lines have not been detected in RX~Lep
(Little et al. 1987; Lebzelter \& Hron 1999), implying that either RX~Lep is
on the early AGB (E-AGB) or that it is still near the beginning of the
thermal pulsing (TP) AGB
phase and has not yet undergone dredge-up
(e.g., Straniero et al. 1997). The  
mass-loss rate and wind outflow speed for the star are also consistent with
values expected near the beginning of the TP-AGB
(Vassiliadis \& Wood 1993). 
As described below
(\S\ref{rxlepdisc}), placement of RX~Lep near the start of the 
TP-AGB is further supported by our new \HI\ observations. 

\begin{figure*}
\centering
\scalebox{0.75}{\rotatebox{-90}{\includegraphics{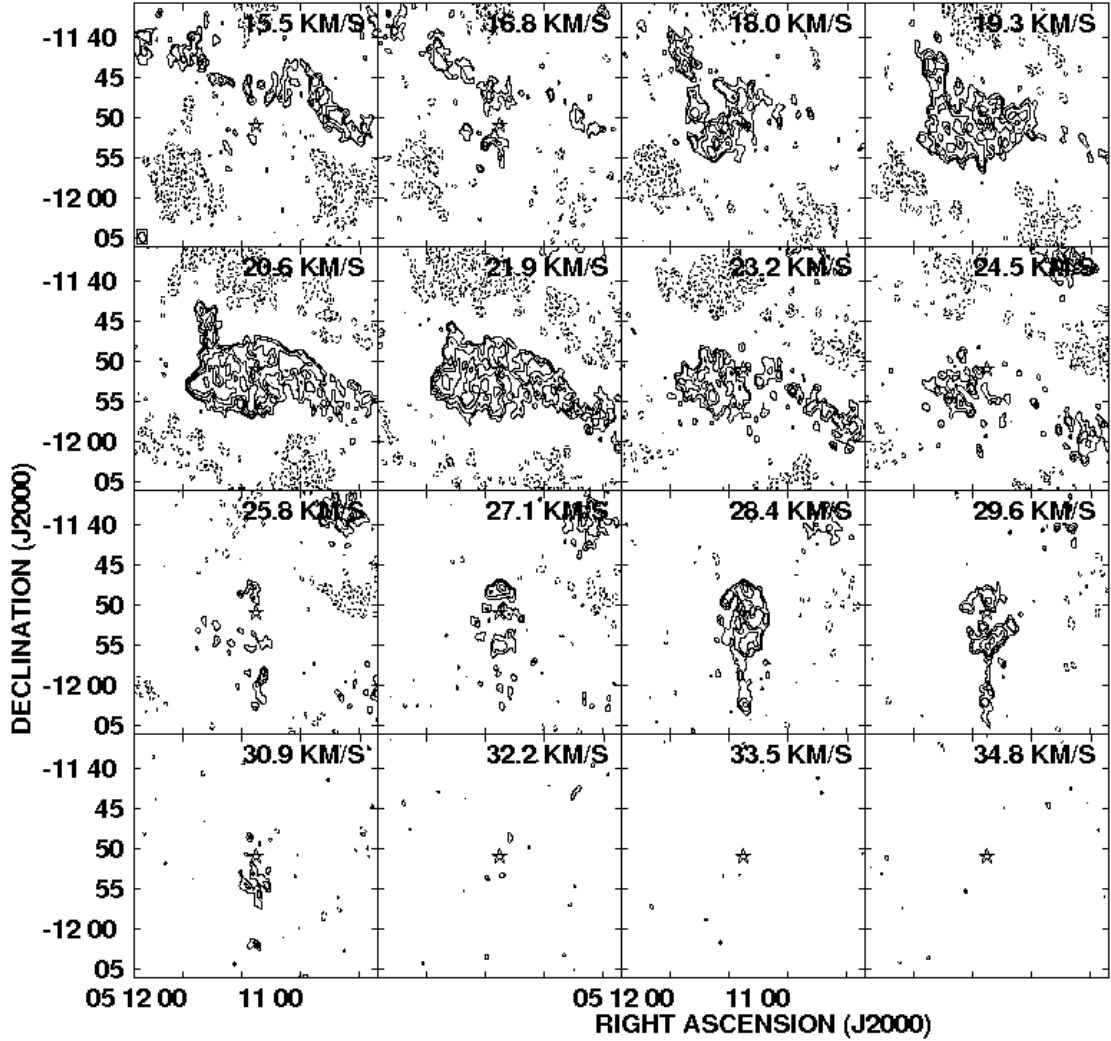}}}
\caption{\HI\ channel maps bracketing the systemic velocity of RX~Lep, derived
from naturally weighted VLA data. The spatial resolution is $\sim76''\times48''$.
A star symbol indicates the stellar position.
Contour levels are 
($-6$[absent],$-$4.2,$-$3,3,4.2,6,8.4,12)$\times$2.0~mJy beam$^{-1}$. The lowest
contour is $\sim3\sigma$. The channels shown correspond
to the range of LSR velocities over which CO(2-1) emission
was detected by Libert et al. 2008. The
field-of-view displayed is comparable to the VLA primary beam. 
  }
\label{fig:rxlepcmaps}
\end{figure*}
\begin{figure*}
\centering
\scalebox{0.75}{\rotatebox{0}{\includegraphics{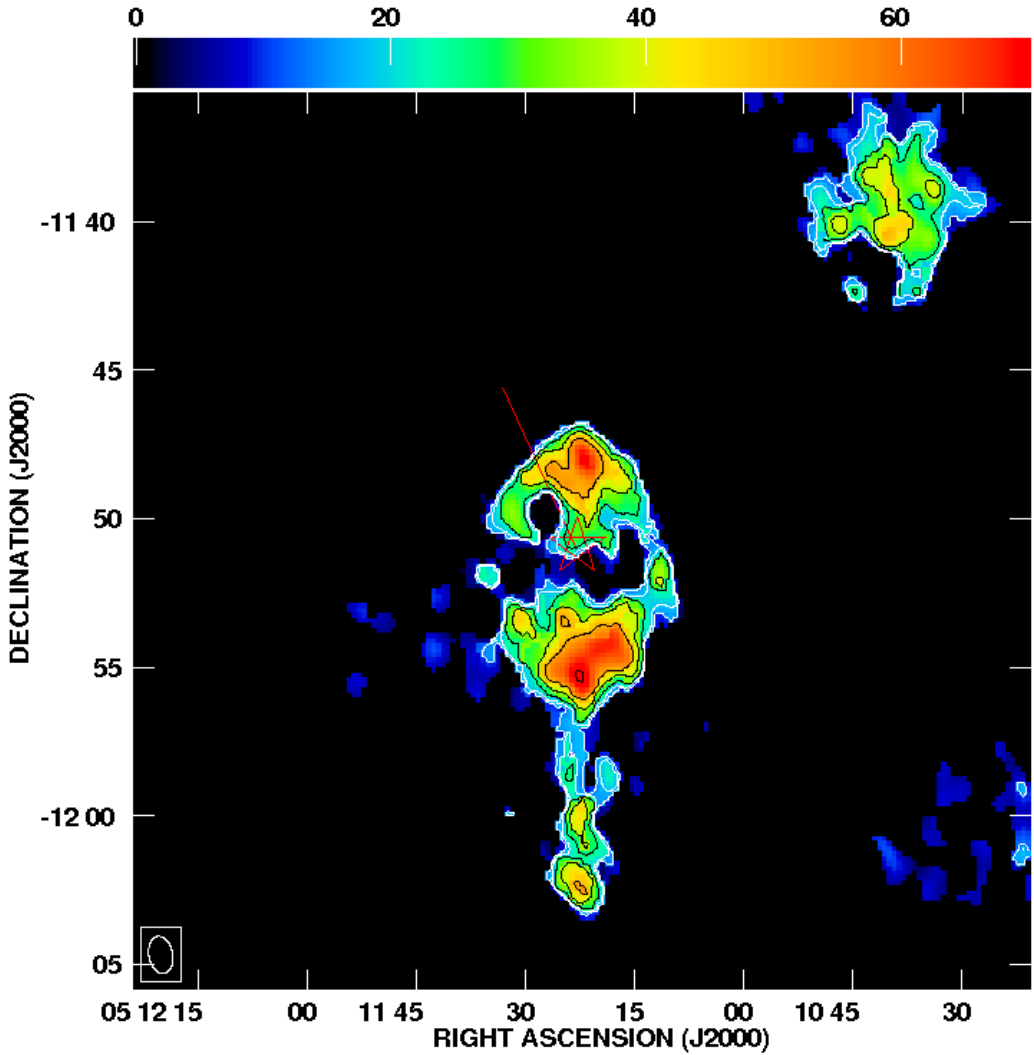}}}
\caption{\HI\ total intensity 
map of RX~Lep derived from naturally weighted data.  
The spatial resolution is $\sim76''\times48''$.  The 
map was constructed from emission spanning LSR velocities
25.8 to 32.2~\kms. The red line indicates the direction
of space motion of the star. Intensity levels are 0 to 70 Jy beam$^{-1}$
m s$^{-1}$ and contour levels are
(1.8,3.5,3.5,7,10)$\times$7.3~Jy beam$^{-1}$ m s$^{-1}$. To minimize noise in the
map, data at a given point were blanked if they did
not exceed a 2.5$\sigma$
threshold after smoothing by a factor of three
spatially and spectrally. }
\label{fig:rxlepmom0}
\end{figure*}
\begin{figure}
\scalebox{0.5}{\rotatebox{0}{\includegraphics{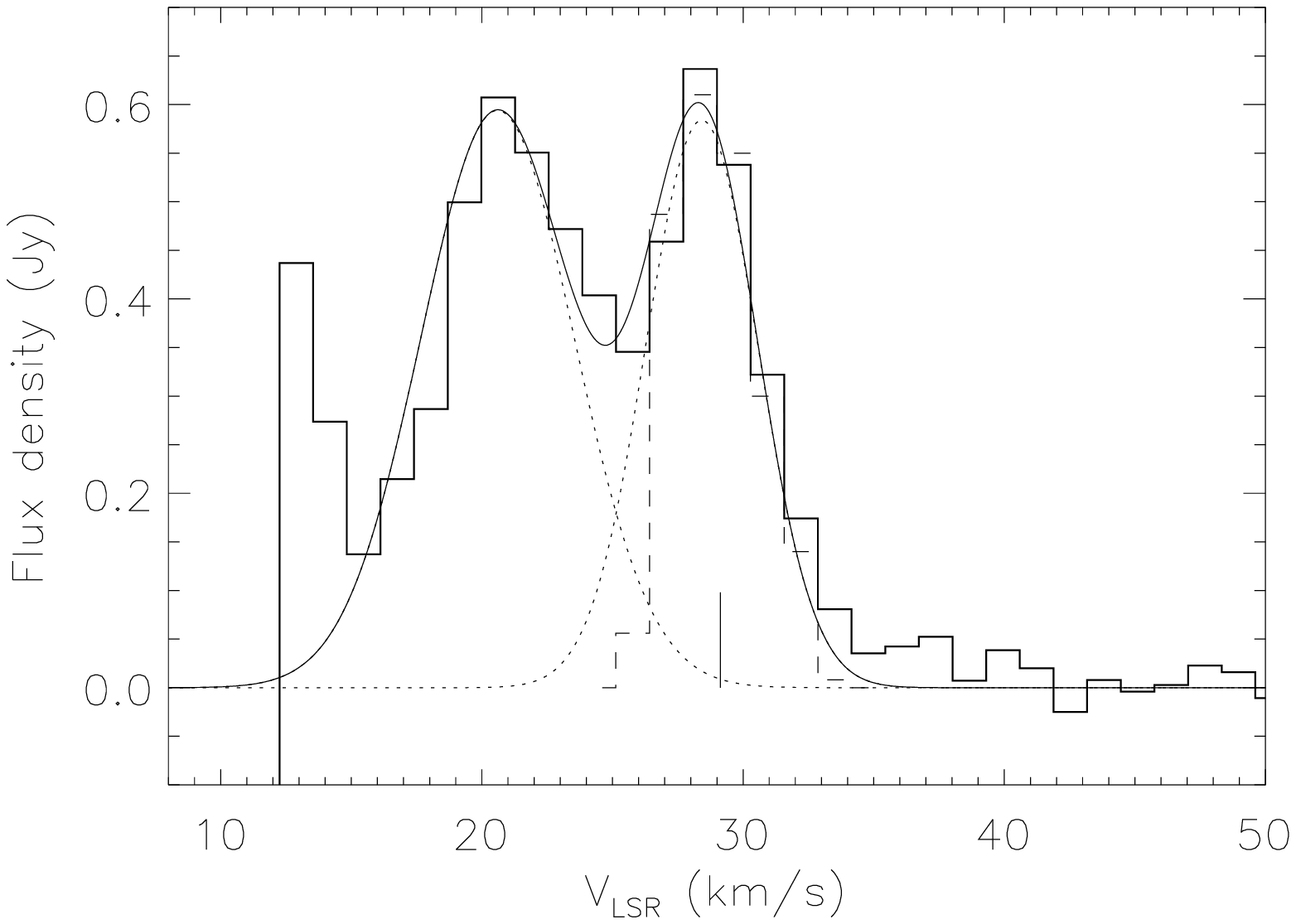}}}
\caption{Spatially integrated 
\HI\ spectrum of RX~Lep.  The stellar systemic
velocity derived from molecular line observations is indicated by a vertical
bar.  The thick histogram shows a spectrum 
derived from naturally weighted VLA data 
(after correction for primary beam attenuation) by integrating the emission in each
velocity channel within a
\am{7}{5}$\times$\am{17}{8} aperture centered at 
$\alpha_{\rm J2000}$=05$^{\rm h}$ 11$^{\rm m}$ 22.5$^{\rm s}$, 
$\delta_{\rm J2000}=-11^{\circ} 55'
01.7''$. The blue edge of the RX~Lep profile is 
blended with Galactic emission. The dotted lines and thin solid line
show the results of a two-Gaussian fit to the blended profile.   The spectrum plotted as a 
dashed histogram was derived by summing the
emission within irregularly  shaped blotches whose outer boundaries
were defined by 3$\sigma$ intensity contours.   }
\label{fig:rxlepspectrum}
\end{figure}

\subsubsection{VLA Results for RX~Lep}
\HI\ channel maps for RX~Lep derived from our VLA
data are presented in Figure~\ref{fig:rxlepcmaps} and an \HI\ total
intensity map derived from the same data is shown in
Figure~\ref{fig:rxlepmom0}.
Together these figures reveal the detection of
an extended \HI\ emission nebula surrounding the
position of RX~Lep whose morphology and
velocity distribution imply an unambiguous 
association with the star.
The \HI\ emission is most prevalent in the two velocity
channels that bracket the stellar systemic velocity  ($V_{\rm
  LSR}=28.4$~\kms\ and 29.6~\kms), but the emission can be traced 
from 25.8~\kms\ to 32.2~\kms. 

Additional \HI\ emission is also detected near the position of RX~Lep
in velocity channels from
 16.8~\kms\ to 24.5~\kms. However, the peak intensity of this
emission is centered a few arcminutes southeast of the star, and it 
reaches its peak spatially-integrated intensity ($\sim$1.2~Jy) 
at $V_{\rm  LSR}=$20.6~\kms\ (see
Figure~\ref{fig:rxlepspectrum}). This is outside the velocity range 
over which molecular line emission
has been detected in the CSE of RX~Lep. Further, the borders of this
emission are not well defined; at low surface brightness levels 
they begin to blend with large-scale bands
of positive and negative emission that stretch across the corresponding
channel images. Together these properties suggest that this emission is not
related to the CSE of RX~Lep, but instead is part of an interstellar
cloud or cloud complex along the
line-of-sight that is
spatially extended and poorly spatially sampled by the VLA. 
This interpretation is consistent
with evidence for confusion over this velocity range seen in the
single-dish spectra of Libert et al. (2008). Based on similar
arguments, 
the \HI\ emission seen northwest of RX~Lep (near
$\alpha_{\rm J2000}=05^{\rm h} 10^{\rm m} 39^{\rm s}$, 
$\delta_{\rm J2000}=-11^{\circ} 39' 16''$) is
also unlikely to be related to the star.

A defining characteristic of the \HI\ emission associated with RX~Lep 
is its ``cracked egg'' morphology. As seen in
Figure~\ref{fig:rxlepmom0}, the velocity-integrated emission 
exhibits a distinctly ovate
shape with its semi-major axis aligned 
north-south. The extent of the ``egg'' 
is $\sim$\am{6}{7} (0.30~pc) east-west by $\sim$\am{9}{8} (0.42~pc) north-south. The
star itself lies $\sim30''$ north of the 
geometric center of the egg. The \HI\ distribution across the
egg appears clumpy, and Figure~\ref{fig:rxlepmom0} reveals 
two enhancements in the \HI\ column density
at its north and south ``poles''. In contrast, a pronounced dearth of emission
is observed along the egg's east-west equator. 

Another striking feature of RX~Lep's circumstellar \HI\
emission is the presence of a narrow, highly
extended ($\sim7'$ or 0.30~pc long) tail of emission that stretches almost due
south of the star. As seen in Figure~\ref{fig:rxlepmom0}, 
the tail includes several compact knots of emission. There are also
hints that the tail may have a
kinked or corkscrew shape, although this cannot be confirmed from the
present data.

Two versions of the
spatially integrated \HI\ spectrum of RX~Lep
are shown in Figure~\ref{fig:rxlepspectrum}. The first (shown as a
solid histogram) is derived by integrating the emission within each
spectral channel within a fixed aperture. Because of the presence of
confusion blueward of the stellar systemic velocity, 
we have also derived a second spectrum (plotted as a
dashed histogram) by summing the emission in irregularly
shaped ``blotches'' surrounding the stellar position in 
each channel over the velocity range at which circumstellar
CO emission has been detected. The blotch peripheries were
defined by 3$\sigma$ contours and encompass the emission comprising
the tail and shell structures 
visible in Figure~\ref{fig:rxlepmom0}. 
 
Based on a two-component Gaussian fit to the ``fixed aperture'' 
spectrum in Figure~\ref{fig:rxlepspectrum}, we derive an \HI\ 
line centroid for the RX~Lep emission of
$V_{\rm LSR}=28.4\pm0.1$~\kms, in good agreement with the value
determined from CO observations. From our Gaussian fits we also find a peak flux density of
0.58$\pm$0.02~Jy and a
FWHM \HI\ linewidth $\Delta V=5.0\pm$0.3~\kms. The latter is 
larger than the value of 3.8$\pm$0.1~\kms\ 
previously reported by Libert et al. (2008). From our fit we
compute an
integrated \HI\ flux density for the CSE of RX~Lep of $\int S_{\nu}d\nu=3.09\pm$0.22~Jy
\kms. Alternatively, using the ``blotch'' spectrum, 
we measure $\int S_{\nu}d\nu=2.77\pm$0.04~Jy
\kms. Because the uncertainty is dominated by the
systemic uncertainties in determining the intrinsic 
line profile shape in the presence of confusion, we adopt as 
our best estimate a mean of our two measurements: $\int S_{\nu}d\nu=2.93\pm0.22$~Jy \kms. 
This translates to an \HI\ mass of $M_{\rm HI}\approx0.015M_{\odot}$. This
is roughly 2.5 times larger than measured by Libert et al. (2008) from
single-dish mapping. The reason for this discrepancy is unclear,
although part of the difference may arise from the manner in which the
RX~Lep line emission was deblended from the neighboring Galactic
signal. Libert et al. used polynomial baseline fits, while here we use
spatial segregation of the emission and Gaussian line decompositions to disentangle the
signals. We
have verified through measurements of continuum sources in the field
(\S\ref{contimaging}) that the discrepancy does not result from an uncertainty in the
amplitude scale of the VLA data. The distinctive morphology of the
circumstellar \HI\ emission seen in Figure~\ref{fig:rxlepmom0} 
also seems to rule out significant
contamination of the VLA flux density measurements from line-of-sight
emission.

\subsubsection{Discussion of RX~Lep Results\protect\label{rxlepdisc}}
Based on single-dish mapping and numerical modeling,
Libert et al. (2008) deduced that
RX~Lep is surrounded by a detached
\HI\ shell comprising material from the stellar wind, augmented by gas
swept up from the local ISM. They also found indications of elongation
of the emission along the direction of space motion of the star. 
Our VLA observations of RX~Lep confirm these general features of the
\HI\ distribution. In addition, the 
wealth of additional detail
revealed by our imaging data  provides important new clues
on the evolutionary history of the star and its interaction
with the local environment. 

%
To first order, the morphological properties of the
\HI\ emission surrounding RX~Lep appear to be significantly influenced by the
interaction between the star and its local interstellar
environment. This is not surprising given the moderately high space
velocity of the star ($V_{\rm space}\approx$57~\kms; Table~5). 
Manifestations of this interaction include the 
elongation of its \HI\ shell, the displacement of the
star from the center of this shell, and the extended tail of gas trailing
to the south. We now discuss each of these properties in turn.

The north-south 
elongation of RX~Lep's egg-shaped shell aligns roughly
with the space motion
vector that we have computed for the star
(PA$\approx25^{\circ}$). This small misalignment between the two may
be due to a
bulk ISM flow in the region, directed roughly perpendicular to the
trajectory of the star. Evidence for similar local flows has recently
been inferred from observations of other evolved stars (Ueta 2008;
Ueta et al. 2010; Wareing 2012; Menten et al. 2012; 
see also \S\ref{txpscdisc}). 

As noted above, RX~Lep is displaced by $\sim30''$ (0.022~pc) north of the geometric
center of its \HI\ shell. If we adopt half of the FWHM linewidth of the \HI\
spectral profile as a measure of the outflow velocity 
(i.e., $V_{\rm out}\approx2.6$~\kms) and a mean shell
radius of 0.15~pc, the dynamical crossing time for the shell is 
$\sim$56,000~yr. If RX~Lep has
traveled from the shell center to its present position in this amount
of time, its implied rate of motion in declination would be only 
$\sim$0.5~mas yr$^{-1}$---significantly less than its {\it Hipparcos}
proper motion along this direction 
(56.93~mas yr$^{-1}$). This implies that the \HI\ shell is not
``free-floating'', but rather that it is  
being dragged along with the moving star.
Furthermore, this suggests that the displacement of the star from the shell center
originates primarily from the lower
pressure encountered in the direction opposite the motion of the star compared
with in the leading direction. 

As described below, our new data support the possibility that 
RX~Lep has only recently
begun a new phase of mass-loss. 
If true, this may help to explain
another prominent feature of its shell: the dearth of emission
along its equatorial region. Villaver et al. (2012) point out that
for mass-losing stars moving supersonically through the ISM,  
an ensuing pressure drop interior to an
existing shell that occurs in-between mass-loss episodes may trigger instabilities,
leading it to break apart. Villaver et al. also
note that as the wind from a new mass
loss episode propagates through previously ejected material, shock
regions are expected to develop. We speculate that the break in
RX~Lep's shell may be linked with such shocks or instabilities. 

Extended gaseous wakes resembling cometary-like tails 
have now been detected in \HI\ 
emission toward several AGB stars, trailing their motions 
through space (Matthews \& Reid 2007; Matthews et al. 2008, 2011b; see also
\S\ref{txpsc}). However, none of the tails discovered to date has
been as highly collimated as the one seen trailing RX~Lep. 
It is likely that this narrow tail results from the confluence 
of the relatively early evolutionary
status of RX~Lep compared with other stars imaged in \HI\ to date, 
coupled with its relatively high space velocity and the star's location
near the Galactic disk (see Table~5).  In
general, higher stellar space velocities allow more efficient
deflection of mass loss debris behind the star to form a trailing wake
(e.g., Villaver et al. 2012), while higher local ISM densities (as are
expected near the Galactic plane) produce
greater external pressure, leading to greater 
collimation (e.g., Wareing 2012). We note however that in simulations,
the adopted cooling function and the temperature assumed for the
stellar wind will also influence the resulting properties of the
cometary tail.

A counterpart to the RX~Lep tail is also visible in {\it IRAS}
60$\mu$m and 100$\mu$m
images of the star, extending $\sim12'$ to the south (see Figure~9
of Libert et al. 2008). 
While Libert et al. cautioned that this
structure could be an artifact resulting from the north-south
scanning direction of the {\it IRAS} satellite, our \HI\ imaging now provides 
corroborating evidence that this feature is likely real. 
The poor angular resolution of the {\it
  IRAS} data preclude a detailed comparison between the FIR and \HI\
emission, but the data are consistent
with the two tracers having comparable linear extents. 

%
To account for their original \HI\ observations of RX~Lep, Libert et
al. (2008) proposed a simple model comprising a freely expanding wind
with a constant outflow speed and mass loss rate, interior to some
radius, $r_{1}$, at which a termination shock occurs owing to the collision
between the freely expanding stellar wind and the surrounding
medium. Exterior to $r_{1}$, the wind sweeps up additional matter from
its surroundings and is decelerated and compressed as a result of this
interaction, leading to the formation of a detached
shell. 

Incorporating new constraints derived from our VLA
measurements together with the molecular wind parameters from Libert
et al. (2008), we have computed a revised version of the Libert et
al. model. Details are described in Appendix~A.  We recognize that 
our model is an oversimplification, since it assumes spherical
symmetry  and does not account for the space motion of RX~Lep.  
The stellar motion will 
result in a higher effective pressure on the leading edge of its \HI\ shell
and a pressure in the trailing direction that is lower than the average ISM
pressure.  These effects will
impart an elongation to the \HI\ shell (as well as a  reduction of the size of the shell in the
leading direction) and lead to the formation of
a trailing wake.
Despite these limitations, our model 
is able to provide a satisfactory fit to the globally averaged
\HI\ line profile of RX~Lep (Figure~\ref{fig:models}a) and
supplies a first order estimate of the duration of
mass loss time history for this star. Based on our best-fit model
(Table~A1), we find a mass loss timescale of $\sim$90,000~yr. Our
derived age is consistent with the hypothesis that RX~Lep  has begun
its life as a TP-AGB star only relatively recently.

\subsection{Y Ursae Majoris (Y UMa)\protect\label{yuma}}
Y Ursae Majoris (Y~UMa) is a semi-regular variable with 
an oxygen-rich chemistry. Its spectral type 
(M7II-III) is later than the vast majority of semi-regular
variables (Dickinson \& Dinger 1982).  Kiss et al. (1999) found this star
to have a triply periodic variability with a dominant period of 162
days, plus additional modulations on time scales of 315 and 324
days. (Another star in our current sample, 
the carbon-rich Y~CVn, is also a triply periodic
semi-regular variable; see \S\ref{ycvn}). Y~UMa has an effective temperature of $\sim$3100~K
based on the temperature scale of normal giants of type later than M6
(Perrin et al. 2004).
We adopt a distance to the star of 385~pc based on the {\it Hipparcos}
parallax of 2.60$\pm$0.59~mas (van~Leeuwen 2007).

Y~UMa was observed in the CO(2-1) line by Knapp et al. (1998). Based
on those data, the authors derived 
a stellar systemic velocity of $V_{\star,\rm LSR}=19.0\pm$0.5~\kms, a wind
outflow speed of $V_{\rm out}=5.4\pm$0.6~\kms, and a mass-loss rate of ${\dot
  M}=2.6\times10^{-7}M_{\odot}$~yr$^{-1}$ (scaled to our
adopted distance). The systemic velocity and outflow velocity are in
agreement with those derived from the thermal SiO $v$=0 $J$=2-1 and
$J$=3-2 lines detected by Gonz\'alez Delgado et al. 
(2003). The {\it Hipparcos} proper motions of $-4.95\pm0.53$~mas
yr$^{-1}$ in right ascension and
2.05$\pm$0.50~mas yr$^{-1}$ in declination imply a peculiar space motion of 19.2~\kms\
along a position angle of \ad{57}{1} (Table~5).

\begin{figure*}
\scalebox{0.74}{\rotatebox{-90}{\includegraphics{f6.ps}}}
\caption{\HI\ channel maps bracketing the systemic velocity of Y~UMa,
  derived from naturally weighted VLA data.  The spatial resolution is $\sim72''\times58''$.
A star symbol indicates the stellar position. 
Contour levels are 
($-4.2$,$-$3,3,4.2,6,8.4)$\times$1.7~mJy beam$^{-1}$. The lowest
contour is $\sim3\sigma$. The
field-of-view shown is comparable to the VLA primary beam. 
  }
\label{fig:yumacmaps}
\end{figure*}
\begin{figure*}
\centering
\scalebox{0.75}{\rotatebox{0}{\includegraphics{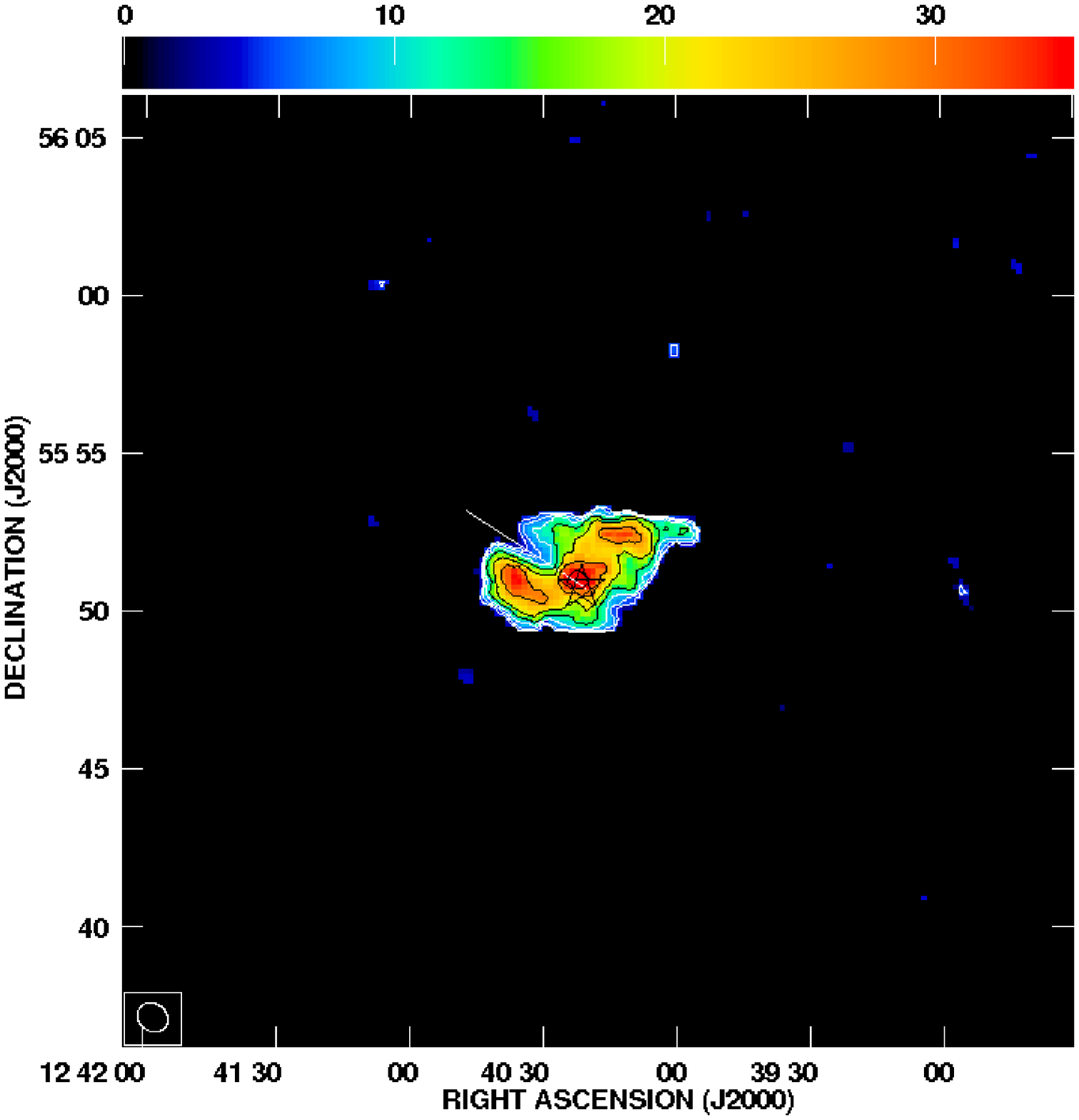}}}
\caption{\HI\ total intensity 
map of Y~UMa derived from data with robust +1 weighting.  The spatial
resolution is $\sim61''\times52''$. The 
map was constructed from emission spanning LSR velocities
13.8 to 22.9~\kms. Contour levels are
(1,1.4,2,2.8,4,5.6,8)$\times$4.8~Jy beam$^{-1}$ m
s$^{-1}$ and the intensity levels are 0 to 35~Jy beam$^{-1}$ m s$^{-1}$. 
To minimize noise in the
map, data at a given point were blanked if they did
not exceed a 1.8$\sigma$
threshold after smoothing by a factor of three
spatially and spectrally.  A star symbol
indicates the stellar position and white line indicates the direction
of space motion. The
field-of-view shown is comparable to the VLA primary beam. }
\label{fig:yumamom0}
\end{figure*}

\subsubsection{VLA Results for Y UMa\protect\label{yumavla}}
We present \HI\ channel maps from our naturally weighted VLA data
cube in
Figure~\ref{fig:yumacmaps}. Emission is detected at the
stellar position in the channel centered at $V_{\rm
  LSR}=19.0$~\kms\ (corresponding to the stellar systemic
velocity) and in several surrounding
channels. Indeed, we detect \HI\ emission at $\ge3\sigma$ that appears
to be associated with the CSE over the velocity range 13.8
to 22.9~\kms.  The velocity extent of the \HI\ emission is thus nearly
identical to that of the CO(2-1) emission detected by Knapp et
al. (1998). Redward of the stellar velocity ($V_{\rm LSR}\ge
19.6$~\kms), the detected emission
is weak and unresolved and lies at or near the stellar
position. However, for velocities $V_{\rm LSR}\le 19.0$~\kms, the
emission is clearly extended, and there are hints of a bifurcation in
the emission structure. This is most clearly seen in the 15.8~\kms\
and 16.4~\kms\ velocity channels. 

In Figure~\ref{fig:yumamom0} we present an \HI\ total intensity map of
Y~UMa. This map was produced from the robust +1 weighted data (Table~4)
to highlight the clumpy nature of the emission and the
existence of a
``notch'' in the CSE to the northeast of the star, along a position angle of
$\sim60^{\circ}$. This position angle closely matches that of the direction of space
motion of the star (Table~5). An analogous feature, also along the direction of
space motion, is seen in the CSE of Y~CVn
(Figure~\ref{fig:ycvnmom0}; \S\ref{ycvndisc}). 
One can also see from Figure~\ref{fig:yumamom0} that the
peak \HI\ column density in the CSE of Y~UMa is displaced by
$\sim20''$ ($\sim$0.037~pc) to
the northeast of the stellar position. 

The morphology of the \HI\ emission surrounding Y~UMa suggests that
the emission in the northwestern part of the CSE has been severely distorted by
its interaction with the surrounding medium or other external
forces. The result is a CSE with a distinctly non-spherical shape, including a small
plume of gas extending from the western edge of the \HI\ nebula.
Despite this distortion, we measure nearly equal \HI\ masses 
on either side of the bisector running along the position angle
of the space motion. Including the extended plume, 
the maximum \HI\ extent of Y~UMa is $\sim$\am{6}{8} (0.76~pc).
\begin{figure*}
\centering
\scalebox{0.6}{\rotatebox{-90}{\includegraphics{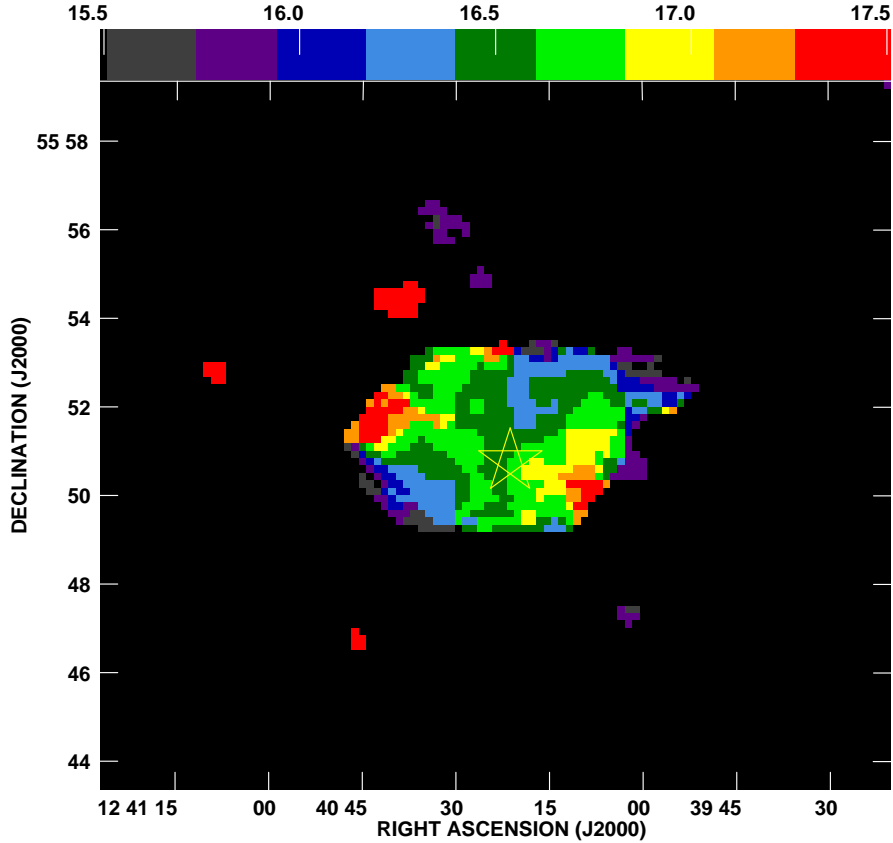}}}
\caption{\HI\ velocity field of Y~UMa derived from naturally weighted
  data.  The spatial resolution is $\sim72''\times58''$. 
The intensity scale indicates LSR radial velocity in \kms.
  }
\label{fig:yumamom1}
\end{figure*}
\begin{figure}
\scalebox{0.5}{\rotatebox{0}{\includegraphics{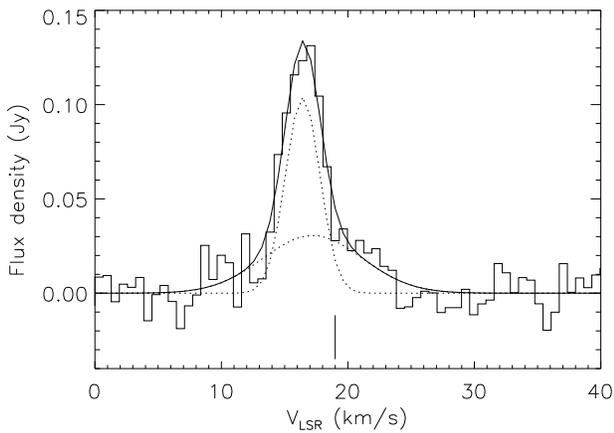}}}
\caption{Spatially integrated 
\HI\ spectrum of Y~UMa. The spectrum was derived from naturally weighted VLA data 
(after correction for primary bean attenuation) by integrating the emission in each
velocity channel within a
\am{8}{5}$\times$\am{7}{0} aperture centered at 
$\alpha_{\rm J2000}$=12$^{\rm h}$ 40$^{\rm m}$ 20.1$^{\rm s}$, 
$\delta_{\rm J2000}=$55$^{\circ}$ 51$'$ 27.6$''$. The
dotted curves and thin solid curve show, respectively, a two-component
Gaussian fit to the line profile and the sum of the two
components. The vertical bar indicates the stellar systemic velocity
as determined from CO observations. }
\label{fig:yumaspectrum}
\end{figure}

\begin{figure*}
\centering
\scalebox{0.6}{\rotatebox{-90}{\includegraphics{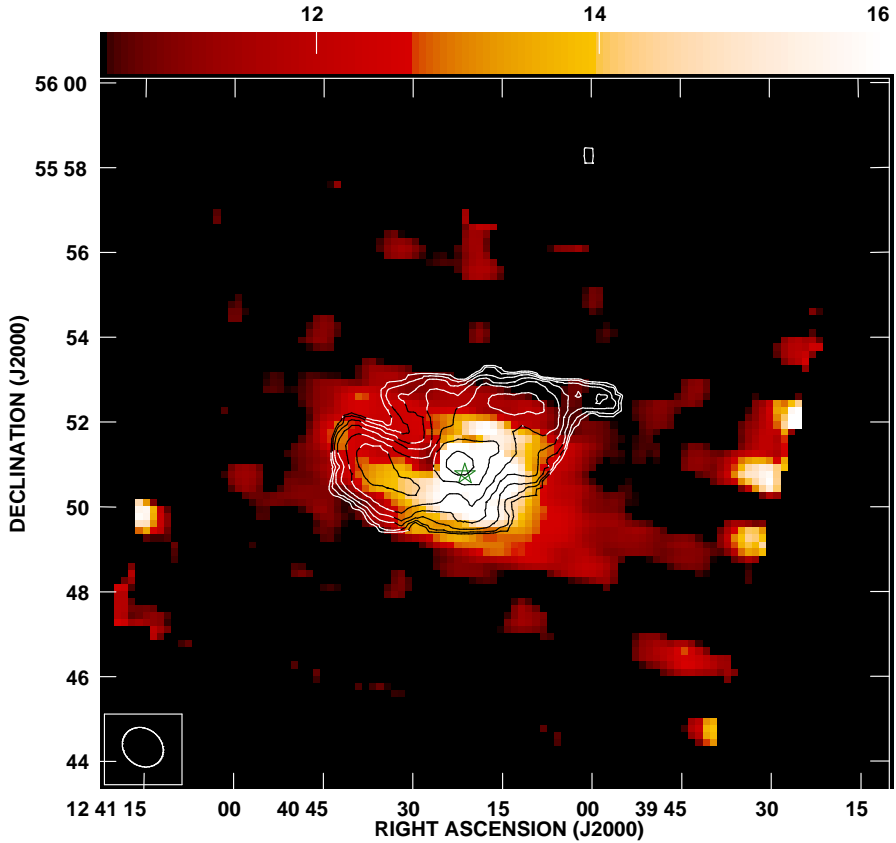}}}
\caption{\HI\ total intensity 
contours (same as in Figure~\ref{fig:yumamom0}) overlaid on an {\it ISO}
65$\mu$m image of Y~UMa. The brightness units of the {\it ISO} image are 
MJy steradian$^{-1}$. }
\label{fig:yumamom0+iso}
\end{figure*}

Figure~\ref{fig:yumamom1} shows an \HI\ velocity field for Y~UMa,
again derived from the robust +1 data.
The \HI\ velocity field is clearly dominated by emission blueward of
the stellar systemic velocity. Interestingly, the velocity pattern
exhibits a 180-degree rotational symmetry, but with the
symmetry point
displaced by $\sim1'$ northeast of the stellar position. 
This displacement is along the 
space motion vector of the star. One can also see in Figure~\ref{fig:yumamom1}
evidence for systematic velocity gradients across the CSE;
along PA$\approx70^{\circ}$ and $250^{\circ}$, the emission
becomes systematically redshifted with increasing distance from the
star, while along PA$\approx110^{\circ}$ and $\approx350^{\circ}$, 
the emission becomes
systematically blueshifted  with increasing distance. These angles
match the bifurcation visible in the channel images in
Figure~\ref{fig:yumacmaps}. Y~UMa is the only star imaged in \HI\ to
date to show these types of features in the velocity field, and their
origin and interpretation are presently unclear.

A spatially integrated \HI\ spectrum of Y~UMa is presented in
Figure~\ref{fig:yumaspectrum}. As expected from the emission
distribution seen in Figure~\ref{fig:yumamom0},
the peak integrated flux density occurs blueward of the stellar
systemic velocity. The spectrum has a two-component
structure similar to what was reported previously by 
G\'erard \& Le~Bertre (2006) based on NRT spectroscopy. Using a
two-component Gaussian fit to the VLA spectrum, we find a narrow component
with a peak flux density of 0.10$\pm$0.01~Jy and a FWHM line width of
3.2$\pm0.4$~\kms, centered at $V_{\rm LSR}=16.4\pm0.2$~\kms, and a weaker,
broad component centered at 
17.2$\pm$0.8~\kms\ with a peak flux density 0.03$\pm$0.01~Jy  and a
FWHM line width of 9.2$\pm$2.3~\kms. Both components are blueshifted
relative to the systematic velocity of the star as determined from CO
observations, although the broad component matches the
systemic velocity determined from near infrared lines by Lebzelter \&
Hinkle (2002) ($V_{\star}=17.7\pm$0.5~\kms). 

The peak
\HI\ flux density measured with the VLA is $\sim$40\% higher than
the original NRT measurement by G\'erard \& Le~Bertre (2006). However, a new analysis
using higher quality NRT data now shows good agreement between the
VLA and NRT measurements, although the NRT data hint that the
source size may be
more extended than we observe with the VLA (up to 16$'$; G\'erard \& Le~Bertre, in
prep.). 

Based on our Gaussian fit to the spectrum in
Figure~\ref{fig:yumaspectrum}, 
we measure an integrated \HI\ flux density associated with the CSE of
Y~UMa of $\int S_{\rm HI}~d\nu=0.63\pm$0.13~Jy \kms, translating to a total \HI\ mass
$M_{\rm HI}\approx 0.022~M_{\odot}$. Although the double Gaussian
provides a good fit to the line profile, the peak of the composite
Gaussian profile is shifted
by $\sim$0.6~\kms\ to the blue of the observed line peak, and there
are hints of residual emission in the blue wing of the line.

\subsubsection{Discussion of Y UMa Results\protect\label{yumadisc}}
From Figure~\ref{fig:yumacmaps} one sees that the emission giving rise
to the broader of the two \HI\ line components in
Figure~\ref{fig:yumaspectrum}  arises from spatially compact
emission lying at or near the position of the star, while the narrower velocity
component comes from material that is more spatially extended. This suggests
that the broad line component may be linked with the freely expanding wind
detected at small radii through its CO and SiO emission. This hypothesis is
consistent with the respective widths of the \HI\ and molecular lines. If we approximate
the  molecular linewidth as twice the CO outflow velocity ($\sim$10.8$\pm$0.6~\kms),
this matches the FWHM of the
broad \HI\ line, although  the line centroids
differ by $\sim$2-3~\kms. 
Higher resolution \HI\ imaging together with CO imaging of
Y~UMa would be of interest for more thoroughly investigating this link.

Y~UMa was reported by Young et al. (1993a) to be an extended infrared
source with a diameter of \am{7}{6} at 60$\mu$m.  The extended nature of
the FIR emission is confirmed by an archival {\it ISO} 65$\mu$m image
shown in Figure~\ref{fig:yumamom0+iso}. 
To our knowledge, these {\it ISO} data
have not  been published previously. The observations were executed on 1996
October 18 as part of program 33700231 (Principal Investigator:
J. Gurtler). We have downloaded the oversampled map as a
``Standard Processed Data'' FITS file from the {\it
  ISO} data archive.\footnote{The archive is hosted by the European
  Space Agency at http://iso.esac.esa.int/ida.} The original image
had a field-of-view of $687''\times687''$ and a pixel size of \as{15}{0}. For display
purposes, the image
was resampled to a pixel size of 10$''$ to match the VLA data. 

While the FIR emission appears to have a roughly comparable spatial extent as
the \HI\ emission detected with the VLA, the two do not correspond 
closely in detail. For example, the FIR emission appears elongated along a
northeast/southwest direction, extending further to the southwest than
the \HI\ emission. Conversely, we 
see that there is little or no FIR emission underlying the
northwestern portion of the \HI\ nebula, including the ridge of
enhanced \HI\ column density and the weaker plume extending from the
northern edge. 
Since in principle, whatever
process is responsible for the distortion on the northwestern side of the
CSE and creation of the ``plume'' observed in \HI\ 
should have acted on both the \HI\ and the FIR-emitting
(presumably dusty) material,
this suggests that either the atomic hydrogen is intrinsically more
extended than the dust or that
dust grains were destroyed during this process. 
The absence of any strong correlations between the atomic gas and the 
FIR emission may also
imply that the dust grains are not uniformly mixed throughout
the CSE. Simulations by van Marle et al. (2011) have shown that such
non-uniform mixing is expected for the largest CSE grains
(with radii $\gsim0.105\mu$m).

As in the case of RX~Lep (\S\ref{rxlepdisc}), we have
derived a simple model for the \HI\ emission of Y~UMa under the assumption
that it arises from an outflowing wind that has been abruptly slowed at a
termination shock where it meets the surrounding ISM, resulting in the
formation of a detached shell (see Appendix~A
and Table~A1 for details). Despite the observed distortions of the Y~UMa \HI\
shell, our simple model provides a reasonable fit to the global (spatially
averaged) \HI\ line profile, including its two-component
structure (Figure~\ref{fig:models}b). Using this model we estimate
a total duration of the mass loss history of the star 
of $\sim1.2\times10^{5}$~yr.

\subsection{R Pegasi (R Peg)\protect\label{rpeg}}
R Pegasi (R~Peg) is an oxygen-rich Mira variable with a mean spectral type
M7e and a pulsation
period of 378 days (Jura \& Kleinmann 1992; Whitelock et al. 2008).
The effective temperature of R~Peg is known to vary significantly
throughout its pulsation cycle; van Belle et al. (1996) measured
$T_{\rm eff}=2333\pm$100~K at minimum  and 
2881$\pm$153~K at maximum. 

We adopt a distance to R~Peg of 400~pc based on the period-luminosity relation
for Miras derived by Jura \& Kleinmann (1992), although we note that
the study of
Whitelock et al. (2008) yielded a value $\sim$25\% higher (490~pc). 
The {\it Hipparcos}
parallax for this star (2.61$\pm$2.03~mas
yr$^{-1}$; Knapp et al. 2003) is too uncertain to provide
any additional meaningful constraints on the distance

Based on the
CO(2-1) observations of Winters et al. (2003),
we assume a stellar systemic velocity of $V_{\star,\rm LSR}$=24.0~\kms,
an outflow velocity of $V_{\rm out}$=5.5~\kms, and 
a mass loss rate of ${\dot
  M}\sim 5.3\times10^{-7}~M_{\odot}$~yr$^{-1}$ (scaled to our adopted
distance). Using the {\it Hipparcos} proper motions from van Leeuwen
(2007) of 12.63$\pm$1.74~mas
yr$^{-1}$ in right ascension and $-9.71\pm1.67$~mas yr$^{-1}$ in declination, 
we derive a peculiar
space motion of $V_{\rm space}$=26.3~\kms\ along a PA=\ad{147}{6} (Table~5).

R~Peg was previously detected in \HI\ by G\'erard \& Le~Bertre (2006),
who found evidence for a two-component line profile comprising a
strong, narrow, roughly Gaussian component superposed on a broader
pedestal. They found a velocity offset of $\sim$2~\kms\ between the
two components, with the two bracketing the systemic velocity of the
star as determined from CO observations.

\subsubsection{VLA Results for R~Peg}
\HI\ channel maps surrounding the stellar systemic velocity of R~Peg
are presented in Figure~\ref{fig:rpegcmaps}. The morphology of the \HI\
emission seen in these channel images is unlike any other star that
has been imaged in \HI\ to date. \HI\ emission is clearly detected near
the stellar position in several velocity channels bracketing
$V_{\star}$. 
However, at $V_{\rm LSR}$=25.1~\kms, the
emission peak is offset by roughly  an arcminute (one beam width) to
the southeast of the stellar position. In the three velocity channels
from 23.2 to 24.5~\kms, multiple emission peaks are present near the
stellar position, but again the peak emission in each channel is
offset to the southwest of the star, with a relative dearth of
emission at the stellar position itself. The morphology in these
channels is somewhat reminiscent of a bipolar flow (but see below). This bifurcation is
even more evident at $V_{\rm LSR}$=22.5~\kms, but here the far reaches
of the two lobes curve to form a horseshoe shape, with an extent of
$>15'$ (1.7~pc). This ``horseshoe''
is also delineated by multiple emission clumps at $V_{\rm LSR}$=21.9~\kms. 
Finally, from $V_{\rm LSR}$=20.0 to 21.3~\kms, the emission clumps
along the horseshoe decrease in number and peak column
density while their projected distances
from the star increase. 

\begin{figure*}
\centering
\scalebox{0.9}{\rotatebox{-90}{\includegraphics{f11.ps}}}
\caption{\HI\ channel maps bracketing the systemic velocity of R~Peg,
  derived from naturally weighted VLA data. The spatial
resolution is $\sim63''\times57''$.
A star symbol indicates the stellar position. 
Contour levels are 
($-4.2$[absent],$-$3,3,4.2,6,8.4)$\times$1.7~mJy beam$^{-1}$. The lowest
contour is $\sim3\sigma$. The
field-of-view shown is comparable to the VLA primary beam. 
  }
\label{fig:rpegcmaps}
\end{figure*}
\begin{figure*}
\centering
\scalebox{0.75}{\rotatebox{0}{\includegraphics{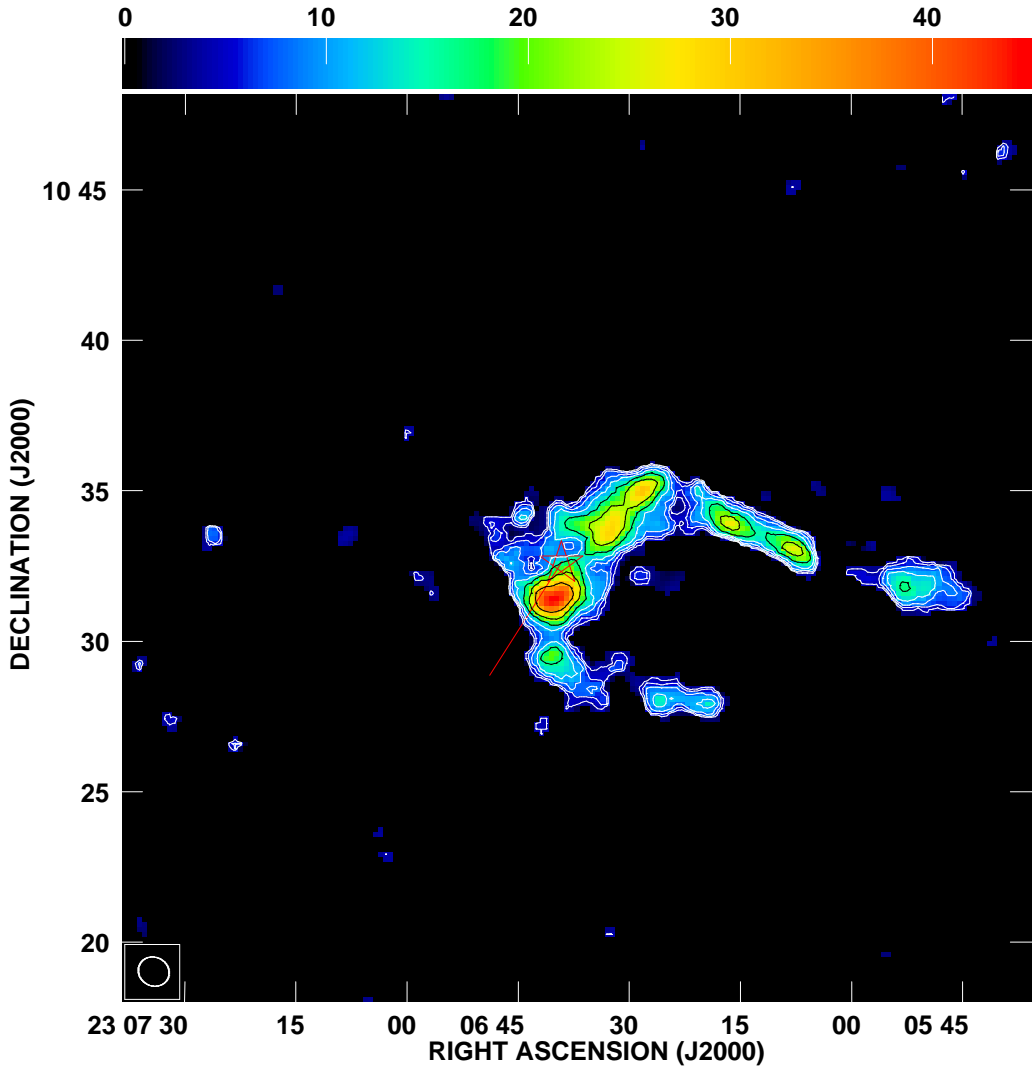}}}
\caption{\HI\ total intensity 
map of R~Peg derived from naturally weighted data. A star symbol
indicates the stellar position and the  red line indicates the
direction of space motion. The spatial
resolution is $\sim63''\times57''$. The 
map was constructed from emission spanning LSR velocities
$20.0$ to $25.1$~\kms. Contour levels are
(1,1.4,2,2.8,4,5.6,8)$\times$4~Jy beam$^{-1}$ m
s$^{-1}$ and the intensity range is 0 to 43~Jy beam$^{-1}$ m s$^{-1}$. 
To minimize noise in the
map, data at a given point were blanked if they did
not exceed a 2$\sigma$
threshold after smoothing by a factor of three
spatially and spectrally. }
\label{fig:rpegmom0}
\end{figure*}
\begin{figure}
\scalebox{0.5}{\rotatebox{0}{\includegraphics{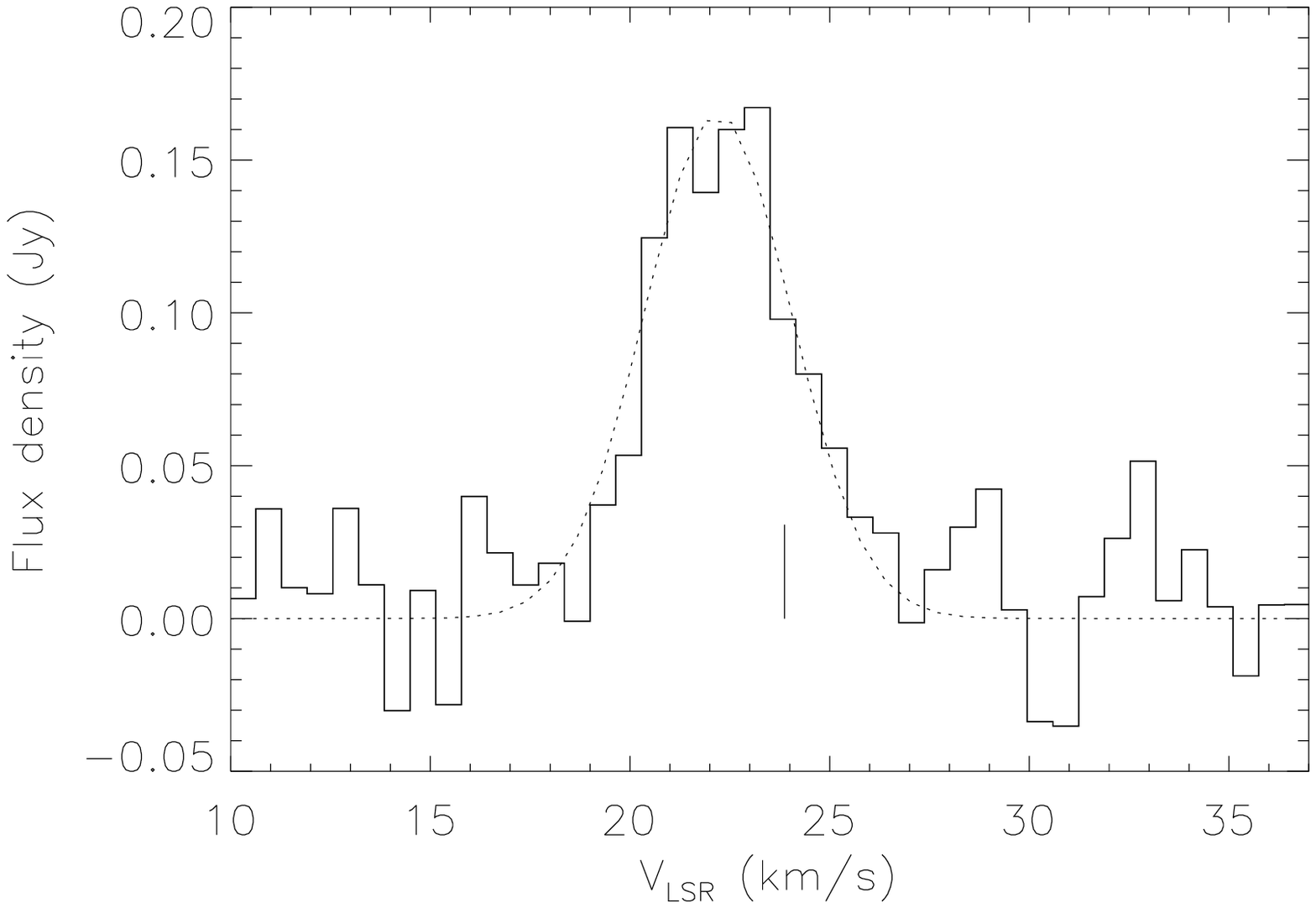}}}
\caption{Spatially integrated 
\HI\ spectrum of R~Peg. The spectrum was derived from naturally weighted VLA data 
(after correction for primary beam attenuation) by integrating the emission in each
velocity channel  within a
\am{15}{5}$\times$\am{8}{3} aperture centered at 
$\alpha_{\rm J2000}$=23$^{\rm h}$ 06$^{\rm m}$ 18.49$^{\rm s}$, 
$\delta_{\rm J2000}=10^{\circ} 31' 06.0''$. The dotted line shows a
Gaussian fit to the line profile. The vertical bar indicates the
stellar systemic velocity as determined from CO measurements. }
\label{fig:rpegspectrum}
\end{figure}


We present an \HI\ total intensity map for R~Peg in
Figure~\ref{fig:rpegmom0}. This map confirms that the peak \HI\ column
density toward R~Peg does not coincide with the stellar position, but
instead lies at a projected distance of $\sim1'$ (0.1~pc) southeast of the
star. 

A global \HI\ spectrum of R~Peg is shown in
Figure~\ref{fig:rpegspectrum}. There are hints of a slight asymmetry
in the line profile, as well as a line peak that is broader and flatter
than the other stars in the current sample. Nonetheless, a single Gaussian (dotted line on
Figure~\ref{fig:rpegspectrum}) provides a reasonable
fit to the data for the purpose of characterizing the global line
parameters. Based on such a fit, we measure
a  peak \HI\ flux density of 0.16$\pm$0.01~Jy, in excellent agreement with
G\'erard \& Le~Bertre (2006). 
We also find  a FWHM line width
of 4.35$\pm$0.39~\kms, centered at $V_{\rm LSR}$=22.4$\pm$0.2~\kms.
The \HI\ line centroid is blueshifted by $\sim$1~\kms\ relative to the
stellar systemic velocity determined from CO observations. 
From our Gaussian fit we also calculate an integrated \HI\ flux density
of $\int S_{\rm HI}~d\nu=$0.76$\pm$0.09~Jy~\kms, translating to an \HI\ mass of
$M_{\rm HI}\approx0.029~M_{\odot}$.
After scaling to account for our different adopted distances, 
the latter agrees closely with the value 
of 0.03~$M_{\odot}$ previously reported by G\'erard \& Le~Bertre
(2006). 

G\'erard \& Le~Bertre (2006) also reported a second weaker, broader
line component centered on the stellar systemic
velocity and with a
velocity extent comparable to that of the CO(2-1) line emission (FWHM $\sim$10~\kms).
New NRT
measurements (G\'erard \& Le~Bertre, in prep.) reaffirm the presence of
this broad 
component with a peak flux density of 0.006$\pm$0.003~Jy, too weak
to have been detected in the current VLA data.

\subsubsection{Discussion of R Peg Results}
Above we noted the presence of an offset of $\sim$0.1~pc to the southeast  between
the peak \HI\ column density observed near R~Peg and the stellar position.
The direction of this
offset is within $\sim10^{\circ}$ of position  angle of the space
velocity vector that we have computed for the star 
(PA$\approx148^{\circ}$; Table~5), 
suggesting that one possible
explanation is that there is an accumulation of material in front of
the star caused by the ``snowplow effect'' (Isaacman 1979). 
In this picture, the gas
to the northwest of the star could comprise a trailing wake of
material, ram pressure-stripped through the interaction between the
CSE and the local ISM (see \S\ref{rxlepdisc}). Consistent
with this hypothesis, over the velocity channels spanning 25.1~\kms\ to
22.5~\kms, the
emission in the putative wake shows a systematic increase in
projected distance from R~Peg with decreasing LSR velocity, as
would be expected for debris that is being slowly decelerated by
dynamical friction from the local ISM (see Matthews et al. 2008, 2011b; Raga
\& Cant\'o 2008). A related possibility is that the lobes of \HI\
emission near the star are the remnants of a wake of debris shed
during a previous cycle of mass-loss, and that the material has since
drifted relative to the stellar position. This could also help to
explain the small blueshift of the \HI\ line centroid relative to
the systemic velocity determined from molecular line observations
visible in Figure~\ref{fig:rpegspectrum}.  However,
an ISM flow originating from a northerly direction 
would seem to be required to account for
the direction of the 
offset between the star and the \HI\ emission, as well as the curving sides
of the horseshoe, which are oriented roughly perpendicular to the direction of
the stellar space motion. 

Despite the bifurcated appearance of the emission in several of
the \HI\ channel maps (Figure~\ref{fig:rpegcmaps}),
we consider the possibility that the \HI\ surrounding
R~Peg results from some type of bipolar outflow to be unlikely. While
bipolar flows are known in the case of a number of AGB stars, with the
exception of the high-velocity flows  ($\sim$100~\kms) 
arising from certain known or suspected binaries (Imai et al. 2002; Sahai et
al. 2003), these outflows do not tend 
to be strongly collimated (e.g., Inomata et al. 2007; 
Libert et al. 2010; Castro-Carrizo et al. 2010).   Further, in
the case of R~Peg, the
two ``lobes'' would have to be oriented close to the plane of the
sky to account for their similar radial velocities.
Finally, the asymmetry of the velocity-integrated emission 
relative to the stellar position  and the 
elongation of the emission parallel to the space motion
vector (Figure~\ref{fig:rpegmom0}) are both effects expected as a result of  interaction
between the ejecta of a moving star and the surrounding ISM.

\subsection{Y Canum Venaticorum (Y CVn)\protect\label{ycvn}}
Y Canum Venaticorum (Y~CVn) is a rare example of a J-type carbon star.
Its exact
evolutionary stage is poorly known (e.g., Dominy 1984; Lambert et
al. 1986; Lorenz-Martins 1996). 
The absence of technetium absorption (Little et al. 1987) and the
lack of enhancement
in $s$-process elements (Utsumi 1985) both suggest that the star has
not yet undergone
a thermal pulse. Some authors have even suggested that Y~CVn is
still on the red giant branch and that its carbon-rich composition may
have resulted
from a core He flash 
(Dominy 1984), although the high derived mass loss rate from the star
appears to preclude this possibility and place the star on the AGB
(e.g., Izumiura et al. 1996).
The bolometric luminosity of Y~CVn also places it on the AGB;
Libert et al. (2007) derived a value of 9652$L_{\odot}$ (scaled to our
adopted distance; see below) using the
bolometric correction of Le~Bertre et al. (2001).

Y~CVn belongs to the class of 
semi-regular variables, having a dominant pulsation period of $\sim$160
days and secondary
periods of 273 and 3000 days (Kiss et al. 1999). (Another star in our sample, 
the oxygen-rich Y~UMa, is also a triply periodic
semi-regular variable; see \S\ref{yuma}). Lambert et al. (1986)
derived an effective temperature for Y~CVn of
$\sim$2730~K. 

From CO(3-2) observations, Knapp et al. (1998) derived a stellar
systemic velocity for Y~CVn of 
$V_{\star,\rm LSR}=21.1\pm$0.9~\kms, an outflow velocity for its
wind of $V_{\rm out}$=7.8$\pm$1.3~\kms, and a mass loss rate ${\dot
  M}=1.7\times10^{-7}~M_{\odot}$~yr$^{-1}$ (scaled to our adopted
distance). We note however that the observations of Jura et al. (1988)
imply a large $^{13}$C abundance in the star, thus the mass-loss
rate derived from $^{12}$CO lines may be underestimated by a factor of
$\sim$1.4 (see Libert et al. 2007).

The distance to Y~CVn is uncertain, with values in the
literature differing by more than $\pm$60~pc. Here we adopt a value of
272~pc based on the re-analysis of the
{\it Hipparcos} parallax measurements by Knapp et al. (2003;
3.68$\pm$0.83~mas).
The proper motion is $-1.40\pm0.31$~mas yr$^{-1}$ in
right ascension and 13.24$\pm0.26$~mas yr$^{-1}$ in declination 
(van Leeuwen 2007), which translates to a peculiar
space velocity of $V_{\rm space}$=31.0~\kms\ at PA=\ad{36}{8}. 

Y~CVn was previously detected in \HI\ with the NRT by Le~Bertre
\& G\'erard (2004) and Libert et
al. (2007) (see \S\ref{ycvndisc} for a discussion). 
Y~CVn has also long been known to be surrounded by extended FIR
emission (Young et al. 1993a; Izumiura et al. 1996; Geise 2011; Cox et
al. 2012a). 
The {\it ISO} 90$\mu$m
data of Izumiura et al. show a central infrared source surrounded by a
quasi-circular shell with an inner radius $r_{\rm in}\sim$\am{2}{8} and
an outer radius $r_{\rm out}\sim$\am{5}{1}. This shell is
displaced slightly relative to the central infrared source. Izumiura
et al. interpreted the properties of the FIR emission as indicating
that Y~CVn has undergone a decrease in mass loss rate of two orders of
magnitude compared to the value at the time of the dust shell
formation. However, Libert et al. proposed an
alternative interpretation, namely that the shell is a
several hundred thousand year-old structure formed by the slowing down of
the stellar wind by the surrounding matter. This does not rule out
that changes in the mass loss rate over time have occurred in Y~CVn, but
it removes the requirement of an abrupt drop in ${\dot M}$ posterior to the
formation of shell in order to explain the central hole seen in the
FIR observations.
As described below, our new data strengthen
the interpretation put forth by Libert et al. 
and permit a more detailed comparison with the
{\it ISO} FIR data.

\subsubsection{VLA Results for Y~CVn}
\begin{figure*}
\centering
\scalebox{0.9}{\rotatebox{0}{\includegraphics{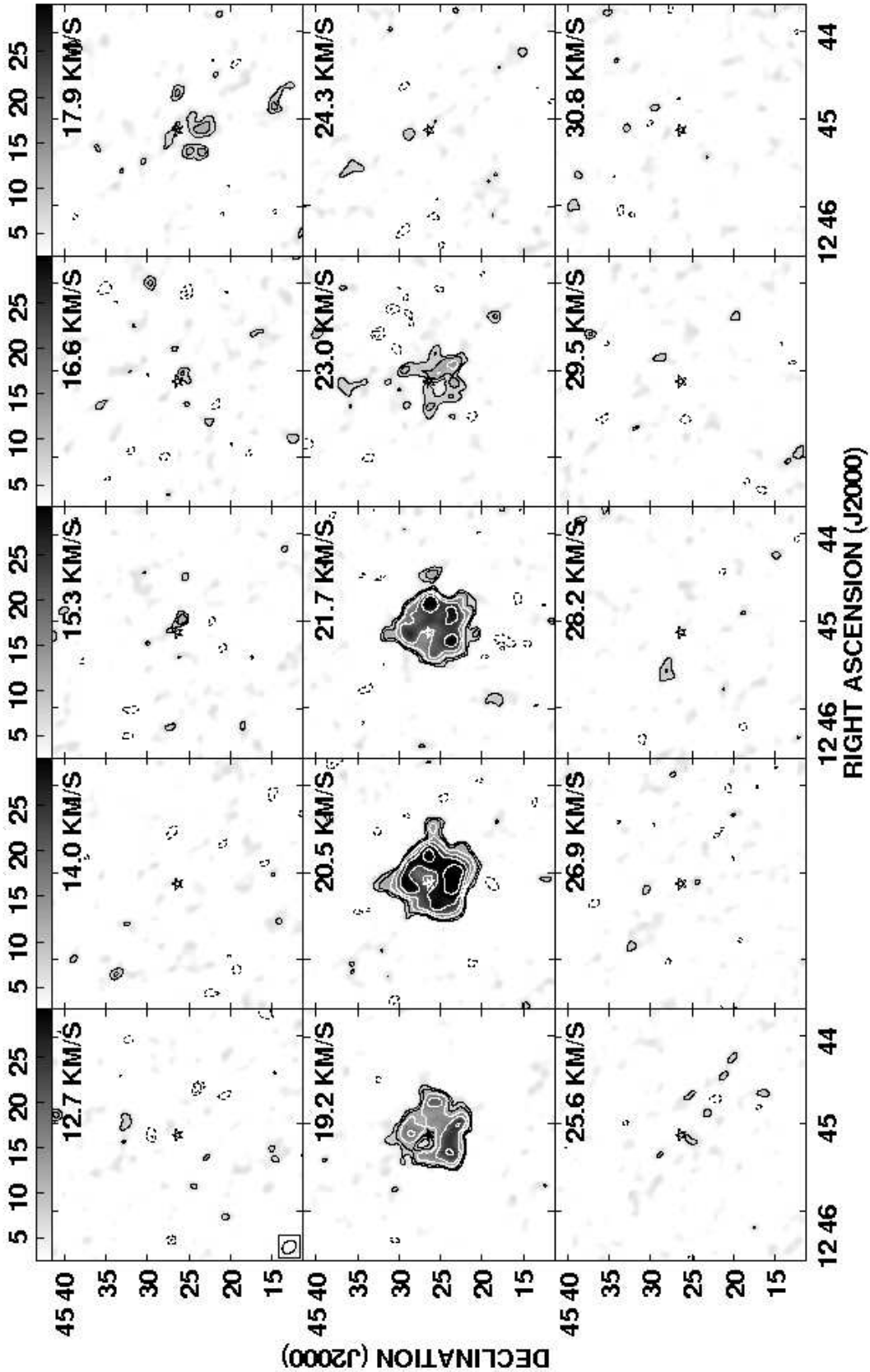}}}
\caption{\HI\ channel maps bracketing the LSR velocity of Y~CVn, derived
from tapered, naturally weighted VLA data. The velocities shown correspond
to the values over which CO(3-2) emission
was detected by Knapp et al. 1998. A star symbol indicates the stellar
position. The spatial
resolution is $\sim107''\times83''$.
Contour levels are 
($-3.5$,$-$2.5,2.5,3.5,5,7.1,10.0,14.1)$\times$2.6~mJy beam$^{-1}$. The lowest
contour is $\sim2.5\sigma$. The greyscale intensity levels are
2.7 to 30.0~mJy beam$^{-1}$. 
  }
\label{fig:ycvncmapstap}
\end{figure*}

\HI\ channel maps for Y~CVn derived from a tapered, naturally weighted version of our VLA
data cube are presented in Figure~\ref{fig:ycvncmapstap}. \HI\
emission associated with
the CSE of Y~CVn is  detected at LSR velocities ranging
from 15.3 to 25.6~\kms. We show a tapered version of our data here 
despite their lower spatial resolution, as they better
highlight the faint emission features present near the stellar
position in the velocity channels from 15.3 to 17.9~\kms\ and from
24.3 to 25.6~\kms. At the velocities closest to the stellar
systemic velocity, the emission
appears extended and clumpy and distributed over a shell-like
structure, while 
a relative dearth of emission is notable directly
toward the stellar position.

\begin{figure*}
\centering
\scalebox{0.75}{\rotatebox{-90}{\includegraphics{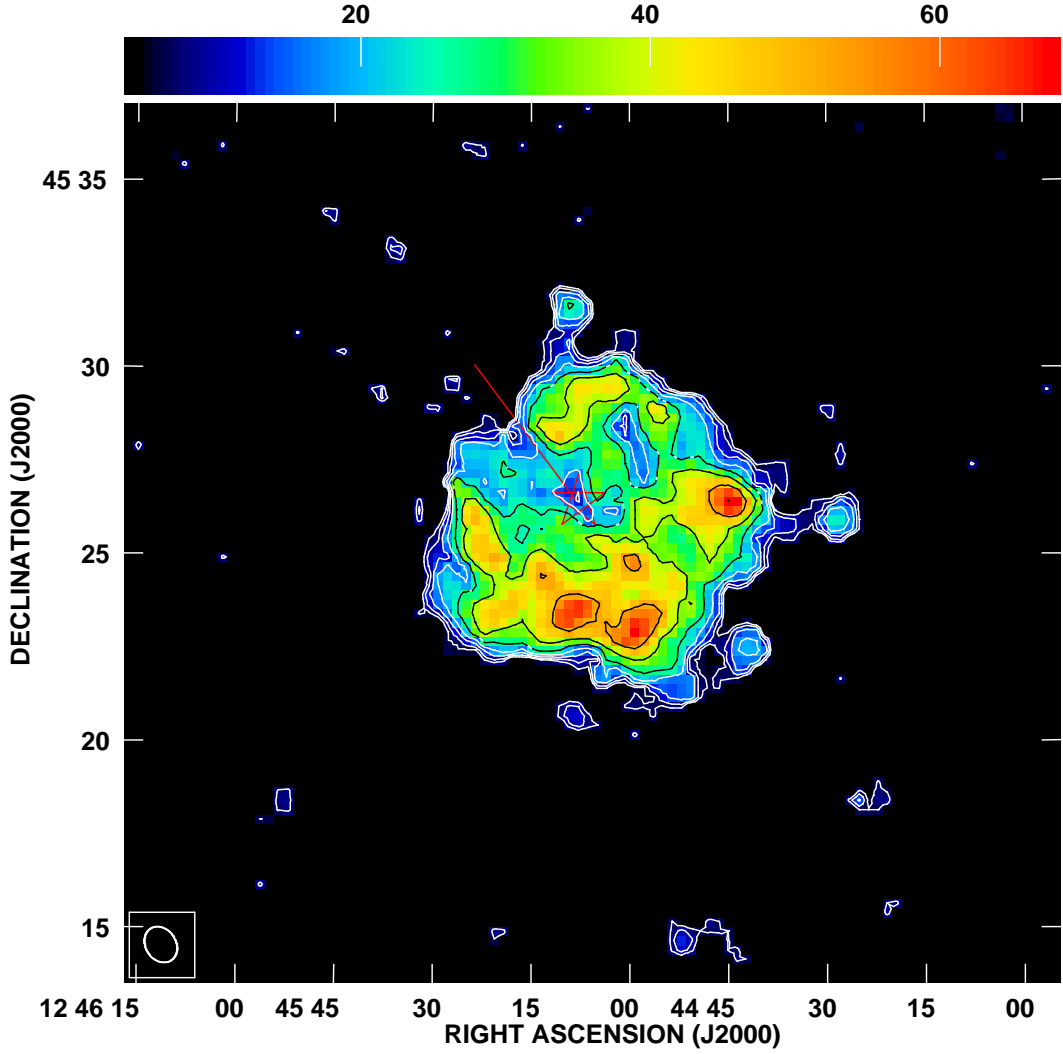}}}
\caption{\HI\ total intensity map  
of Y~CVn
derived from naturally weighted VLA data. The spatial
resolution is $\sim61''\times49''$. The 
map was constructed from emission spanning LSR velocities
17.8 to 23.0~\kms. Contour levels are
(0.5,0.7,1,1.4,2,2.8,4,5.6,8)$\times$13~Jy beam$^{-1}$ m
s$^{-1}$.  To minimize noise in the
map, data at a given point were blanked if they did
not exceed a 2$\sigma$
threshold after spatially smoothing the data by a factor of three. 
Intensity levels are 4 to 68 Jy m s$^{-1}$.
A star symbol indicates the stellar position, and the red line
indicates the direction of space motion. }
\label{fig:ycvnmom0}
\end{figure*}

An \HI\ total intensity map of Y~CVn derived from naturally weighted
data is presented in
Figure~\ref{fig:ycvnmom0}. The stellar position is offset
$\sim45''$ to the northeast relative to the geometric center of the
\HI\ emission distribution.  The \HI\ nebula is
roughly symmetric about the position angle of this offset ($\sim$42$^{\circ}$), which
corresponds to the direction of space motion of the star
(Table~5). 
In contrast to the quasi-spherical appearance of the CSE in previous FIR images
(Izumiura et al. 1996), the integrated \HI\ 
distribution exhibits a more boxy shape, with the
extent of the northeastern edge of the nebula 
being somewhat broader ($\sim9'$) than the southwest side
($\sim7'$). This boxiness is also visible in the {\it Spitzer}
70$\mu$m image of Geise (2011). 

The \HI\ distribution of Y~CVn is quite clumpy down to the
resolution limit of our data, with two of the brightest clumps
lying at the southern and western ``corners'' of the nebula. Several
additional compact, quasi-spherical clumps of gas are also detected outside the main
emission distribution and appear to be connected with the CSE by faint
($\sim3\sigma$) tendrils of gas. The nature of these structures is
unclear, but they appear to be circumstellar in nature.

The depression in the total \HI\ column density toward the stellar
position seen in the individual \HI\ channel maps is also apparent in
the \HI\ total intensity map. In addition, the \HI\ total intensity map
reveals a ``notch'' extending from the star to the northeastern edge of
the nebula along PA$\approx 42^{\circ}$ where relatively little \HI\ emission is
detected. The nature of this feature is discussed further below. 

\begin{figure}
\scalebox{0.5}{\rotatebox{0}{\includegraphics{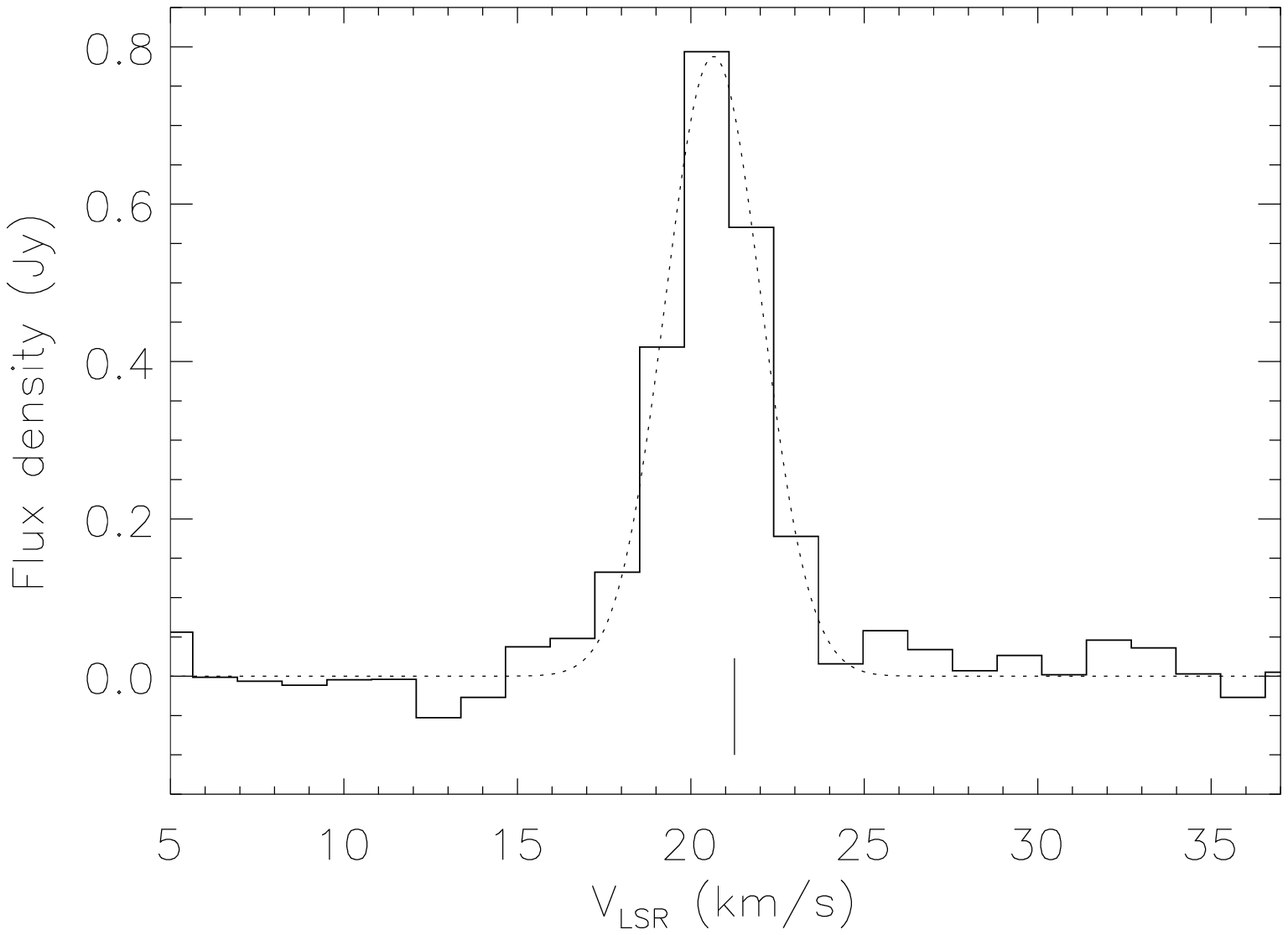}}}
\scalebox{0.5}{\rotatebox{0}{\includegraphics{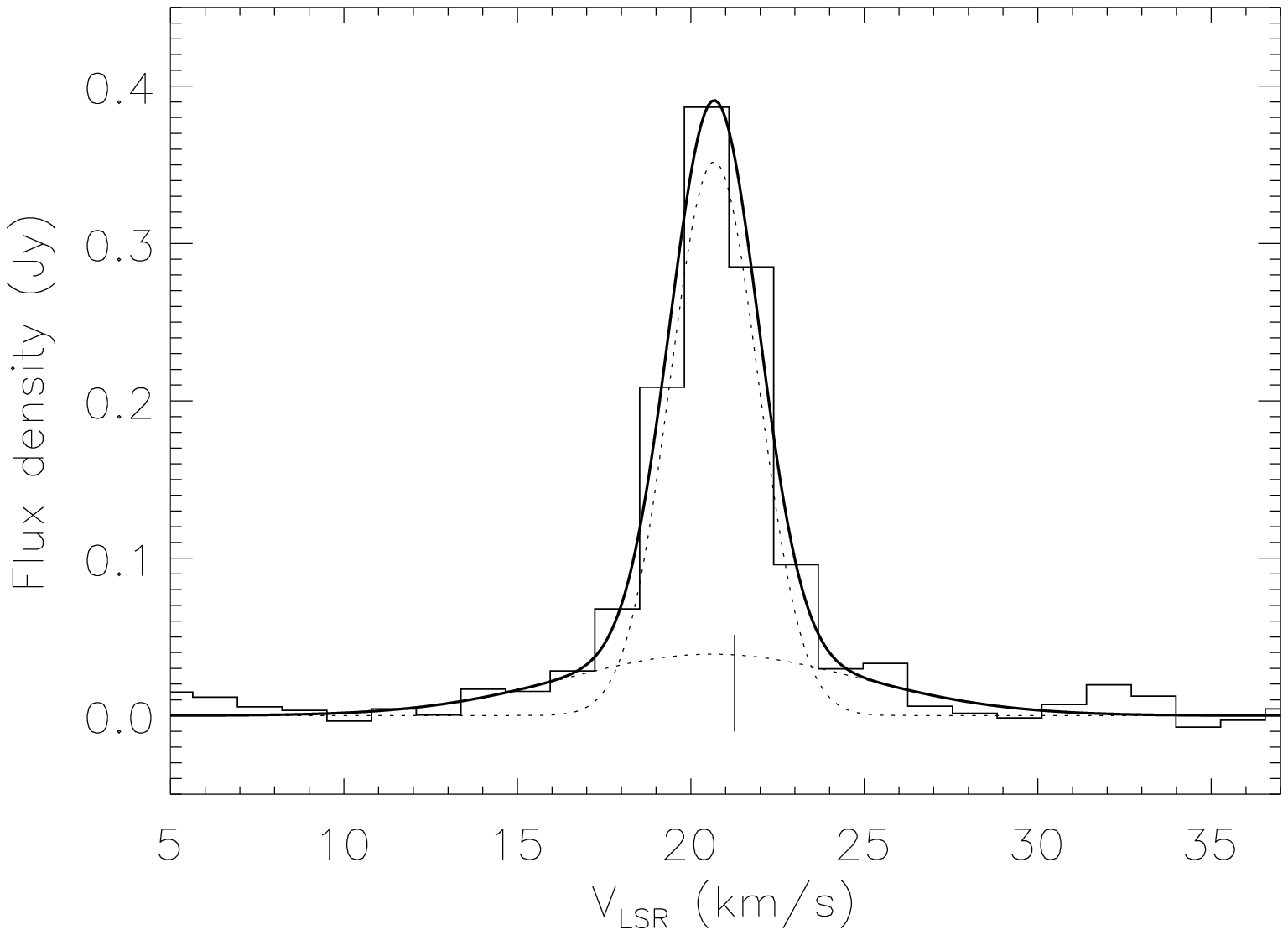}}}
\caption{Spatially integrated 
\HI\ spectra of Y~CVn, derived from naturally weighted VLA data (after
correction for primary beam attenuation). {\it Upper panel:} Spectrum obtained
by integrating the emission in each
velocity channel within a
\am{12}{0}$\times$\am{12}{2} aperture centered at 
$\alpha_{\rm J2000}$=12$^{\rm h}$ 45$^{\rm m}$ 01.2$^{\rm s}$, 
$\delta_{\rm J2000}=45^{\circ} 25' 39.9''$. The dotted line shows a
Gaussian fit to the spectrum and the vertical bar indicates the
stellar systemic velocity as determined from CO observations. {\it
  Lower panel:} Spectrum integrated within a circular aperture of radius
\am{2}{8} centered on the star. 
The dotted lines show a two-component
Gaussian fit and the thick solid line shows their sum (see text for
details). }
\label{fig:ycvnspectrum}
\end{figure}

The spatially integrated \HI\ spectrum of Y~CVn 
(Figure~\ref{fig:ycvnspectrum}) displays a narrow, roughly Gaussian shape,
similar to what we observe for V1942~Sgr (\S\ref{v1942disc}). Based on a single Gaussian
fit (dotted line on Figure~\ref{fig:ycvnspectrum}) 
we measure a peak flux density of 0.79$\pm$0.02~Jy and
a FWHM linewidth of 3.2$\pm$0.1~\kms\ centered at an
LSR velocity of  20.65$\pm$0.05~\kms. From this fit we derive an
integrated \HI\ flux density of
$\int S_{\rm HI}~d\nu=$2.68$\pm$0.11~Jy \kms, translating to an
\HI\ mass $M_{\rm HI}\approx0.047~M_{\odot}$.

\begin{figure*}
\centering
\scalebox{0.75}{\rotatebox{-90}{\includegraphics{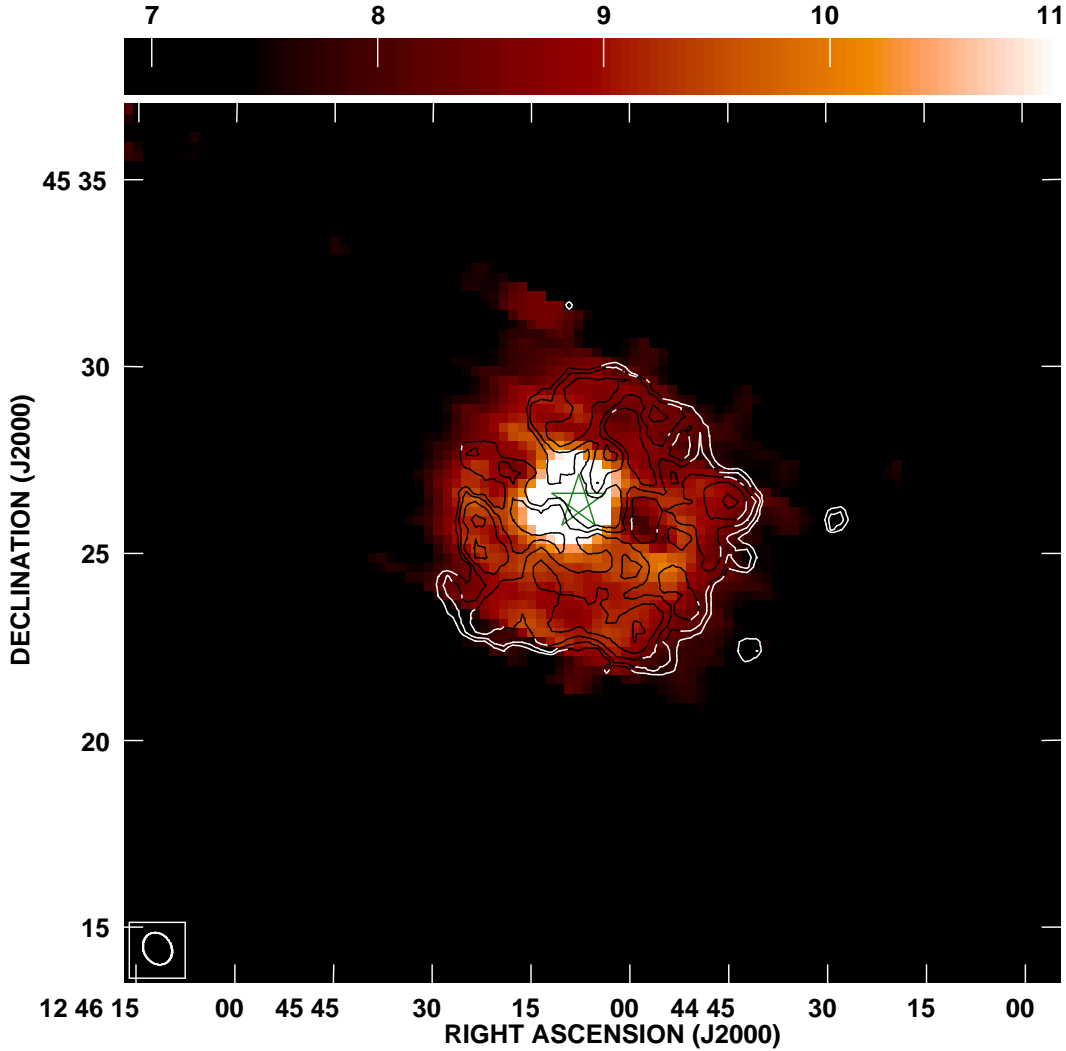}}}
\caption{\HI\ total intensity 
contours of Y~CVn derived from VLA data with robust +1 weighting, overlaid on an
{\it ISO} 90$\mu$m image of the star from Izumiura et al. 1996.  The
\HI\ data have a resolution of $\sim54''\times44''$. 
The contour levels are
(1,1.4,2,2.8,4,5.6,8)$\times$13~Jy beam$^{-1}$ m
s$^{-1}$. The {\it ISO} image has brightness units of MJy sr$^{-1}$.
  }
\label{fig:ycvnisomom0}
\end{figure*}

The \HI\ mass and global
line width that we derive  from the VLA data
are in excellent agreement with those derived previously from
NRT observations by Le~Bertre
\& G\'erard (2004), although our integrated \HI\ line flux is $\sim25$\% smaller than
reported by Libert et al. (2007). Further comparisons with prior NRT results are
described below.

\subsubsection{Discussion of Y~CVn Results\protect\label{ycvndisc}}
Above we drew attention to the presence of a ``notch'' (depression in the \HI\
column density) in the shell of Y~CVn,
lying ahead of the star and running parallel to its direction of space motion.
A similar feature was also seen in the CSE of
Y~UMa (\S\ref{yumadisc}). In both cases, a corresponding depression is also
present in the FIR light (Figure~\ref{fig:yumamom0+iso} and
Figure~\ref{fig:ycvnisomom0}, 
discussed below).  The origin of these notches is
unclear, but one possibility is that they are linked to instabilities along the leading
edge of the CSE. For example, in the hydrodynamic simulations
of mass-losing stars moving through the ISM performed by Wareing
et al. (2007a, b) and Villaver et al. (2012), 
instabilities may develop under certain conditions 
that lead to a bending of the
leading edge of the CSE (bow shock) back
toward the star.  Unfortunately, none of the currently published simulations are
closely matched to the current age, mass-loss parameters, and space
motion parameters of Y~CVn and Y~UMa.

Another interesting phenomenon seen in the \HI\ images of Y~CVn is the presence
of multiple compact 
knots of emission lying just exterior to the CSE at several
locations along both its leading and trailing edges. These knots are
detected at high significance and appear to be linked to the 
Y~CVn shell by tendrils of gas, suggesting that they are quite likely
to be  circumstellar in nature. One possibility is that these knots may
arise from thermal instabilities  (e.g., Zucker \& Soker 1993) 
along the interface between the Y~CVn shell
and the surrounding ISM.

In Figure~\ref{fig:ycvnisomom0} we present \HI\ contours derived from
the robust +1 version of our 
VLA data overlaid on the {\it ISO} 90$\mu$m data from Izumiura et
al. (1996). We downloaded the archival data for Y~CVn as described in \S\ref{yumadisc}.
The original {\it ISO} image has a pixel size of \as{15}{0} but was
rebinned to 10$''$ pixels for display purposes.

The shape and extent of the \HI\  and FIR emitting shells are
quite similar along the northwest and southwest edges. However,
along the southeast portion of the CSE, the \HI\  contours extend outside
the FIR emission detected by {\it ISO}. At the same time, an absence of \HI\
emission is apparent along the northeast (leading) 
side of the FIR-emitting shell. The gas along this edge
may have been ionized by shocks owing to the interaction with the ISM. However, 
there is no obvious enhancement in FIR surface brightness or change in
morphology along the
leading edge of the shell that would indicate the presence of a strong
bow shock capable of ionizing hydrogen. 
Similarly, no clear evidence of a FIR bow shock is seen in the more
recent, higher resolution observations from {\it Herschel} (Cox et al. 2012a). 
We have also examined
archival {\it Galaxy Evolution Explorer} ({\it GALEX}) 
images of the region around the star to search
for FUV emission from a possible bow shock as has been
detected in the case of at least three evolved stars (Martin et
al. 2007; Sahai \& Chronopoulos 2010; Le~Bertre et
al. 2012), but no extended FUV emission is visible.

Izumiura et al. (1996) noted that the western part of the FIR shell of Y~CVn
has a slightly lower surface brightness in both 90$\mu$m and 160$\mu$m
images compared with the eastern side. Similarly, we find that the \HI\
emission is slightly enhanced on the eastern side of the shell
compared with the western side (cf. Figures~\ref{fig:ycvnmom0} \&
\ref{fig:ycvnisomom0}). 
After applying a multiplicative correction of 1.4 to account for the mass
of helium, the mass we derive for the circumstellar shell of Y~CVn 
($\approx0.066~M_{\odot}$) is consistent with the
value derived previously from FIR observations 
($M_{\rm shell}\approx$0.047 to 0.16~$M_{\odot}$, scaled to our adopted
distance; 
Izumiura et al. 1996). Since
the latter determination depends on the adopted dust-to-gas ratio,
implicit to this agreement 
is that the dust-to-gas ratio of the shell is typical of
other carbon-rich stars ($\sim4.5\times10^{-3}$; Jura 1986). It also
assumes that the gas in the shell is predominantly atomic, although as
noted by Libert et al. (2007), the temperature and density
conditions within Y~CVn's shell may be conducive to molecule formation
through a non-equilibrium chemistry.  

Our VLA imaging confirms that the narrow \HI\ line component
detected by Le~Bertre \& G\'erard (2004) and Libert et al. (2007) is
closely linked with material distributed within the extended FIR-emitting shell surrounding
Y~CVn. However, both of these studies also reported a
second broad (FWHM$\sim$16~\kms) line component that they attributed
to a counterpart to the freely expanding wind traced by
molecular lines at small
radii. 
This broad component is not evident in the global \HI\
profile presented in Figure~\ref{fig:ycvnspectrum}. However, we tentatively
identify a counterpart to this component 
in Figure~\ref{fig:ycvnspectrum} (lower panel), where we plot a spectrum
derived from only the emission confined within a radius of
\am{2}{8}. (This radius was adopted based on a model of the shell
presented in Appendix~A; see also below). 
We find that a single Gaussian fit to the spectrum in the lower panel
of Figure~\ref{fig:ycvnspectrum}
leaves statistically significant residuals in the line wings. Adopting instead
a two-Gaussian fit, we find a narrow component
with $F_{\rm peak}=0.35\pm0.02$~Jy and a FWHM line width of 3.0~$\pm$0.1~\kms, centered
at $V_{\rm LSR}=20.68\pm$0.04~\kms\ (i.e., consistent in position and width with
the fit to the 
global spectrum 
in the upper panel of Figure~\ref{fig:ycvnspectrum}) and a broad component centered
at 20.6$\pm$0.5~\kms\ with
$F_{\rm peak}=0.04\pm0.02$~Jy and a FWHM line width of
10.0$\pm$1.8~\kms. The line width we derive for the broader component is smaller
than the values reported by Le~Bertre \& G\'erard (14.3~\kms) and by Libert et
al. ($\sim$16~\kms). However, Libert et al. fitted a rectangular rather
than a Gaussian line profile, and moreover, the single-dish determinations 
are sensitive to uncertainties
in baseline fitting. Thus accounting for systematic uncertainties, it
appears that all three measurements of the broad \HI\
component are roughly
consistent with the CO(3-2) line, for which Knapp et al. 1998 measured a 
FWHM$\sim$12~\kms\ and derived
$V_{\rm out}=7.8\pm$1.3~\kms. This supports the suggestion that this material is linked with
Y~CVn's freely expanding wind. Corroborating evidence is also seen in
our \HI\ channel maps (Figure~\ref{fig:ycvncmapstap}), where weak
emission ($\ge2.5\sigma$) is detected near the stellar position at velocities offset
by $\sim\pm$4 to 6~\kms\ from the stellar systemic velocity. However,
we note that there is
evidence that the \HI\ emission at these velocity extremes 
is spatially extended on scales $>1'$ and thus must be extended well
beyond the outer boundary of the detected CO(1-0) emission
($r\sim$\as{6}{5}) and the
molecular dissociation radius ($r\sim20''$; Neri et al. 1998). Additionally,
the peaks of this emission near the velocity extrema in some channels appear offset from the
stellar position by $\sim1'$. 

Our new VLA measurements allow us to provide improved constraints on the numerical model of
the \HI\ shell of Y~CVn presented by Libert et al. (2007) (see Appendix~A
for details). In Figure~\ref{fig:models}c we show our model
spectrum overplotted on the global VLA \HI\ spectrum of Y~CVn. Based on this
model, we estimate an age for  the \HI\ shell 
of $\sim 4.5\times10^{5}$~yr. This value is consistent
with the previous determination of Libert et al., but we note that we
have adopted a larger distance to the star than those authors as well
as slightly different model parameters. Our value is more than 6 times
larger than the age for the circumstellar shell previously derived by
Young et al. (1993b) based on FIR data.

\subsection{V1942 Sagittarii (V1942 Sgr)\protect\label{v1942sgr}}
V1942~Sagittarii (V1942~Sgr) has been previously classified an N-type
carbon star and a
long period irregular (Lb) variable. However, Miller et
al. (2012) recently reported evidence 
that V1942~Sgr belongs to the class of DY~Persei-type
stars.  DY~Persei stars, only a handful of which have been identified
to date,
are thought to be either a cooler subclass of R~Coronae Borealis
stars (see Clayton 2012) or normal carbon stars that have experienced ejection events
(Tisserand et al. 2009). DY~Persei-type stars generally 
exhibit C-rich spectra similar to N-type carbon stars and show
detectable $^{13}$C. They are also characterized by large-amplitude,
irregular variations in brightness  ($>$1.5 magnitudes). 
In the case of V1942~Sgr, Miller et al. reported evidence for 
multiple periodicities in the light curve variations (120, 175, and
221 days) that appear to change on time scales of 1-2 years.   
However, Soszy\'nski et al. (2009) 
found no evidence for a  marked discontinuity between the variations observed
among DY~Persei stars and other carbon-rich long-period variables.

V1942~Sgr has an effective temperature
$T_{\rm eff}\approx2960$~K (Bergeat et al. 2001) and a bolometric
luminosity of 5200$L_{\odot}$ (Libert et al. 2010a). Its properties
are believed to place V1942~Sgr on the TP-AGB (e.g., Bergeat et
al. 2002), 
although its new classification 
as a DY~Persei-type star suggests that its evolutionary status and/or
history could be somewhat
different from the other carbon stars previously studied in \HI. 
We adopt a distance
to V1942~Sgr of 535~pc based on the 
{\it Hipparcos} parallax of 1.87$\pm$0.51 mas (van Leeuwen 2007).

{\it IRAS} 60$\mu$m data show V1942~Sgr to be surrounded by an
extended infrared shell of radius \am{3}{2} ($\sim$0.5~pc; Young et
al. 1993a). 
V1942~Sgr exhibits a featureless mid-infrared (8-22$\mu$m) spectrum, consistent
with the absence of current mass loss (Olnon \& Raimond 1986). However, the changes in its
variability and its DY~Persei classification suggest that regular ejection
events may be taking place (Miller et al. 2012). Furthermore, 
Libert al. (2010a) found evidence of ongoing mass loss from CO
observations. 

The CO(1-0) and CO(2-1) profiles of V1942~Sgr obtained by Libert et
al. (2010a) exhibit a two-component structure comprising broad and narrow
components.  In several cases, such line profiles have been 
found to be the hallmark of bipolar
outflows in AGB stars (e.g., Kahane \& Jura 1996; Bergman et al. 2000; 
Josselin et al. 2000; Libert et al. 2010b; Castro-Carrizo et al. 2010), 
although CO imaging observations are needed to
confirm this interpretation for the case of V1942~Sgr. Based on the
CO(1-0) line, Libert et al. (2010a) derive a stellar systemic velocity
$V_{\star,\rm LSR}=-33.75\pm$0.25~\kms,  outflow  velocities for the
broad (narrow) line components of $V_{\rm out}=17.5\pm$0.5~\kms\
(5.0$\pm$0.5~\kms), and mass loss rates for the broad (narrow) 
components of ${\dot M}=6.1\times10^{-7}~M_{\odot}$~yr$^{-1}$ 
($1.0\times10^{-7}~M_{\odot}$~yr$^{-1}$). Results from the CO(2-1)
line were comparable to within uncertainties. We note that the outflow velocity
of V1942~Sgr is unusually large for an Lb variable (e.g., Olofsson et
al. 2002), a fact that
may be linked with its DY~Persei status.

The proper motions derived from {\it Hipparcos} measurements of V1942~Sgr are
10.98$\pm$0.59~mas yr$^{-1}$ in right ascension and 
$-5.10\pm0.43$~mas yr$^{-1}$ in declination. From these values we
derive a Galactic
peculiar space velocity
$V_{\rm space}$=40.2~\kms\ at a position angle of \ad{95}{1} (Table~5).

\subsubsection{VLA Results for V1942~Sgr}
\HI\ channel maps for V1942~Sgr derived from our naturally weighted VLA
data are presented in Figure~\ref{fig:v1942cmaps}. An \HI\ total
intensity map derived from the same data is shown in
Figure~\ref{fig:v1942sgrmom0}. 
As seen  in Figure~\ref{fig:v1942cmaps} (and in 
Figure~\ref{fig:v1942spectrum}, discussed below), 
V1942~Sgr lies at a velocity that suffers
very low contamination from \HI\ emission along the line-of-sight (see
also Libert et al. 2010a). We are able to
trace  circumstellar \HI\ emission at or near the stellar 
position at $\ge3\sigma$ significance
across five contiguous spectral channels ($V_{\rm LSR}=-35.2$ to $-30.1$~\kms).
The emission distribution at each of these velocities appears somewhat
amorphous, but with a sharp drop-off in \HI\ surface
density at its boundaries. Versions of our \HI\ data cube made using
$u$-$v$ tapering (not shown) do not reveal any emission extended beyond what is
seen in Figure~\ref{fig:v1942cmaps}.

\begin{figure*}
\centering
\scalebox{0.9}{\rotatebox{0}{\includegraphics{f18.ps}}}
\caption{\HI\ channel maps bracketing the LSR velocity of V1942~Sgr, derived
from naturally weighted VLA data. The spatial
resolution is $\sim89''\times49''$.
A star symbol indicates the stellar position.
Contour levels are 
($-$4.2[absent],$-$3,3,4.2,6,8.4)$\times$2.7~mJy beam$^{-1}$. The lowest
contour is $\sim3\sigma$. 
The field-of-view shown is comparable to the VLA primary beam. 
  }
\label{fig:v1942cmaps}
\end{figure*}
\begin{figure*}
\scalebox{0.75}{\rotatebox{-90}{\includegraphics{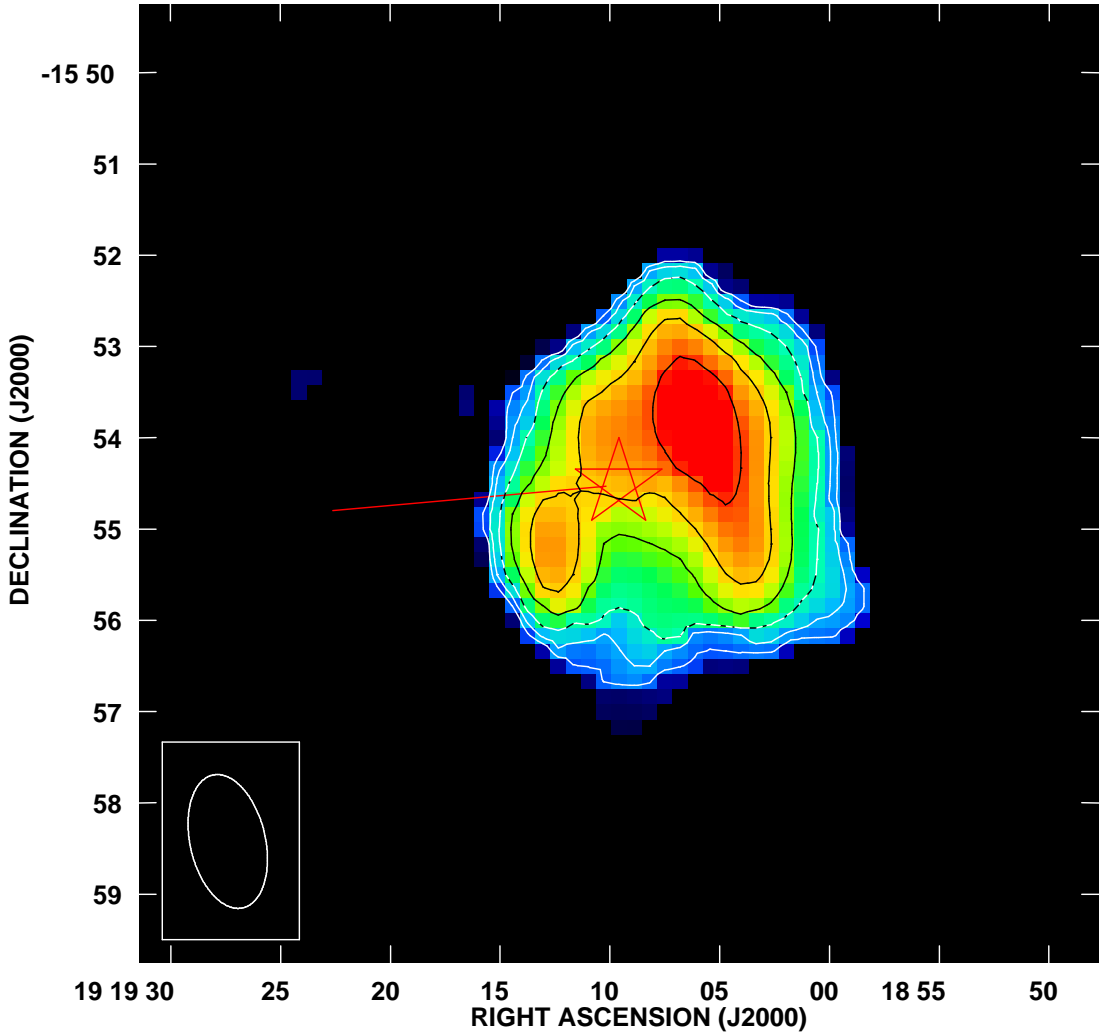}}}
\caption{\HI\ total intensity 
map of V1942~Sgr derived from naturally weighted VLA data. The spatial
resolution is $\sim89''\times49''$. A star symbol
indicates the stellar position and the red line indicates the direction of
space motion.  The 
map was constructed from emission spanning LSR velocities
$-35.2$ to $-30.1$~\kms. Intensity levels are 0 to 120 Jy beam$^{-1}$
m s$^{-1}$ and contour levels are
(1.4,2,2.8,4,5.6,8)$\times$14.1~Jy beam$^{-1}$ m
s$^{-1}$. To minimize noise in the
map, data at a given point were blanked if they did
not exceed a 2$\sigma$
threshold after smoothing by a factor of three
spatially and spectrally. }
\label{fig:v1942sgrmom0}
\end{figure*}

\begin{figure}
\scalebox{0.5}{\rotatebox{0}{\includegraphics{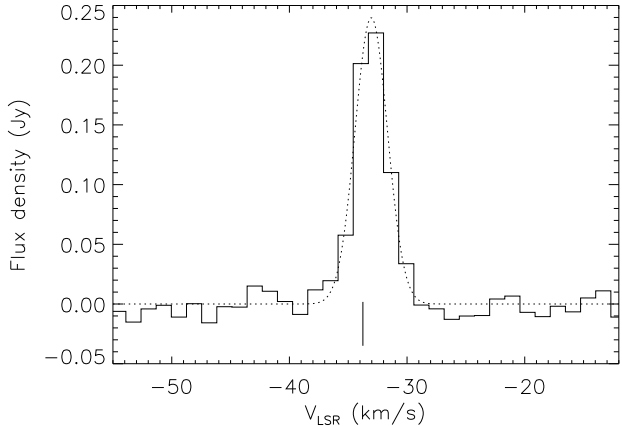}}}
\caption{Spatially integrated 
\HI\ spectrum of V1942~Sgr. The spectrum was derived from naturally weighted VLA data 
(after correction for primary beam attenuation) by integrating the emission in each
velocity channel within a
\am{4}{7}$\times$\am{5}{5} aperture centered at 
$\alpha_{\rm J2000}$=19$^{\rm h}$ 19$^{\rm m}$ 06.5$^{\rm s}$, 
$\delta_{\rm J2000}=-15^{\circ} 54'$ \as{45}{0}. The vertical line
indicates the stellar systemic velocity derived from CO
observations. The dotted line shows a Gaussian fit. }
\label{fig:v1942spectrum}
\end{figure}

To first order, the velocity-integrated \HI\ emission distribution of
V1942~Sgr (Figure~\ref{fig:v1942sgrmom0}) has a shell-like morphology, 
reminiscent of Y~UMa
(\S\ref{yuma}) and Y~CVn (\S\ref{ycvn}). Similar to those other two
stars, the outer boundaries of the shell are irregular.
The total extent of the \HI\ emission of V1942~Sgr is roughly \am{4}{0}
($\sim$0.6~pc) east-west 
and \am{4}{6} ($\sim$0.7~pc) north-south, as measured through the position of the
star. The stellar position is offset $\sim30''$ to the east of the geometric
center of the \HI\ nebula. Other noteworthy features are the 
two enhancements in \HI\ surface brightness that
are observed southeast and northwest of the star, respectively (see also
Figure~\ref{fig:v1942cmaps}). The northwestern
feature is the brighter of the two and lies atop a spatially extended ridge of
enhanced surface brightness. Both of the \HI\ enhancements
lie along a  line that bisects the
stellar position at PA$\approx126^{\circ}$, although the significance
of this is unclear, and 
the two surface brightness enhancements do not show any obvious
relationship to the position angle of the peculiar space motion of the star 
(PA$\approx95^{\circ}$). 

A Gaussian fit to the integrated \HI\ 
spectrum in  Figure~\ref{fig:v1942spectrum} yields a peak flux density
of 0.240$\pm$0.009~Jy, a FWHM line width of
3.23$\pm$0.13~\kms, and a line centroid of $V_{\rm
  LSR}=-33.04\pm0.06$~\kms. From these fit parameters, we derive
an integrated \HI\ flux density 
$\int S_{\rm HI}~d\nu=0.82\pm$0.04~Jy~\kms, translating to a total \HI\ mass 
$M_{\rm HI}\approx0.055~M_{\odot}$.  
Our measurements are in
good agreement with the \HI\ line parameters reported by Libert
et al. (2010a).

\subsubsection{Discussion of V1942~Sgr Results\protect\label{v1942disc}}
The maximum angular extent of the \HI-emitting shell that we observe surrounding
V1942~Sgr ($\sim$\am{4}{6}
across) is markedly smaller than the FIR diameter derived by Young et
al. (1993a) from {\it IRAS} data ($\sim$\am{6}{4}).  Furthermore, the good agreement
between the \HI\ size and \HI\ line parameters that we derive in the
present study compared with those derived from the NRT by Libert et al. (2010a)
also suggests that V1942~Sgr has little or no circumstellar \HI\ emission extended
beyond what is detected in our VLA images. Thus in contrast to the
carbon star Y~CVn (\S~\ref{ycvn}), our findings suggest that the \HI\
shell of V1942~Sgr may be
significantly more compact than the FIR-emitting shell. Simulations by van Marle et
al. (2011) have shown that this type of segregation can occur
preferentially for
large dust grains (radius $\lsim$0.105$\mu$m). Higher resolution 
observations would be of considerable interest for confirming the FIR
extent of V1942~Sgr and
for examining in detail the
relationship between the FIR and \HI\ emission. 

Libert et al. (2010a) found evidence in their integrated \HI\ line
profile of V1942~Sgr for a weak ($\sim6\pm$2~mJy) pedestal of emission extending
from $-$39 to $-27$~\kms. Based on the results of numerical modeling, 
they attribute this weaker,
broader line component to a counterpart to 
the freely expanding wind at small radii (as traced by the
narrower of the two CO line components detected by these authors). They
interpret the dominant narrow \HI\ line component as arising from a
detached shell that has been decelerated as a result of its
interaction with the ISM. Thus the interpretation of the line profile is analogous to the
cases of RX~Lep, Y~UMa, and Y~CVn (see Appendix~A).

We do not find any evidence for a counterpart to the broad \HI\ emission component in
our integrated VLA line profile (Figure~\ref{fig:v1942spectrum}) or in our \HI\
channel maps (Figure~\ref{fig:v1942cmaps}). In the channel maps, we
would expect the broad component to give rise to 
emission at or near the stellar position at
velocities of roughly $V_{\star}\pm5$~\kms. 
However, the absence of this
component in our data is not surprising. Since its expected peak line strength of
only $\sim6\pm2$~mJy,  it is near the limits of detectability
in our present data (where RMS noise per channel is $\sim$2.7~mJy
beam$^{-1}$; Table~4).

We have also searched our \HI\ data cube for a counterpart to the
broad CO line component detected by Libert et al. (2010a). Consistent
with Libert et al., we find no evidence of \HI\ emission at velocities 
near $V_{\star}\pm$17.5~\kms. Libert et al. suggested that this
high-velocity outflow is likely to result from very
recent mass-loss ($\lsim10^{4}$~years) and that it may be a signpost
of the star being close to the end of its life on the AGB. This
possibility is interesting in light of the newly established DY~Persei
classification of V1942~Sgr, since some workers have suggested that
these may
be highly evolved stars nearing the planetary nebula stage 
(see Miller et al. 2012 for discussion). 
\HI\ and CO observations of additional
DY~Persei stars would be valuable for establishing whether
the CSE properties of V1942~Sgr are typical of its class.

We have estimated an age for the \HI\ shell surrounding V1942~Sgr
using a simple numerical model (see Appendix~A and Table~A1 for
details). The resulting fit to the global \HI\ spectrum is presented
in Figure~\ref{fig:models}d.
We estimate that a duration of $\sim7.2\times10^{5}$~yr
was required to form the observed shell. This is $\sim$20\% higher
than the
previous estimate from Libert et al. (2010a) based on constraints from
single dish data and more than 6 times larger than the CSE age derived
by Young et al. (1993b) based on FIR data.

\subsection{AFGL~3099\protect\label{crl3099}}
AFGL~3099 (=CRL~3099=IZ~Peg) is a carbon star and Mira-type variable with a period
of 484 days (Le~Bertre 1992). This star has a very low effective
temperature (1800-2000~K;
G\'erard \& Le~Bertre 2006) and is the coolest AGB star observed in \HI\
with the VLA to date. We adopt a distance to the star of 1500~pc based
on the period-luminosity relation of Groenewegen \& Whitelock
(1996). 

AFGL~3099 is thought to be in a highly evolved evolutionary
state (Gehrz et al. 1978). It is similar to the prototypical
carbon star IRC+10216 in that
one of the few known examples of Galactic carbon stars to show regular,
large-amplitude variability and exhibit a pulsation period longer than normal
(optically selected) carbon Miras (Feast et al. 1985). 

Knapp \& Morris (1985) detected CO(1-0) emission from AFGL~3099 from
which they derived a rather high mass-loss rate of
1.3$\times10^{-5}~M_{\odot}$~yr$^{-1}$ (after scaling to our adopted distance).
They also found a stellar systemic velocity of $V_{\star,\rm
LSR}=46.6\pm0.4$~\kms\ and a wind outflow speed of $V_{\rm
    out}=10.1\pm$0.5~\kms. 
The extremely red color of AFGL~3099 implies it is surrounded
by a dense circumstellar dust shell (Gehrz et al. 1978). 
AFGL~3099 was not found to be an extended FIR source based on {\it
  IRAS} observations (Young et al. 1993a), although 
this may have been a consequence of the relatively large stellar distance.

\subsubsection{VLA Results for AFGL~3099}
\HI\ channel maps of the AFGL~3099 field are presented in
Figure~\ref{fig:crl3099cmaps} and a global \HI\ spectrum coincident
with the stellar position is presented in
Figure~\ref{fig:crl3099spectrum}. 
We find no evidence for \HI\ emission
associated with the CSE of AFGL~3099 despite the presence of negligible 
interstellar contamination. No emission is seen coincident with the
stellar position at $\ge 2.5\sigma$ in any spectral channel within the
band. Assuming a Gaussian line profile with FWHM equal to twice the
outflow velocity measured in CO, we derive a 3$\sigma$ upper limit for
the \HI\ mass of AFGL~3099 within a single synthesized beam centered on
the star of $M_{\rm
  HI}<0.058M_{\odot}$. This is consistent with the 1$\sigma$ limit
previously reported by G\'erard \& Le~Bertre (2006).

\begin{figure*}
\centering
\scalebox{0.9}{\rotatebox{-90}{\includegraphics{f21.ps}}}
\caption{\HI\ channel maps bracketing the systemic velocity of AFGL~3099
  derived from naturally weighted VLA data. The spatial resolution is $\sim64''\times52''$.
A star symbol indicates the stellar position. 
Contour levels are 
($-$3,3,4.2[absent])$\times$1.7~mJy beam$^{-1}$. The lowest
contour is $\sim3\sigma$. The
field-of-view shown is comparable to the VLA primary beam. 
  }
\label{fig:crl3099cmaps}
\end{figure*}

\begin{figure}
\scalebox{0.5}{\rotatebox{0}{\includegraphics{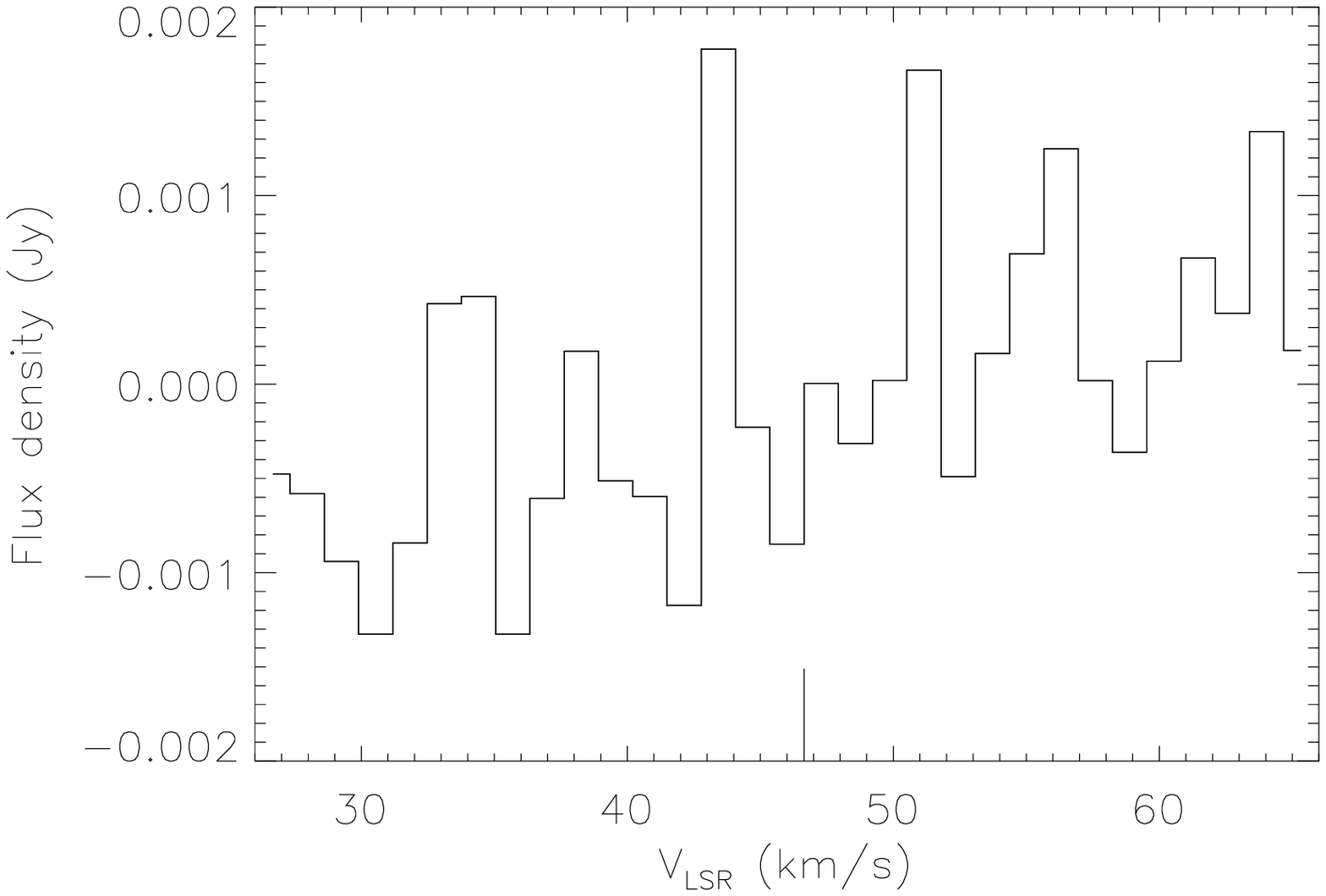}}}
\caption{VLA \HI\ spectrum of AFGL~3099 obtained by averaging over a
  single synthesized beam centered on the star.  The vertical line indicates the
  systemic velocity of the star as determined from CO observations.
  }
\label{fig:crl3099spectrum}
\end{figure}

\subsubsection{Discussion of AFGL~3099 Results}
The absence of detectable \HI\ emission directly toward the position of AFGL~3099
despite its high mass loss rate 
is not entirely surprising. AFGL~3099 is significantly more distant
than other stars in our sample (Table~1), and
the extremely low effective temperature of the
star suggests that the bulk of its wind may be molecular
(Glassgold \& Huggins 1983) or simply too cold to emit efficiently in the
21-cm line. However,  another  similarly cool and
distant carbon
Mira, AFGL~3068, was detected in \HI\ by G\'erard \& Le~Bertre
(2006). Based on the data of Knapp \& Morris (1985), 
AFGL~3099 has a $\sim$3 times weaker integrated CO line intensity
compared with AFGL~3068, but its peak \HI\ intensity is more than 8 times weaker.

In the case of the cool and highly evolved carbon Mira
IRC+10216 ($T_{\rm eff}\sim$2200~K), Matthews \& Reid (2007) 
found no \HI\ emission coincident with the position of the
star (see also Le~Bertre \& G\'erard 2001),
but detected arcs of \HI\ emission displaced from the
central position. These arcs were later found to be associated with the
interface between the CSE and the ISM and to lie along an
astrosheath and vortical tail visible in the FUV 
(Matthews et al. 2011a). We do
not find any sign of analogous
features in the AFGL~3099 field, although the large distance of this
star from the Galactic plane suggests that the local ISM density is 
likely to be significantly lower than in the case of IRC+10216.

Another possible explanation for the absence of extended \HI\ emission
around AFGL~3099 is that the star has only recently began undergoing mass loss at a
high rate (cf. Libert et al. 2010a).  We further note that 
the outflow velocity for AFGL~3099 is
surprisingly low for a star this red (e.g., Loup et al. 1993, their
Figure~8), suggesting that its
mass-loss properties may be atypical.


\subsection{TX Piscium (TX Psc)\protect\label{txpsc}}
The N-type carbon star TX~Piscium (TX~Psc) is frequently classified as an
irregular (Lb) variable, although evidence for a period of $\sim$220 days has
been reported by Wasatonic (1995, 1997).  TX~Psc was among the first
evolved stars in which 
technetium was detected (Merrill 1956) and it is thought to be in
a post-Mira evolutionary phase (Judge \& Stencel 1991). 
TX~Psc is the warmest carbon
star that has been imaged in \HI\ to date, with $T_{\rm eff}=3115\pm$130~K
(Bergeat et al. 2001).
Based on the {\it Hipparcos} parallax of 3.63$\pm$0.39~mas (van
Leeuwen 2007) we adopt a distance to TX~Psc of 275~pc. A recent
analysis by Klotz et al. (2013) suggests that the initial mass of
TX~Psc was in the range 1-3~$M_{\odot}$.

Using the CO(1-0) data of Eriksson et al. (1986), 
Loup et al. (1993) 
derived  an outflow velocity for TX~Psc's wind of $V_{\rm
out}=10.7\pm1.9$~\kms\ and a mass loss rate ${\dot
    M}=(3.1\pm1.6)\times10^{-7}~M_{\odot}$~yr$^{-1}$ (scaled to our
  adopted distance).  The stellar systemic velocity as derived from CO
observations varies among different authors depending on the telescope
and the transition used; here we adopt the mean of
the values quoted by Olofsson et al. (1993): $V_{\star,\rm
  LSR}=12.2\pm1.1$~\kms, where the quoted uncertainty is based on the
dispersion in the measurements. The {\it Hipparcos}
proper motions of $-33.68\pm0.40$~mas yr$^{-1}$ in right ascension and $-24.49\pm0.32$~mas
yr$^{-1}$ in declination imply a peculiar space velocity of 
$V_{\rm space}=66.4$~\kms\ along a
position angle of 248$^{\circ}$ (Table~5).

\subsubsection{VLA Results for TX~Psc}
\begin{figure*}
\centering
\scalebox{0.9}{\rotatebox{-90}{\includegraphics{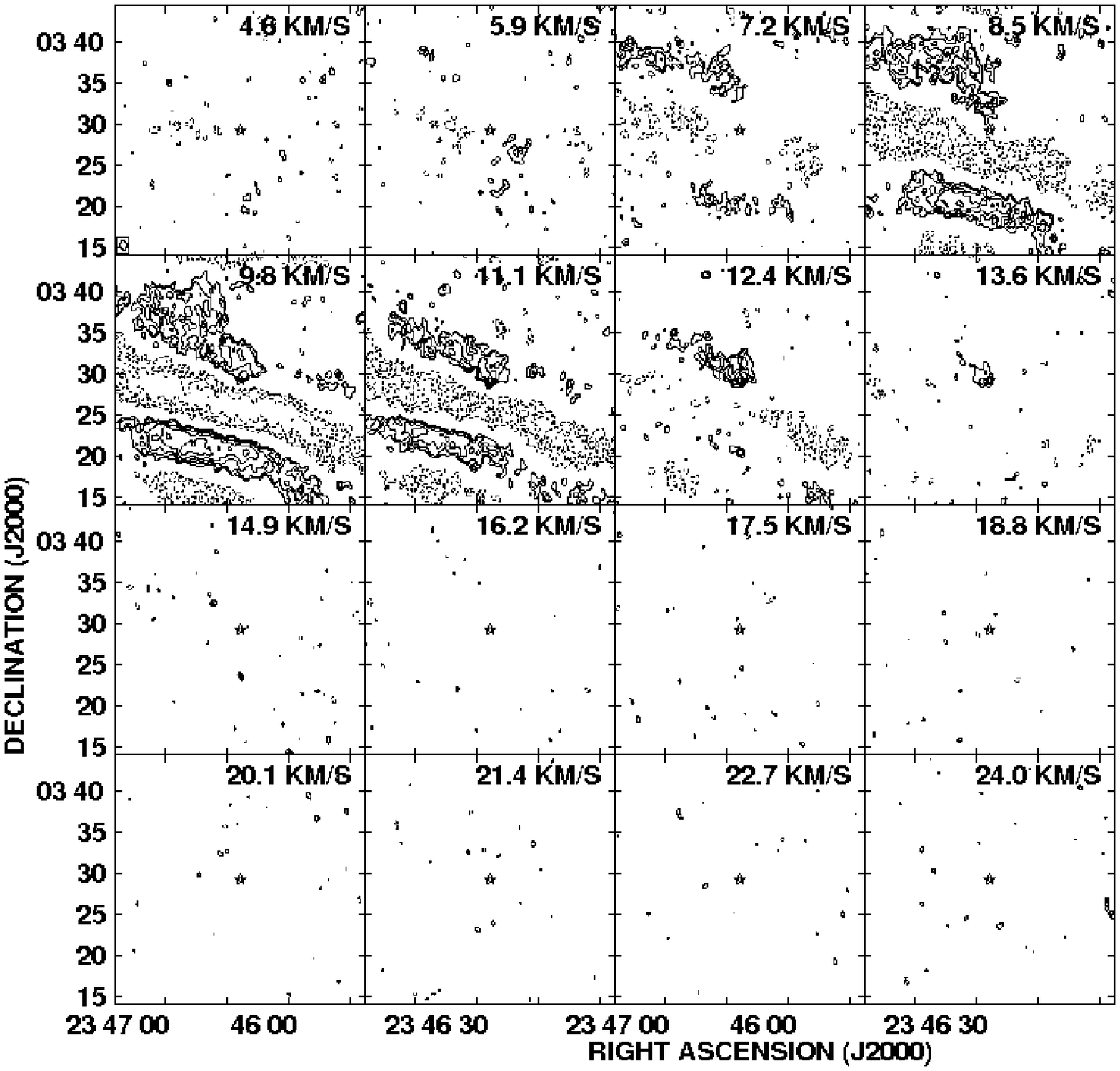}}}
\caption{\HI\ channel maps bracketing the systemic velocity of TX~Psc
  derived from naturally weighted VLA data. The spatial
resolution is $\sim72''\times49''$.
A star symbol indicates the stellar position. 
Contour levels are 
$(-6,-4.2,-3,3,4.2,...12)\times$2~mJy beam$^{-1}$. The lowest
contour is $\sim3\sigma$. The
field-of-view shown is comparable to the VLA primary beam. 
  }
\label{fig:txpsccmaps}
\end{figure*}

We present \HI\ channel maps for TX~Psc  in
Figure~\ref{fig:txpsccmaps}. In the spectral channel centered at
$V_{\rm LSR}=$12.4~\kms\ (which is the nearest to the systemic
velocity of the star), \HI\ emission  is detected at the
position of TX~Psc and is seen extending several arcminutes to the
northeast. 
Spatially resolved emission is also clearly detected
toward the stellar position at $V_{\rm LSR}=13.6$~\kms. The spectral channels
with $V_{\rm LSR}\ge 12.4$~\kms\ appear to 
suffer only minimally from line-of-sight 
confusion, implying that
the detected emission almost certainly arises from the CSE of TX~Psc. 
The distribution of this material is consistent with a
gaseous wake trailing the star. 
The \HI\ emission structure in these velocity channels also resembles 
the morphology and orientation of the faint
FIR-emitting wake seen in the {\it Herschel} images of Jorissen et
al. (2011) and the {\it Spitzer} 70$\mu$m image of Geise (2011).

\begin{figure*}
\centering
\scalebox{0.75}{\rotatebox{0}{\includegraphics{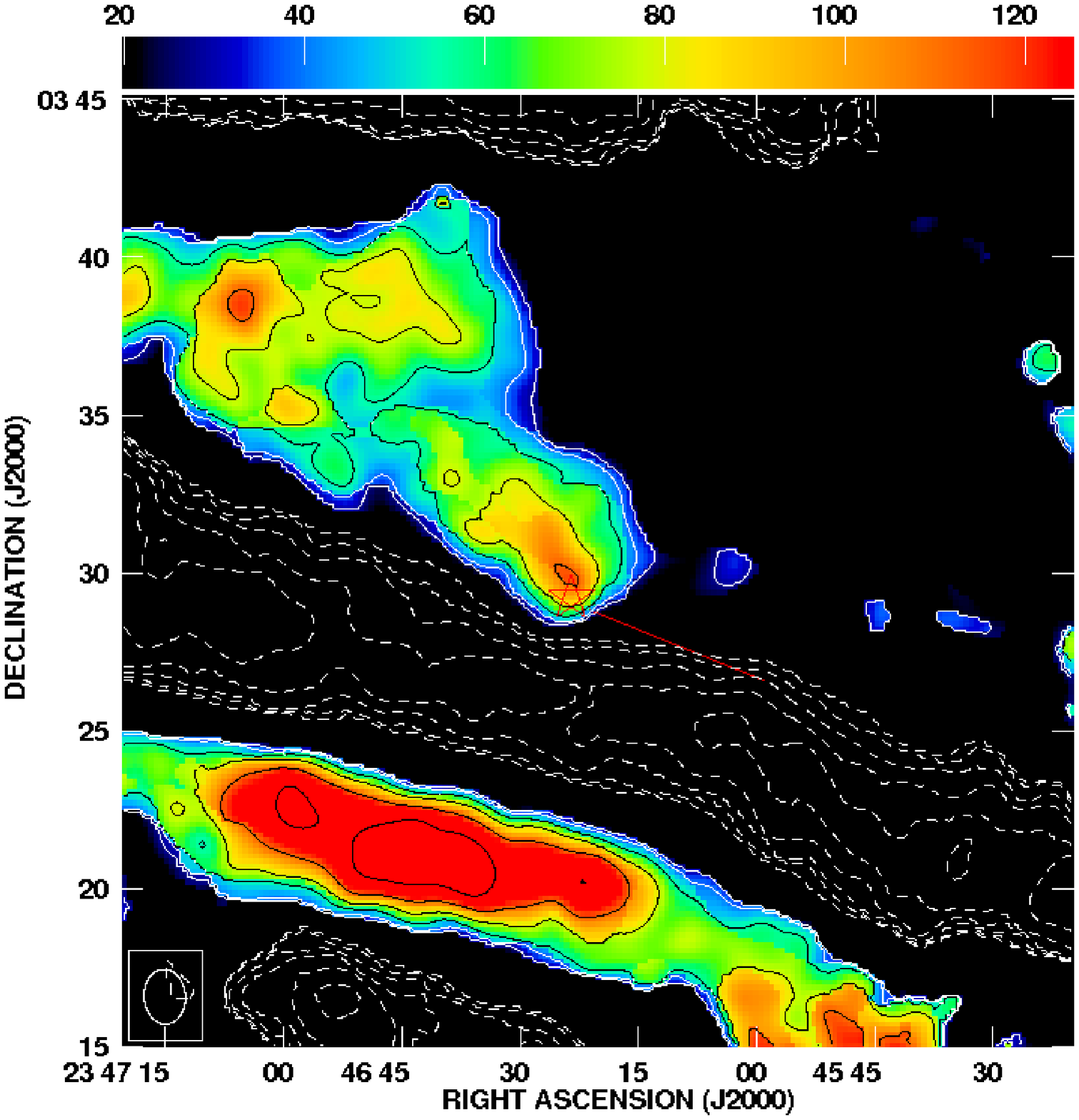}}}
\caption{\HI\ total intensity map of TX~Psc,
  derived from tapered VLA data, incorporating
  emission at LSR velocities from 8.5 to 13.6~\kms. The spatial
resolution is $\sim104''\times79''$. A star indicates the stellar
  position and the red line indicates the direction of space motion. The contour levels are
  ($-5.6,-4,-2.8,-2,-1.4,-1,1...5.6$)$\times$28~Jy beam$^{-1}$ m s$^{-1}$ and intensity
  levels are 20 to 125~Jy beam$^{-1}$ m s$^{-1}$. The data have been
  corrected for primary beam attenuation. The emission associated with
  the CSE of TX~Psc exhibits a
  head-tail morphology, although the
  bulk of the emission
east of $\alpha_{\rm J2000}=23^{\rm h} 46^{\rm m} 48^{\rm s}$ 
and north of $\delta_{\rm
  J2000}=03^{\circ} 35' 35.8''$ is probably unrelated to the
star (see text). To minimize noise in the
map, data at a given point were blanked if
  their absolute value did
not exceed a 3$\sigma$
threshold after spatially smoothing the data by a factor of three in
  velocity and factor of 5 spatially. The positive and negative 
bands running across the image arise from large-scale interstellar 
emission that is poorly spatially
  sampled by the VLA. 
  }
\label{fig:txpscmom0}
\end{figure*}

Looking toward more blueshifted velocities ($V_{\rm LSR}=$7.2 to 11.1~\kms),
we see evidence of an increasingly
extended trail of emission stretching behind TX~Psc. This emission 
appears well-collimated out to roughly $8'$ from the star. 
However, within this velocity range there
are also clear signatures of large-scale emission that is poorly
sampled by the VLA. The incomplete spatial sampling of this 
large-scale emission results in the occurrence of
positive and negative bands  across the field (see also
\S\ref{txpscdisc}).  These artifacts make
it difficult to accurately and unambiguously determine the full extent of the \HI\
wake associated with TX~Psc. However, a careful examination of the
data does provide some important clues.

In the spectral channel centered at $V_{\rm LSR}=11.1$~\kms, a 
``kink'' is visible in the emission trailing to the northeast of
TX~Psc, near 
$\alpha_{J2000}=23^{\rm h} 46^{\rm m} 48^{\rm s}$,  
$\delta_{J2000}=03^{\circ} 33' 02''$.
The presence of this discontinuity raises some doubt that the material east of
this position is related to the circumstellar wake. Furthermore,
in the 
$V_{\rm LSR}=$8.5 and 9.8~\kms\ velocity channels, an extended
\HI\ ``cloud'' is observed to the northeast of the star whose characteristics
suggest that it is not associated with the
CSE of TX~Psc. This becomes
further evident in the \HI\ total
intensity map in Figure~\ref{fig:txpscmom0}.

\begin{figure}
\scalebox{0.5}{\rotatebox{0}{\includegraphics{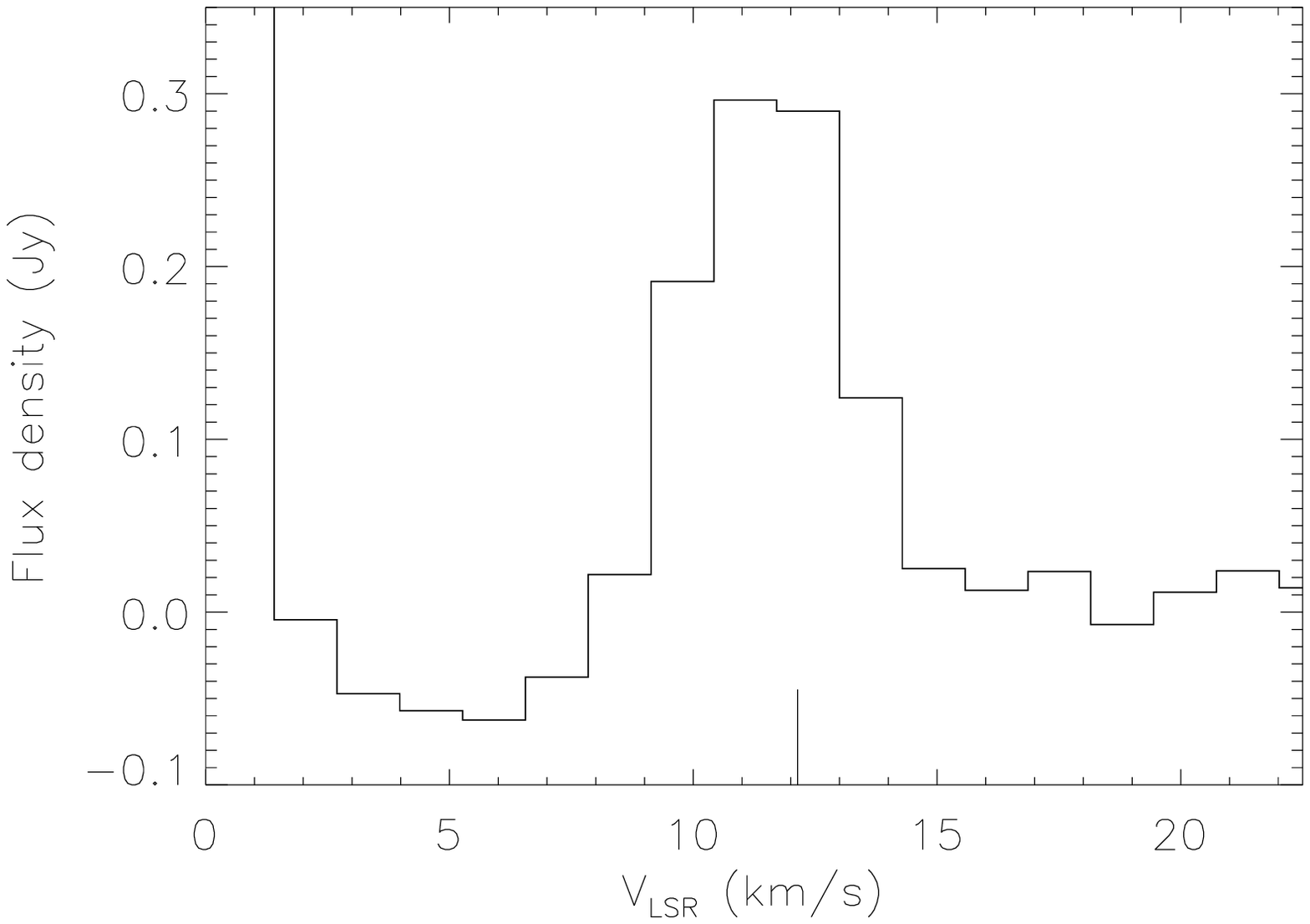}}}
\caption{Spatially integrated \HI\ spectrum of TX~Psc derived from
  naturally weighted VLA data (after correction for primary beam
  attenuation). Emission integrated over two adjacent rectangular apertures was summed to
  produce the final spectrum; the first aperture was
  \am{6}{2}$\times$\am{6}{8}, centered 
at $\alpha_{\rm J2000}=23^{\rm h} 46^{\rm m} 29.8^{\rm s}$,
  $\delta_{\rm J2000}=03^{\circ} 30' 27.5''$; the second was
  \am{2}{8}$\times$\am{4}{0}, centered 
at $\alpha_{\rm J2000}$=23$^{\rm h}$ 46$^{\rm m}$ 47.9$^{\rm s}$,
  $\delta_{\rm J2000}=03^{\circ} 32' 42.5''$. The portion of the spectrum
  blueward of $V_{\rm LSR}\approx$8~\kms\ is affected by sidelobes 
from large-scale Galactic emission. The vertical bar indicates the
  systemic velocity of the star as determined from CO observations.
  }
\label{fig:txpscspectrum}
\end{figure}

Figure~\ref{fig:txpscmom0} highlights a tapering of the
emission trailing TX~Psc, occurring roughly $8'$ behind the star. This
is analogous to
what has been observed in the \HI\ wakes of other AGB stars (Matthews
et al. 2008, 2011b). However, beyond this, there
is a dip in the \HI\ column density followed by an abrupt change in the
opening angle of the trailing emission. The marked discontinuity in
column density is evident even in
the $u$-$v$ tapered version of our data  
where we expect improved sensitivity to
extended, low surface brightness emission. Furthermore,
the overall morphology of this emission beyond this point (i.e., at
$r\gsim8'$ from the star) appears to be generally
inconsistent with a circumstellar 
wake (cf. Villaver et al. 2002, 2012; Wareing et
al. 2007a, b, c). While wide-angle
circumstellar wakes
may indeed arise from mass-losing stars 
under certain conditions (e.g., Comer\'on \& Kaper 1998; 
Wareing et al. 2007a; Sahai \& Chronopoulos 2010; Matthews et
al. 2012), a change from narrow
collimation near the star to a broad, wide-angle wake at larger
projected distances would be difficult to explain. For these reasons, we
conclude that only the gas within $\lsim 8'$ from TX~Psc can be
convincingly associated with a trailing circumstellar wake. While we
cannot rule out the presence of some
additional circumstellar debris beyond this radius (see below), the
dominant emission in our VLA maps at these projected distances from
the star appears
to be interstellar in origin.
Under these assumptions, we have derived the spatially integrated \HI\
spectrum of TX~Psc shown in (Figure~\ref{fig:txpscspectrum}). Based on
a Gaussian fit to this spectrum (including only data with $V_{\rm
  LSR}\ge8.5$~\kms), we derive a peak \HI\ flux density of
0.35$\pm$0.02~Jy, a line centroid of $V_{\rm LSR}=11.4\pm$0.1~\kms,
and a FWHM line width of 4.0$\pm$0.2~\kms. From this we measure an
integrated \HI\ flux density of 
$\int S_{\rm HI}~d\nu=1.48\pm0.11$~Jy \kms, translating to
an \HI\ mass $M_{\rm HI}\approx0.026~M_{\odot}$. 

\begin{figure*}
\centering
\scalebox{0.9}{\rotatebox{0}{\includegraphics{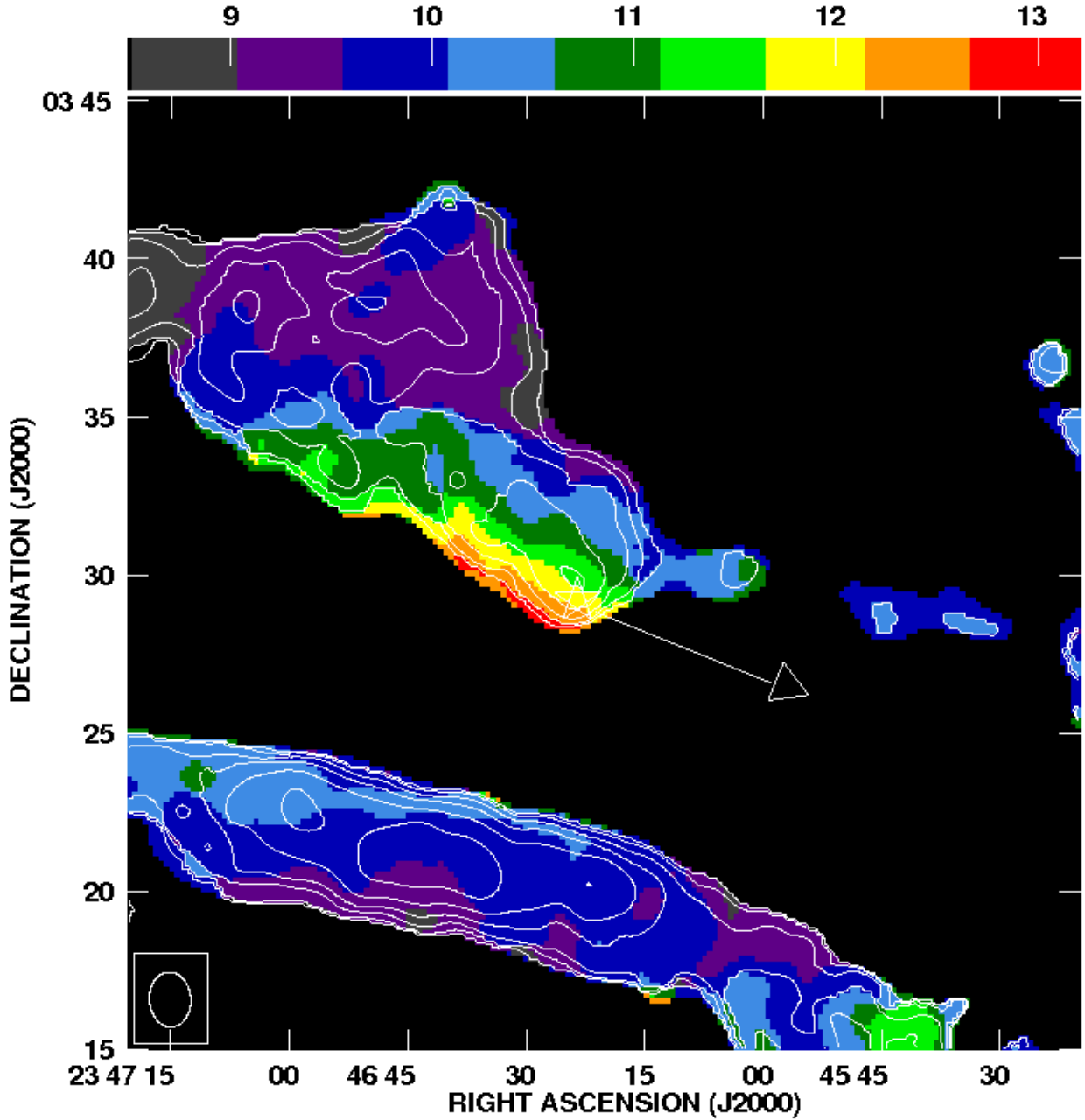}}}
\caption{Intensity-weighted \HI\ velocity field of TX~Psc,
  derived from tapered VLA data. The spatial
resolution is $\sim104''\times79''$. The arrow
  indicates the direction of space motion of the star.
Note the presence of velocity gradients both parallel 
and perpendicular to the direction of motion. The \HI\ total intensity
  contours from Figure~\ref{fig:txpscmom0} (positive values only) are overplotted for
  reference.  The color bar shows LSR radial velocity in units of \kms.
  }
\label{fig:txpscmom1}
\end{figure*}
\begin{figure}
\scalebox{0.5}{\rotatebox{0}{\includegraphics{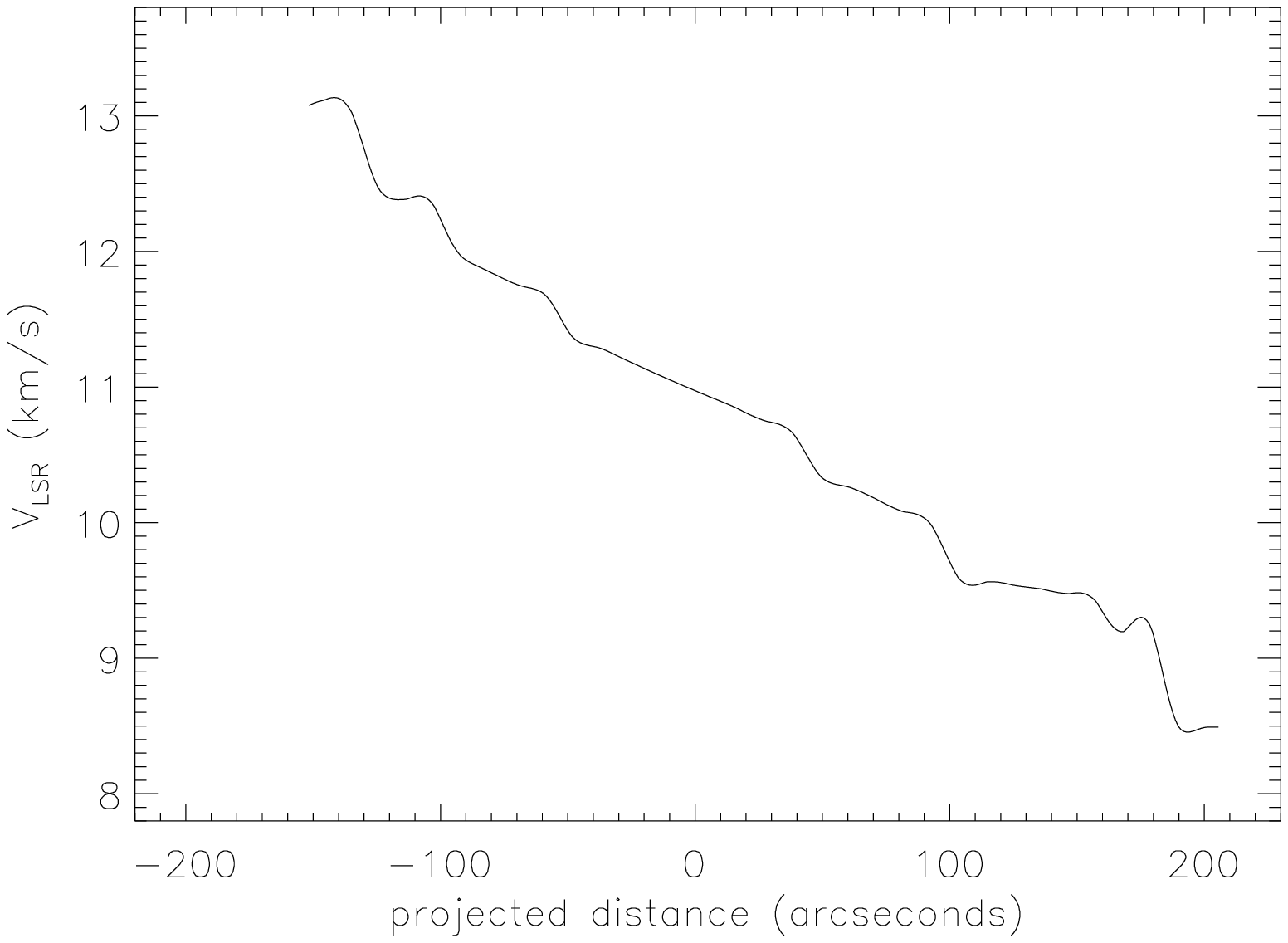}}}
\caption{Position-velocity cut perpendicular to the direction of space
  motion of TX~Psc, derived from velocity field shown in
  Figure~\ref{fig:txpscmom1}. The cut was extracted along a position angle  of
  158$^{\circ}$ and is centered northeast of the stellar position, 
at $\alpha_{\rm J2000}$=23$^{\rm h}$ 46$^{\rm m}$ 26.8$^{\rm s}$,
  $\delta_{\rm J2000}=03^{\circ} 31' 17.5''$. Negative displacements are to the
  southeast and positive displacements to the northwest.
  }
\label{fig:txpscPV}

\end{figure}

The wake of emission trailing behind TX~Psc lies along a
position angle of $\sim45^{\circ}$, and is thus displaced by
$-23^{\circ}$ compared with a bisector defined by the 
space motion vector of
the star (see Table~5). This suggests the presence of a transverse flow in
the local ISM (see below).

Perhaps the most striking feature of the \HI\ emission toward TX~Psc
is its velocity field (Figure~\ref{fig:txpscmom1}). 
While the velocity field map contains a great deal of
spurious structure resulting from undersampled large-scale 
emission, some noteworthy
features are nonetheless apparent, including velocity gradients both
parallel and perpendicular to the space motion of the star. These are
discussed in detail below.

\subsubsection{Discussion of TX~Psc Results\protect\label{txpscdisc}}
%
%
TX~Psc was previously observed in \HI\ by G\'erard \& Le~Bertre
(2006). Its CSE was clearly detected, although the blue edge of the
line profile suffered from confusion. G\'erard \& Le~Bertre
found evidence that the \HI\ emission associated with TX~Psc is quite
extended, with a radius of $\sim12'$. 
This is significantly larger than the
60$\mu$m radius of \am{3}{1} reported by Young et
al. (1993a) based on {\it IRAS} data. 
Similarly, based on higher resolution {\it Herschel} 70$\mu$m
data, Jorissen et al. (2011) found a source with a total extent of
roughly $3'$ (as estimated from their Figure~6), which includes a broad,
low surface brightness trail of emission extending toward the
northeast. Geise (2011) measured a slightly larger angular extent
($\sim$\am{4}{6}) from a {\it Spitzer} 70$\mu$m image. 

Examination of \HI\ survey spectra from Kalberla et al. (2003) toward
the direction of TX~Psc reveals that there is a line component
with peak brightness temperature $\sim$2~K centered near $V_{\rm
  LSR}\approx$10~\kms, very near the systemic velocity of TX~Psc. This
emission is too strong to be plausibly associated with the CSE of
TX~Psc, and instead is likely to be interstellar in origin. This
emission reveals its presence in our VLA maps as the positive and negative
banding seen in our channel images from $V_{\rm LSR}=8.5$ to 11.1~\kms\
(Figure~\ref{fig:txpsccmaps}) implying that it is extended over
scales of $\gsim15'$ (the maximum angular scales that can be sampled with
the VLA in D configuration). The presence of this additional
emission component so close in projected position and velocity to TX~Psc suggests 
that the amount of
extended emission in the CSE may have been overestimated from the
NRT observations. Indeed,  
the \HI\ mass for the CSE of TX~Psc derived from the NRT data is
nearly five times
larger than we derive from the VLA. While we
cannot rule out that the \HI\ wake of TX~Psc may include a low surface
brightness extension that cannot be readily detected with the VLA (analogous to
the one confirmed by the NRT toward Mira; see 
Matthews et al. 2008), our VLA maps suggest that 
confusion may preclude a definitive 
test of this possibility.

%
A systemic decrease in the radial velocity of gas in extended
circumstellar wakes parallel to the direction of motion 
is expected as a consequence of the gradual
deceleration of the wake material owing to its interaction with the
ISM (e.g., Wareing et al. 2007c; Raga \& Cant\'o 2008). Indeed, such
gradients have  been detected in the extended wakes
of Mira (Matthews et al. 2008) and X~Her (Matthews et al. 2011b).
In the case of TX~Psc, we likewise see evidence for a 
systematic blueshifting of the ``tail''
gas with increasing distance from the star in the
\HI\ channel maps
(Figure~\ref{fig:txpsccmaps}) and \HI\ velocity field (Figure~\ref{fig:txpscmom1}). 
The gradient is maximized along a
position angle of $\sim58^{\circ}$ and has a magnitude of
$\sim$1-2~\kms\ projected in the plane of the sky.

While in principle, quantifying the velocity gradient along a gaseous
wake provides a means of age-dating the stellar mass loss history
(Matthews et al. 2008, 2011b; Raga \& Cant\'o 2008), our ability to do
this for TX~Psc is compromised by the presence of a second prominent
velocity gradient {\em perpendicular} to the direction of space motion
of the star.
The magnitude of
this gradient (projected in the plane of the sky) is $\sim$4.5~\kms\ across the 
width of the \HI\ envelope ($\sim6'$ or 0.48~pc). This is
further illustrated in Figure~\ref{fig:txpscPV}, where we show a
position-velocity cut across the emission.     This perpendicular
velocity gradient appears to be real and not an artifact
caused by unresolved large-scale structure or its sidelobes in our
maps. The most compelling evidence of this is that the velocity
gradient persists even at velocities
that are uncontaminated by confusion ($V_{\rm LSR}\ge 12$~\kms). It
also persists if we impose $u$-$v$ restrictions on the data during
imaging in order to suppress the short spacings ($u,v<350\lambda$)
that give rise to 
the spurious large-scale structure in the images. Unfortunately, 
such $u$-$v$ restrictions also suppress much of
the large-scale emission that appears to be genuinely associated with TX~Psc.

For the semi-regular variable X~Her,
Matthews et al. (2011b) also reported evidence of a velocity
gradient orthogonal to the direction of space motion in their \HI\ data. 
In the X~Her case, this orthogonal axis 
corresponds to the symmetry axis 
of a bipolar outflow previously detected at smaller radii through
molecular line observations. 
Matthews et al. suggested that
the \HI\ may therefore be tracing an extension of this outflow out
beyond the molecular dissociation radius. Might a similar situation
apply to the case of TX~Psc?

TX~Psc has no published molecular line imaging observations, so it is unknown
whether the star has an axisymmetric molecular outflow.
Single-dish mapping by Heske et al. (1989) revealed
hints of deviations from spherical symmetry, and the CO profile 
published by Heske (1990)
shows a two component line structure, which is frequently a hallmark of stars with
bipolar outflows (e.g., Kahane \& Jura 1996; Bergman et al. 2000; 
Josselin et al. 2000; Libert et al. 2010b).
The CO mapping data of Heske et
al. also show evidence for a spatial offset between the blue and red
portions of the line emission along a position angle roughly parallel 
to the direction
of space motion of the star. 
However, this velocity shift 
has the opposite sign to the gradient
seen in the \HI\ along this direction 
(i.e., with blueshifted CO emission peaking to the southwest and
redshifted  CO emission to the northeast).\footnote{We assume that right ascension
  increases to the left in Figure~3 of Heske et al. 1989. However,
  the signs of the offsets in the $x$-axis labeling create an
  ambiguity.}  Another significant problem with this explanation for the
velocity gradient transverse to the TX~Psc wake is that the velocity
gradient appears to persist well beyond the vicinity of the
star, where the gas should be dominated by a turbulent flow of ram
pressure stripped material, not the
stellar outflow.

An alternative
explanation for the velocity gradient orthogonal to its \HI\ wake 
is the influence of a transverse flow in the local ISM. Evidence
for local
transverse flows has now been seen in the neighborhood of several
other evolved stars 
(e.g., Ueta et al. 2008; Ferguson \& Ueta
2010; Ueta et al. 2010; see also \S\ref{rxlepdisc}), 
although the origin of these flows remains
unclear. In the case of TX~Psc, such a transverse
flow 
could also offer a natural explanation for the offset in  position
angle between the observed \HI\ wake and the derived space motion
vector in Figure~\ref{fig:txpscmom0}. 

We can attempt to quantify the magnitude and direction of a local ISM
flow following the procedure outlined in Ueta et al. (2008).
Wilkin (1996) derived an analytic expression that relates the 
shape of a bow shock surface to its standoff
distance, $R_{0}$:

\begin{equation}
R(\theta) = R_{0}{\rm csc}~\theta\sqrt{3(1 - \theta~\rm{cot}\theta)}
\end{equation}

\noindent where $\theta$ is the angular displacement from the bow shock
apex as measured from the position of the star, and:

\begin{equation}
R_{0} = \sqrt{\frac{{\dot M}V_{\rm out}} {4\pi\mu m_{H}n_{\rm H}V^{2}_{\rm space}}} .
\end{equation}

\noindent Here $m_{\rm H}$ is
the mass of the hydrogen atom, 
$n_{\rm H}$ is the \HI\ number density of the local ISM, and $\mu$ is
the mean molecular weight of the gas (assumed to
be 1.3). $R_{0}$ 
effectively defines the locus of ram pressure balance between
the internal pressure of the CSE surrounding a
mass-losing star and the ram pressure exerted by the
ISM due to the star's supersonic space motion. 

Based on a three-dimensional fit of Equation~1 to the {\it Herschel}
FIR data of TX~Psc, Cox et al. (2012a)  derived a deprojected standoff distance, $R_{0}=37''$
($1.52\times10^{17}$~cm), with a position angle and inclination for
the bow shock of 238$^{\circ}$ and 11$^{\circ}$,
respectively. Inverting Equation~2 and using the fitted $R_{0}$ value
together with ${\dot M}$ and $V_{\rm out}$ from Table~1, one may then
derive the peculiar space velocity of the star with respect to its
local ISM: $V'_{\rm space}\approx 55.6 n_{\rm
  H}^{-0.5}$~\kms. 

Adopting the position angle and inclination of the bow shock from
Cox et al.,
$V'_{\rm space}$ can now be decomposed 
into its equatorial space-velocity components:

\begin{equation}
\mbox{$V'_{\rm space}$}=
\left[
\begin{array}{c}
V_{r}\\
V_{\alpha}\\
V_{\delta}
\end{array}
\right]_{\star,\rm ISM} =
\left[
\begin{array}{c}
10.6\\
-46.3\\
-28.9
\end{array}
\right]n_{\rm H}^{-0.5}~\kms
\end{equation}

\noindent From the measured proper motion of TX~Psc (see above), the
heliocentric Galactic space velocity of the star in equatorial coordinates (see
Table~5) can be expressed as

\begin{equation}
\mbox{$V_{\rm space}$}=\left[
\begin{array}{c}
V_{r}\\
V_{\alpha}\\
V_{\delta}
\end{array}
\right]_{\star,\odot} =
\left[
\begin{array}{c}
10.8\\
-60.6\\
-24.9
\end{array}
\right]~\kms
\end{equation}

\noindent The local ISM flow in the vicinity of TX~Psc may then be
obtained as the difference between $V'_{\rm space}$ and $V_{\rm space}$:

\begin{equation}
\mbox{$V_{\rm flow} = V_{\rm space} - V'_{\rm space}$ }=\left[
\begin{array}{lcr}
10.8 & - & 10.6n_{\rm H}^{-0.5}\\
-60.6 & - & -46.3n_{\rm H}^{-0.5}\\
-24.9 & - & -28.9n_{\rm H}^{-0.5}
\end{array}
\right]~\kms
\end{equation}

Based on the distance of TX~Psc from the Galactic plane ($-212$~pc; Table~5),
the expected ISM density in the vicinity of TX~Psc is 
$n_{\rm  H}\sim$0.2~cm$^{-3}$ (Table~5). Adopting this value yields 
$V_{\rm flow}\approx 60$~\kms\
along a position angle PA=338$^{\circ}$. While the true value of
$n_{\rm H}$ is uncertain, the nominal value predicts a
position angle for the flow that corresponds precisely to the direction of the vertical
velocity gradient across the TX~Psc envelope
(Figure~\ref{fig:txpscPV}). It thus seems quite plausible that the
velocity gradient across the TX~Psc shell results from a transverse
ISM flow rather than outflow from the star itself. The direction of
the observed velocity gradient is also consistent with this
interpretation. 

%
\begin{deluxetable*}{lcccccccccc}
\tabletypesize{\tiny}
\tablenum{5}
\tablecaption{Derived Properties of the Sample}
\tablehead{
\colhead{Name} & \colhead{$S_{\rm peak}$} & \colhead{$\int S {\rm
    d}\nu$} & \colhead{$V_{\rm LSR,HI}$} &
\colhead{$\Delta V_{\rm HI}$} & \colhead{$M_{\rm HI}$} & \colhead{$\theta_{\rm H}$}
&  \colhead{$V_{\rm
    space}$} & \colhead{PA} & \colhead{$z$} & \colhead{$n_{\rm HI}$}
  \\
\colhead{}  & \colhead{(Jy)} & \colhead{(Jy \kms)} & \colhead{(\kms)} & \colhead{(\kms)}  
& \colhead{$(M_{\odot}$)}    & \colhead{($'$)}
& \colhead{(\kms)} & \colhead{($^{\circ}$)} & \colhead{(pc)} & \colhead{(cm$^{-3}$)}\\
  \colhead{(1)} & \colhead{(2)} & \colhead{(3)} & \colhead{(4)} & \colhead{(5)}
& \colhead{(6)} & \colhead{(7)} & \colhead{(8)} & \colhead{(9)} &
  \colhead{(10)} & \colhead{(11)} }

\startdata

\tableline
\multicolumn{11}{c}{Oxygen-rich stars} \\
\tableline

IK Tau$^{*}$ & $\le$0.006 & $\le$0.10 & ... & ... & $<0.001$  & ... &
$>$34.6 & ... & $-113$ & 0.65\\

RX Lep &0.58$\pm$0.02 & 2.93$\pm$0.22 & 28.4$\pm$0.1 &5.0$\pm$0.3 & 0.015 & \am{17}{5} 
& 56.6 & 24.6 & 54 & 1.16\\ 

Y UMa &0.10$\pm$0.01,0.03$\pm$0.01 & 0.63$\pm$0.13 &16.4$\pm$0.2,17.2$\pm$0.8 
&3.2$\pm$0.4,9.2$\pm$2.3 & 0.022 & \am{6}{8} &  19.2 & 57.1 & 352 & 0.06\\

R Peg &0.16$\pm$0.01 & 0.76$\pm$0.09 &22.4$\pm$0.2 &4.4$\pm$0.4 & 0.029 & $\gsim15'$ 
& 26.3 & 147.6 & $-266$ & 0.14\\

\tableline
\multicolumn{11}{c}{Carbon-rich stars}\\
\tableline

Y CVn &0.79$\pm$0.02 & 2.68$\pm$0.11 &20.65$\pm$0.05 &3.2$\pm$0.1 &
0.047 & $9'$ & 31.0 & 36.8 & 273 & 0.13\\ 

V1942 Sgr &0.240$\pm$0.009 & 0.82$\pm$0.04 &$-33.04\pm0.06$ &3.2$\pm$0.1 & 0.055 &
\am{4}{6} & 40.2 & 95.1 & $-107$ & 0.69\\

AFGL3099$^{*}$ & $<$0.0024 & $<$0.11 & ... & ... & $<0.058$ & ... &
$>$46.6 & ... & $-1081$ & 0.00\\

TX Psc &0.35$\pm$0.01 & 1.48$\pm$0.11 &11.4$\pm$0.1 & 4.0$\pm$0.2& 0.026 & $\sim 8'$ &
66.4 & 
248.0 & $-212$ & 0.24\\

\enddata

\tablecomments{Explanation of columns: (1) target name; (2) peak \HI\
  flux density in the spatially integrated line profile; upper limits
  are 3$\sigma$; (3) integrated flux density of the \HI\ line
  profile, based on a Gaussian fit; (4) LSR
  velocity of the \HI\ line center based on a Gaussian fit; two values
  are quoted for multi-component lines (5) FWHM
  linewidth of the \HI\ profile based on a Gaussian fit; two values
  are quoted for multi-component lines; (6) \HI\ mass
  of the CSE derived from the current VLA observations; upper limits
  are 3$\sigma$;
(7) maximum angular extent of the observed circumstellar
  \HI\ emission; (8)  peculiar space velocity (in LSR frame); (9) position angle
  of space motion projected on the plane of the sky (measured east from north),
derived using the prescription of Johnson \& Soderblom
  1987 and the solar constants from Sch\"onrich et
  al. 2010; (10) distance from the Galactic plane,
  estimated as $z\approx d {\rm sin}b+15$~pc, where values of distance $d$ and
 Galactic latitude $b$ are taken from Table~1; (11) local ISM number
  density, esimated from $n_{\rm H}(z)=2.0e^{-\frac{|z|}{100 {\rm
  pc}}}$ (Loup et al. 1993), where $z$ is taken from column 10. }
\tablenotetext{*}{Proper motion measurements are unavailable; no space
  motion vector was derived.}

\end{deluxetable*}


\section{General Discussion\protect\label{discussion}}
\subsection{A Synopsis of \HI\ Imaging Results to Date}
The present study now brings the total number of red giants
successfully imaged in
the \HI\ 21-cm line to 12.  These comprise 11 AGB stars and the red
supergiant $\alpha$~Ori (see also Bowers \& Knapp 1987, 1988; Matthews \& Reid
2007; Matthews et al. 2008, 2011a, b; Le~Bertre et al. 2012). 
Including the current study, five additional stars have also been observed
with the VLA but were
not detected (see also Hawkins \& Proctor 1993; Matthews \&
Reid 2007).\footnote{Knapp \& Bowers 1983 presented upper limits on
  the circumstellar \HI\ content of 16 evolved stars based on  VLA snapshot
  observations; however, several of these stars have been subsequently
  detected, and the  upper limits for the remainder cannot rule out the presence of
circumstellar \HI\ at comparable levels.} 
This is only a small fraction of the total
sample of evolved stars now surveyed in the \HI\ line from single-dish
surveys. For
a more extensive discussion of \HI\ detection statistics as a function
of stellar properties, we therefore direct the
reader to the single-dish surveys of 
G\'erard \& Le~Bertre (2006) and G\'erard et al. (2011b). 
Nonetheless, the imaged sample is now large
enough to enable investigation of some emerging trends  and
to begin to explore how \HI\ properties correlate with
other properties of the stars.  

We emphasize that the current
sample is a biased one in that imaging studies to date have
largely (although not exclusively)
targeted stars that were
already securely detected in \HI, based on single-dish surveys (see
\S2; Matthews \& Reid 2007). 
Additionally, since a portion of the \HI\ imaging sample was chosen to be displaced
from the predominant Galactic \HI\ emission along the line-of-sight,
the sample may contain a disproportionate number of 
stars with moderately high space velocities, whose ages and
metallicities may not be representative of mean
Galactic values. 

One unifying trend among the stars successfully imaged in 
\HI\ so far is that in  all
cases, the \HI\ images and spatially resolved kinematics display
evidence for interaction
between the CSE and surrounding medium. This local medium includes the 
ISM, but may also include material shed during
earlier mass-loss episodes. Manifestations of this interaction come in
several forms, including: (1)
{\em extended wakes, trailing the motions of the stars through
space} (hereafter ``Category 1''); (2) {\em quasi-stationary detached shells},
whose properties can be modeled as the slowing of ejected wind
material through its interaction with the ISM (hereafter
``Category~2''); 
(3) {\em arcs or clumps of gas displaced from the stellar
position} that
correlate with bow shocks or other signatures of interaction with the ISM
seen at other wavelengths (hereafter ``Category~3''). 
A few stars display features of more than one of these
classifications. 

To date, five stars have been imaged that display the trailing \HI\
wakes characteristic of Category~1: RX~Lep, TX~Psc, X~Her, Mira, and
RS~Cnc (see \S\ref{rxlep}; \S\ref{txpsc}; Matthews \& Reid 2007; Matthews et al. 2008, 2011b;
Libert et al. 2010b).
Among these stars, four have in common  relatively high space velocities
($V_{\rm space}\gsim$57~\kms). 
This is consistent with the results of hydrodynamic simulations 
(e.g., Wareing et al. 2007a; Villaver et al. 2012) that show
that high space velocity is one of the
primary criteria leading to the development of extended gaseous tails. The
one exception is RS~Cnc, with $V_{\rm
  space}\approx$12.4~\kms. This star is located relatively
close to the Galactic plane ($z\approx$97~pc), so its \HI\ tail may
be enhanced through the effects of a high local ISM density.
Simulations also show that cometary tails may form very early in the
AGB stage
even for stars with low space velocities (Villaver et
al. 2012). Furthermore, since the proper motion values
for this star  have large uncertainties (van Leeuwen 2007), the
space velocity may be underestimated.

One feature in common to all of the stars so far found to have
extended gaseous wakes is that their mean effective temperatures are
high enough ($T_{\rm eff}>$2500~K) such 
that their winds are predicted to be dominated by atomic
hydrogen (Glassgold \& Huggins 1983). In addition, all five stars have
 mass loss rates within a narrow range of values ${\dot M}\approx
(1-3)\times10^{-7}M_{\odot}$~yr$^{-1}$, despite a range of outflow
velocities and variability classes. 

The current study has identified four stars belonging to Category~2
(i.e., that appear to be surrounded by quasi-stationary detached shells):
RX~Lep (\S\ref{rxlep}), Y~UMa (\S\ref{yuma}), Y~CVn (\S\ref{ycvn}),
and V1942~Sgr (\S\ref{v1942sgr}). The red supergiant $\alpha$~Ori (Le~Bertre et al. 2012)
also falls into this category. All  four of the AGB stars  in this category
have mass-loss rates in the same narrow range as the
stars with \HI\ tails  [${\dot M}\approx
(1-3)\times10^{-7}M_{\odot}$~yr$^{-1}$]. However, the space
velocities of these stars (Table~5) are on average lower, ranging from 19.2~\kms\
for Y~UMa to 56.6~\kms\ for RX~Lep (which also exhibits an extended
\HI\ tail). Similarly, $\alpha$~Ori has $V_{\rm space}\approx$31~\kms\
(Le Bertre et al. 2012).
Our imaging confirms that detached \HI\ shells with similar
properties can
occur in both carbon-rich and oxygen-rich stars. As already suggested
by Libert et al. (2007), 
the formation mechanism for these \HI\
shells is almost certainly different from the thin molecular shells seen
around some carbon stars (cf. Sch\"oier et al. 2005; Olofsson et
al. 2010). 
Some key differences are that the
molecular shells are extremely thin and nearly perfectly circular, whereas the \HI\
shells are geometrically thick and have shapes ranging from irregular to boxy to
egg-shaped. 

We have presented the results of simple numerical
modeling of the four AGB \HI\ shells from which we estimate ages in the range
$9.0\times10^{4}$ to $7.2\times10^{5}$~yr  (Appendix~A; see also
Le~Bertre et al. 2012 for the case of $\alpha$~Ori). For Y~UMa, the
age that we derive for the CSE is comparable to that previously
derived from FIR observations by Young et al. (1993b). However, 
for Y~CVn and V1942~Sgr (both carbon stars) the duration of the mass loss history
inferred from our new \HI\ observations is
nearly 7 times larger than derived previously from FIR
observations. Indeed, as more \HI\ measurements of the
extended CSEs become available, they continue to point
toward a past systematic underestimation of the mass loss timescales
for many
red giants (see also Matthews et al. 2008, 2011b; Le~Bertre et al. 2012).

One of the stars recently identified as belonging to Category~3 (i.e.,
stars surrounded by 
arcs or clumps of gas displaced from the stellar position) is 
$\alpha$~Ori (Le~Bertre et al. 2012), 
while the other two stars we place in this class, IRC+10216 and R~Cas
(Matthews \& Reid 2007; Matthews et al. 2011a), are
  Mira variables. All three of these stars have high mass-loss rates (${\dot
  M}>1\times10^{-6}M_{\odot}$~yr$^{-1}$), although $\alpha$~Ori has a
  much higher effective temperature ($T_{\rm
    eff}=3641\pm53$~K; Perrin et al. 2004) compared with the two Miras 
($T_{\rm eff}\lsim 2500$~K). For both IRC+10216 and R~Cas, 
the \HI\ imaged by the VLA comprises isolated arcs of
emission displaced from the stellar position,
 although the two stars differ dramatically in the size
scales of their CSEs (cf. Matthews \& Reid 2007), and they are believed 
to be at very different
evolutionary stages. In the case of IRC+10216, it is suspected that
the observed \HI\ emission is partly the result of swept-up
interstellar material, although there is evidence that it is
augmented by photodissociated H$_{2}$ (Matthews et al. 2011a). 

Only one star imaged in \HI\ to date, the Mira variable R~Peg,  
does not fall neatly into any of
the three classifications defined above. 
While its \HI\ mass and integrated \HI\ profile  are similar to the
other detected stars, its morphology is rather peculiar
(\S\ref{rpeg}). R~Peg has a higher mass loss
rate than any
of the other VLA sample stars where \HI\ has been detected directly toward
the 
stellar position
(${\dot M}\approx5\times10^{-7}M_{\odot}$~yr$^{-1}$), suggesting that
it may be somewhat more evolved. In this case, one
possibility is that the ejecta that we observe in \HI\ are left over
from previous cycles of mass loss and have been subsequently distorted by
transverse flows in the surrounding ISM. 

For all stars detected in \HI\ so far, we have estimated the
number density of hydrogen atoms in the local ISM using the expression from Loup et al. (1993):  
$n_{\rm H}(z)=2.0e^{-\frac{|z|}{100 {\rm
  pc}}}$ cm$^{-3}$, where $z$ is the distance from the Galactic Plane in pc,
estimated as $z\approx d {\rm sin}(b) + 15$~pc. Here $d$ is the
distance to the star and $b$ is the Galactic latitude.  For
the full VLA-detected sample,
the values range from $n_{\rm H}\sim0.059$~cm$^{-3}$
for Y~UMa (Table~5) to $n_{\rm H}\sim1.7$~cm$^{-3}$ for
$\alpha$~Ori and R~Cas. For the Category~1 stars, values of $n_{\rm
  H}$ range from 0.24~cm$^{-3}$ for TX~Psc to 1.2 cm$^{-3}$
for RX~Lep (Table~5). Thus there is no evidence that trailing \HI\ wakes
are preferentially found associated with stars in dense local
enviroments, although as noted in \S\ref{rxlepdisc}, the moderately high local ISM
density 
surrounding RX~Lep  
may help to account for its highly collimated wake. We also
emphasize that the
$n_{\rm H}$ values computed using the above formula are 
rather uncertain owing to their dependence on the adopted Galactic model
and small-scale fluctuations in the ISM density across the Galaxy.

The sample of five stars so far undetected in \HI\ with the VLA is still
too small to draw any firm conclusions, but here we briefly review their
properties to help guide future studies. 
First, we point out that the two undetected stars from the current study, IK~Tau
and AFGL~3099, have temperatures and mass loss rates comparable to
two previously detected Category~3 Miras: R~Cas and IRC+10216.
However,  G\'erard et al. (2011b) have found that in general, cooler stars
  with higher mass loss rates are less frequently detected in
  \HI. This may be due to a combination of a predominantly molecular composition in
  the CSEs, their geometrically large sizes, and/or temperatures in
  the CSE that are too cool to allow the detection of \HI\ in emission
  (see \S\ref{iktaudisc}).

Having a low
  mass loss rate does not necessarily imply that a star should be
  more difficult to detect in \HI, since available ${\dot M}$ estimates
  generally represent only recent mass loss and are derived from
  tracers at much smaller radii than the extended circumstellar
  material traced by \HI. However, to date, the two stars observed
  with the VLA having the lowest mass loss rates both remain
  undetected.
The Mira R~Aqr (${\dot
    M}=6\times10^{-8}~M_{\odot}$~yr$^{-1}$; Spergel et al. 1983) 
is part of a symbiotic binary
and is surrounded by a hot ionized region. Matthews \& Reid (2007)
  placed an upper limit on its \HI\ mass of $M_{\rm
  HI}<4.9\times10^{-4}~M_{\odot}$.   
The undetected semi-regular variable W~Hya (Hawkins \&
Proctor 1993) also has a comparably small mass loss rate (${\dot
  M}\approx7\times10^{-8}~M_{\odot}$~yr$^{-1}$; Olofsson et
  al. 2002). In this case, the lack of detection might be related
  to the moderately cool stellar temperature
($T_{\rm eff}\approx$2450~K; Feast 1996) or the difficulty of
disentangling weak and clumpy CSE emission from line-of-sight
  contamination (see Hawkins \& Proctor 1993). 

The remaining star so far undetected in \HI\ by the VLA is the semi-regular variable 
EP~Aqr, whose mass loss rate is an order of magnitude higher than
R~Aqr or W~Hya
(${\dot M}\approx6\times10^{-7}~M_{\odot}$~yr$^{-1}$; Winters et
al. 2007). It poses an interesting case, since it 
is not a known binary, and its effective temperature is
high enough to predict the presence of atomic hydrogen in its 
atmosphere ($T_{\rm eff}\approx 3240$~K; Dumm \& Schild 1998). 
The non-detection of this star
with the VLA is particularly puzzling, as Le~Bertre
\& G\'erard (2004) reported an \HI\ detection of EP~Aqr with the NRT,
and the star is also an extended FIR source (Young et al. 1993a; Cox
et al. 2012a). However,
Matthews \& Reid (2007) found only a few isolated clumps of \HI\
in the region, none of which could be unambiguously linked with the
CSE. Deeper observations of this star with improved $u$-$v$ coverage
might be enlightening, as some of the clumps detected by
Matthews \& Reid might turn out to be density enhancements in an
extended, diffuse envelope or pockets of photodissociated gas
within an ancient molecular envelope.

\subsection{Future Prospects}
It is clear that
\HI\ imaging studies of larger samples of stars are needed to enable 
better statistical comparisons between the
properties of the stars, their CSEs, and their interstellar
neighborhoods.  However,
despite the small size of the current VLA \HI\ sample, certain trends
are already clear. The results
to date underscore initial findings from single-dish surveys and
unambiguously establish the importance of the interstellar
environment on the large-scale properties of CSEs. While
additional examples of stars interacting with their environments have
also been identified in recent FIR and FUV surveys
(e.g., Martin et al. 2007; Stencel 2009; Sahai \& Chronopoulos 2010; 
Ueta 2011; Cox et al. 2012a), \HI\ data have the
unique advantage that they
supply {\em kinematic} information on the extended circumstellar
material. As demonstrated in
the present study, 
such data in turn permit numerical modeling of the
interaction between circumstellar shells and their surroundings and
the determinations of
the duration of the stellar mass loss history (see also Matthews et
al. 2008, 2011b; Libert et al. 2007, 2010a, b). 
\HI\ data
can also provide crucial morphological and kinematic
information for testing hydrodynamic
simulations of individual mass-losing stars moving through the ISM---a
potential that has largely gone untapped
(cf. Villaver et al. 2012; Wareing 2012; Mohamed et al. 2012). 

Another important lesson that has emerged from \HI\ imaging surveys 
to date is that no single tracer
can provide a complete picture of the complex circumstellar
environments of mass-losing evolved stars. If one hopes to understand
the complete stellar mass loss history, it is critical to
study the CSE on both large and small spatial scales. Moreover, it has
become clear that even when different tracers span overlapping spatial
scales (e.g., \HI\ and
FIR-emitting dust), each supplies
unique yet
complementary information that can be used together to provide a more
comprehensive picture of the kinematics,
physical conditions, and chemistry across the entire CSE.

\section{Summary}
We have presented \HI\ 21-cm line imaging observations of a sample of eight 
AGB stars obtained with the VLA. 
We have unambiguously detected \HI\ emission associated with the CSEs of
six of the stars. All six of these stars were already known to have
atomic hydrogen in their CSEs
based on the results of single-dish surveys. However, our imaging
results reveal a wealth of new information on their spatially resolved
morphologies and kinematics. In all cases we identify clear
evidence that the properties of the extended CSEs are significantly
affected by interaction with their local interstellar environments. The detected
stars have circumstellar \HI\ masses in the range $M_{\rm HI}\approx
0.015-0.055M_{\odot}$ and the size scales of their circumstellar \HI\
emission range from $\sim$0.6 to 1.7~pc.

Four stars in the present study (RX~Lep, Y~UMa, Y~CVn, and V1942~Sgr) are
found to be surrounded by quasi-stationary detached hydrogen shells whose origin can be
explained as a result of the stellar outflow being abruptly
decelerated at a termination shock that occurs at the interface
between the circumstellar and interstellar environments.  
These shells are seen around both oxygen-rich
and carbon-rich stars, and appear to have a
different origin from the thin detached molecular shells seen around
some carbon stars. All four \HI\ shell stars have similar mass-loss rates
[${\dot M}\approx(1-3)\times10^{-7}M_{\odot}$~yr$^{-1}$] and
low-to-moderate space velocities,
but otherwise show a range in properties, including  chemistries,
wind outflow velocities, and variability classes.  
All of the shells have a clumpy rather than smooth
structure, and none of the observed \HI\ shells are spherical; they range from
egg-shaped (RX~Lep) to boxy
(Y~CVn) to amorphous (Y~UMa and V1942~Sgr). In the case of Y~UMa,
there is evidence that one edge of the shell was distorted through its
interaction with the local ISM. 
The shell of
RX~Lep exhibits a pronounced elongation along the direction of space motion of the
star, as well as a ``cracked egg'' morphology, with a
dearth of emission along its equatorial region. 

Based on simple numerical
models we have estimated ages for the four shells in the range
$9.0\times10^{4}$ to $7.2\times10^{5}$~yr  (Appendix~A). These results
support previous  suggestions that RX~Lep is in a relative early evolutionary
state compared with the other three stars. For the carbon stars Y~CVn
and V1942~Sgr the derived ages also imply that mass loss time scales previously
derived from FIR measurements have been underestimated by a factor of $\sim$7.

For both Y~UMa and Y~CVn,
high-resolution FIR images are available that reveal counterparts
of the \HI-emitting shells. In both cases, the
extent of the FIR shell is comparable to that of the \HI\ shell,
but the two tracers show notable differences, perhaps
due to effects such as ionization of the \HI\ from shocks or local destruction of
dust grains.  Both Y~UMa and Y~CVn also exhibit ``notch''-like
depressions in their \HI\ and FIR intensity along the leading edge of
their \HI\ shells and running parallel to their direction of space motion.
These notches may be signatures of instabilities along the leading
edge of the CSE. While the \HI\ shells of Y~UMa and Y~CVn are similar in
terms of their sizes, morphologies, and line widths, the Y~UMa shell
contains only half the \HI\ mass as that of Y~CVn and also shows a
distinct rotationally symmetric pattern in its \HI\ velocity field
whose origin is unclear.

The two highest velocity stars in our study (RX~Lep and TX~Psc) were found to have
extended gaseous wakes trailing the motions of the stars through the
Galaxy. Such wakes are formed by the effects of ram pressure on the
circumstellar ejecta as the stars move supersonically through the
ISM. The wake of RX~Lep (which stretches $\sim$0.36~pc) 
is more highly collimated than any \HI\ wake detected to date, a fact that is
likely due to a combination of the relatively early evolutionary status of the
star on the TP-AGB and the high density of its interstellar surroundings. 
For TX~Psc, the orientation of the wake (which stretches
$\sim 0.6$~pc) appears to be affected by a local transverse flow in the
ISM. This flow could also explain a velocity gradient observed across
the CSE, perpendicular to the direction of space motion.

We find the oxygen-rich Mira R~Peg to be surrounded by a peculiar ``horseshoe''-shaped
\HI\ nebula that extends over $\sim$1.7~pc on the sky. This morphology
may result in part from the effects of 
local ISM flows. However, the CSE morphology close to
the stellar position is also peculiar, with two asymmetric lobes of
emission stretching roughly parallel to the direction of space motion
of the star. These lobes differ from the head-tail structures seen
toward other stars in that the ``head'' of the nebula lies at a
projected distance of $\sim$0.1~pc southeast of the star. One possible
explanation is that the observed \HI\ structures are the vestiges of
an earlier cycle of mass-loss.

The two undetected stars in the present sample (IK~Tau and AFGL~3099)
are both relatively cool Mira variables with high mass-loss rates
[${\dot M}\approx(4.2, 13)\times10^{-6}M_{\odot}$~yr$^{-1}$]. Both stars
were also undetected in previous single-dish studies. For these stars,
hydrogen in the circumstellar environment may be too cold to detect
via 21-cm line emission or may be largely in  molecular form.

\acknowledgements
LDM acknowledges support for this work from
grant AST-1009644 from the National Science Foundation. 
MCJ was supported through a National Science
Foundation Research Experience for Undergraduates 
grant to MIT Haystack Observatory. 
The observations presented here were part of NRAO programs AM887 and
AM1001 and archival program AK237.
This research has made use of the SIMBAD database,
operated at CDS, Strasbourg, France.

\appendix
\section{Age Estimates for the Detached \HI\ Shells Based on Simple
  Numerical Models}
As a mass-losing star becomes progressively surrounded by
an expanding 
shell of circumstellar material, the collision of the
supersonic stellar wind with the surrounding medium (local ISM and/or
debris from earlier episode of mass loss) produces a slowly 
expanding shell of denser material
(Lamers \& Cassinelli 1999). Young et al. (1993b) developed a
model to simulate the formation of these ``detached shells'' as seen
in infrared emission, and Libert et al. (2007) subsequently
adapted this type of modeling to circumstellar hydrogen. 
 
In this picture, a detached shell results from a stellar outflow that
is abruptly slowed down at a termination shock. This first shock
defines the inner boundary of the detached shell ($r_{1}$). The outer limit
($r_{2}$) is defined by the leading edge (bow shock) where external matter
is compressed by the expanding shell (see Figure~5 of Libert et
al. 2007). Between these two
limits, the detached shell is composed of compressed circumstellar
and interstellar matter separated by a contact discontinuity at
a radius $r_{f}$. The circumstellar matter is decelerated and heated when crossing
the shock at $r_{1}$. As it cools from a temperature $T^{+}_{1}$
(at $r_{1}$) 
to $T_{f}$ (at $r_{f}$), the
expansion velocity decreases, from $v^{+}_{1}$ to $v_{f}$, and the density
increases from $n^{+}_{1}$ to $n^{-}_{f}$. Adopting an arbitrary temperature profile, 
one can numerically solve the equation of motion between
$r_{1}$ and $r_{f}$. 

Here we have applied a similar method to modeling
the \HI\ shells of RX~Lep, Y~UMa, Y~CVn, and V1942~Sgr. 
Models for three of the stars (RX~Lep, Y~CVn, and V1942~Sgr) 
have previously been published based on
single-dish \HI\ observations  (Libert et al. 2007, 2008, 2010a), but
our new imaging observations provided improved constraints on the
model parameters. 

We have not attempted to compute models for either TX~Psc or  
R~Peg. In the case of TX~Psc, the relative high space velocity and 
large fraction of the circumstellar
ejecta present in its extended trailing wake makes it a poor candidate for a
stationary and spherically symmetric 
model. For R~Peg, the
complex \HI\ morphology and line shape also imply that a simple model fit
of the type described here is unlikely to be meaningful.

Following Libert et al. (2008),
the temperature profile in the detached shell of the four modeled
stars is
constrained by the observed \HI\ line profile. For the external part
of the detached shell (between $r_{f}$ and $r_{2}$), we assume
a density profile that falls off as $1/r^{2}$, with the condition that the
total mass ($M_{\rm DT,CS}$) is given by the equivalent ISM mass enclosed in a
sphere of radius $r_{2}$. Inside $r_{1}$, we assume a spherical wind
in uniform 
expansion, whose properties (mass-loss rate, outflow velocity, and
systemic velocity) match those derived from CO observations
(Table~1). We assume mass lost from the star comprises
90\% H atoms and 10\% He atoms by number (Glassgold \& Huggins 1983). 
The outer radius of the shell, $r_{f}$, is
constrained by the VLA images.

Based on our models, the detached shell is expected to produce a 
narrow, quasi-Gaussian emission peak  at or near 
the stellar systemic velocity in the spatially integrated spectrum, while the freely
expanding wind should produce a weaker, broader line component 
with peaks located on either side of the stellar systemic velocity, near
$V_{\star}\pm V_{\rm o}$ (see Le~Bertre et al. 2012). In general, the
shell component dominates the global spectral profile.

\begin{figure}
\scalebox{0.45}{\rotatebox{0}
{\includegraphics{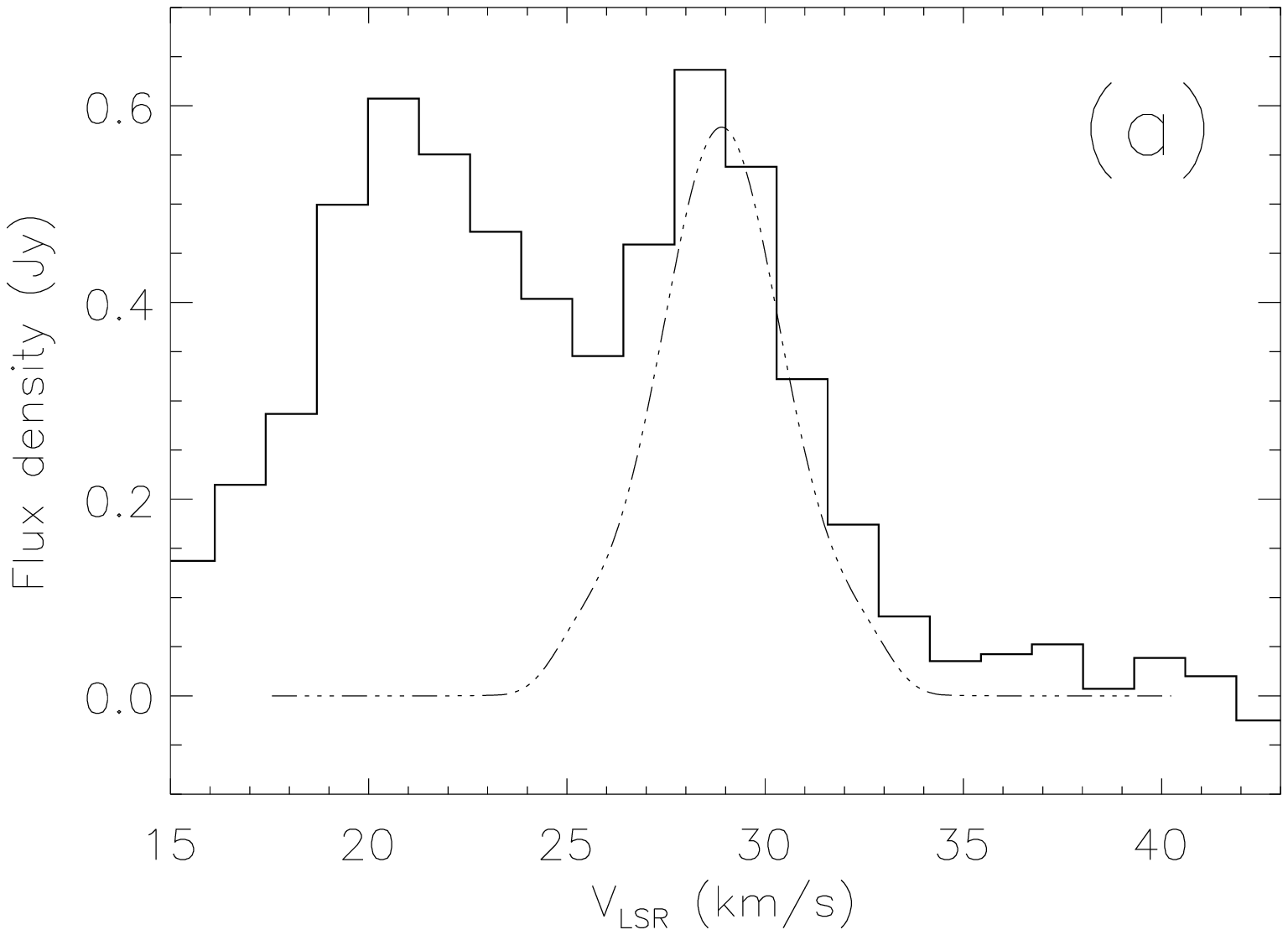}
\includegraphics{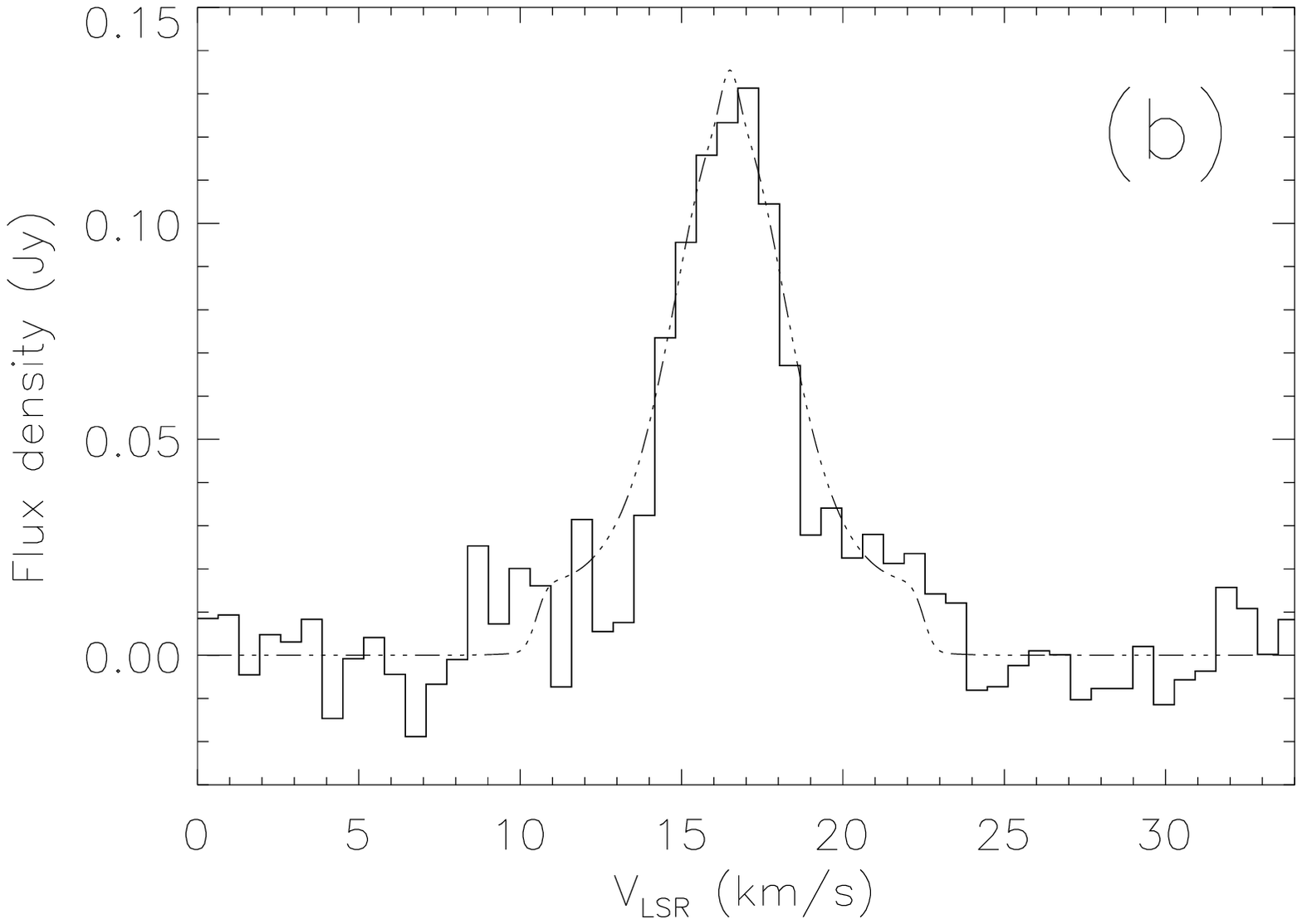}
}}
\scalebox{0.45}{\rotatebox{0}
{\includegraphics{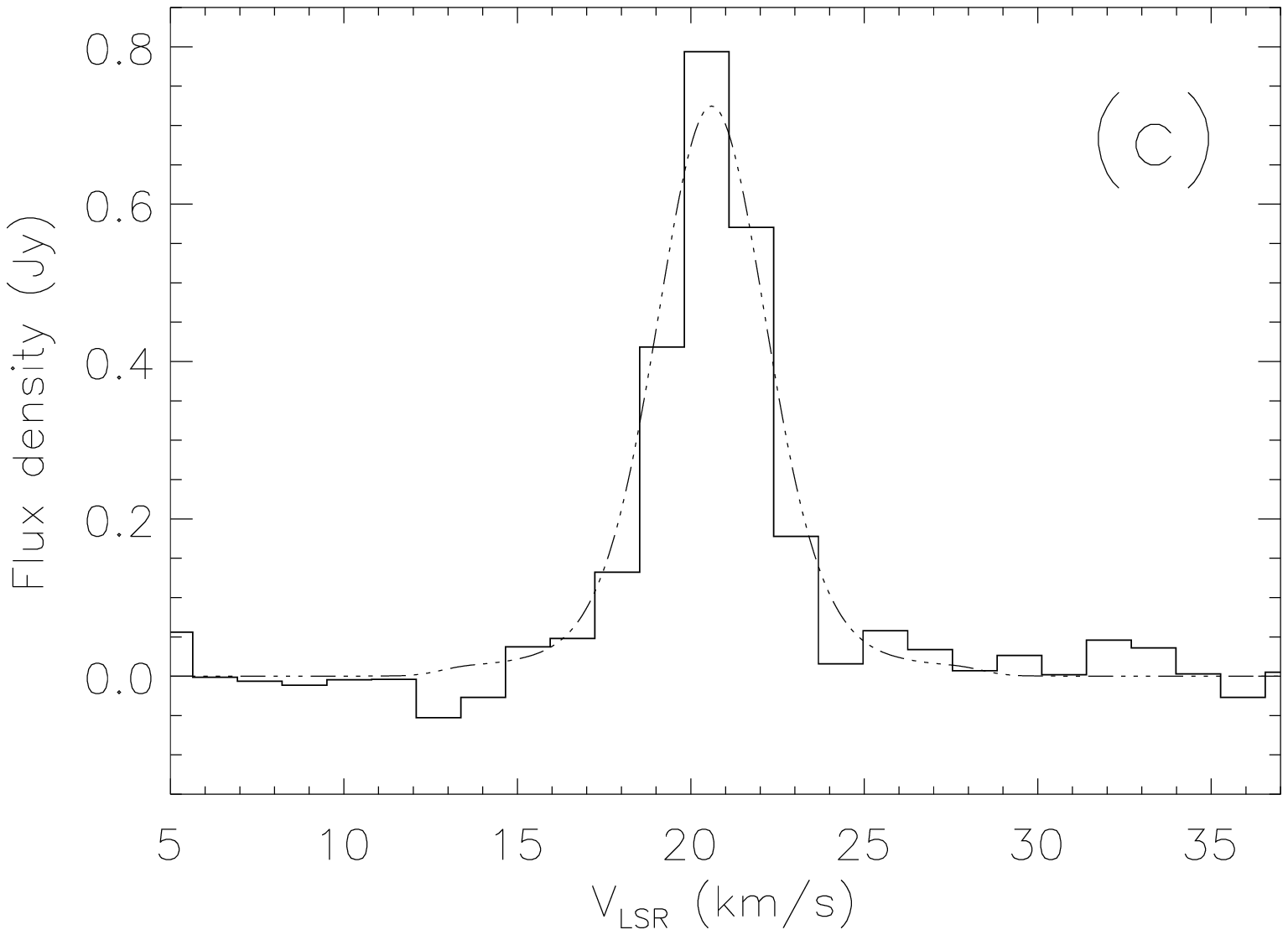}
\includegraphics{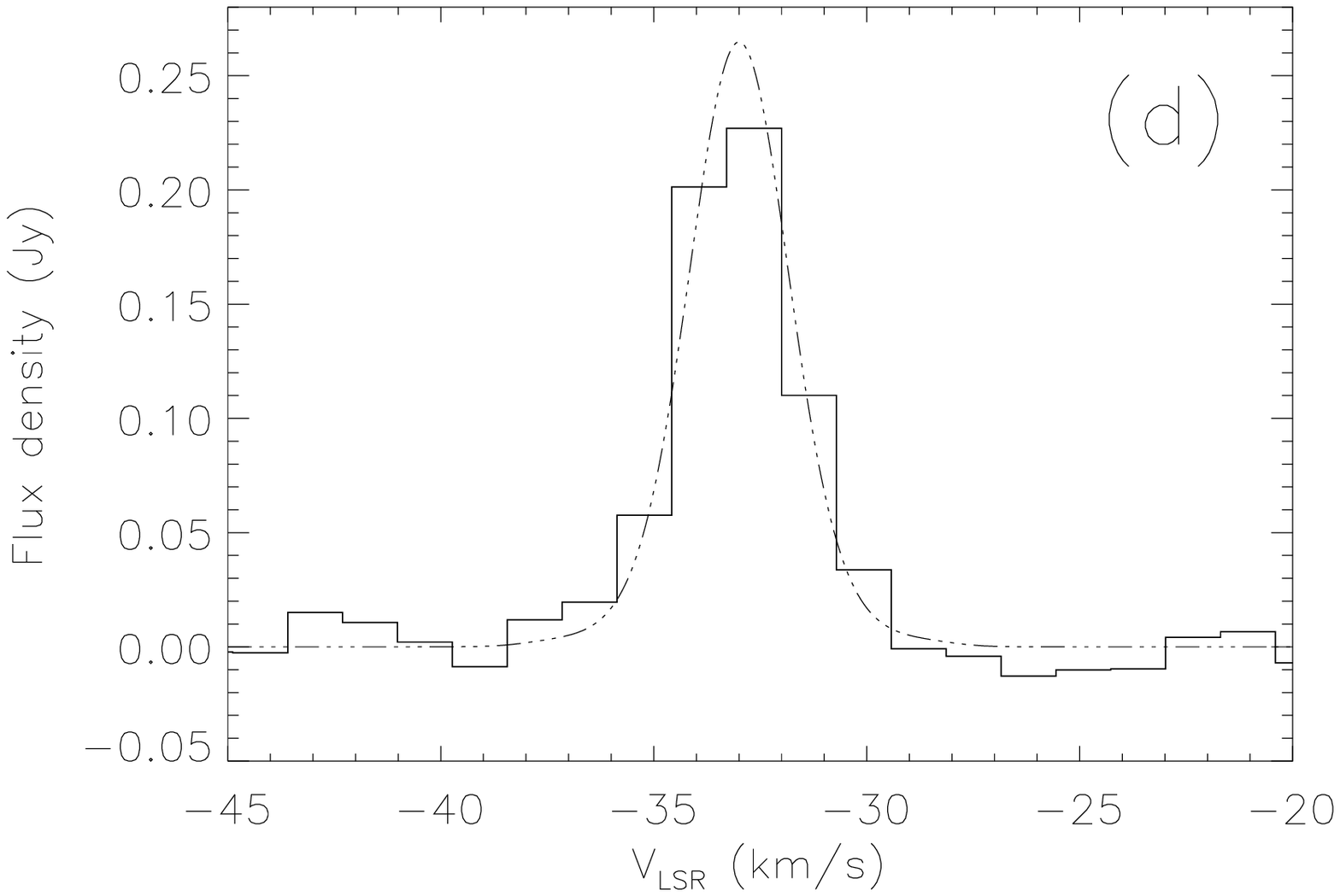}
}}
\caption{Global (spatially integrated) \HI\ spectra for four stars
  from the current sample (solid histograms), with spectra derived
  from numerical models overplotted (dot-dash lines; see Appendix~A): (a) RX~Lep;
  (b) Y~UMa; (c) Y~CVn; (d) V1942~Sgr. The model spectra have been
  smoothed to match the spectral resolution of the observations. For RX~Lep, the
  spectral peak blueward of the stellar emission results from
  interstellar contamination.}
\label{fig:models}

\end{figure}

Our best-fit parameters for the four stars that we have modeled 
are summarized in Table~A1, and model fits are overplotted on the
observed spectra in Figure~\ref{fig:models}.  In all cases, our
toy model provides a reasonable fit to the strength and width of the
global \HI\ line profile. 

For the present study, the most useful outcome of
our models is an estimate for the age of each of the shells, which can be
obtained from a sum of the parameters $t_{1}$ and $t_{\rm DS}$ in
Table~A1. Not surprisingly, we find the smallest age ($\sim$90,000~yr)
for RX~Lep, a star that is believed to be in the early stages of the
TP-AGB (see \S\ref{rxlep}). For the other three stars, our models
imply that the \HI\ data are tracing more than $10^{5}$ years of
stellar mass loss history. In the case of Y~UMa, our derived shell age
is comparable to the age of the FIR-emitting shell previously derived
by Young et al. 1993b ($9.2\times10^{4}$~yr), while in the cases of
the carbon stars
Y~CVn and V1942~Sgr, our new values are 6.8 and 6.6 times larger, respectively.

To obtain a more complete picture of these four stars, future
models will ultimately need to take into account the motion of the
stars through the ISM and the resulting distortions that this produces
on the \HI\ shells. However, this will require sophisticated
hydrodynamic simulations that are beyond the scope of the present
paper. Furthermore, we note that our model fits break down at small
radii ($r\lsim r_{1}$), systematically predicting more flux than is
observed in both the shell component (central peak) of the line
profile and the freely
expanding wind 
(``horns''). This may be due to an uncertainty in the temperature profile adopted in our
models, short-term fluctuations in the mass-loss
rate, asymmetric or non-spherical mass loss, 
and/or a breakdown in our assumption that the wind is largely
atomic as it leaves the star. 
These possibilities will be investigated in a future work.
Fortunately the ages that we derive for the detached shells (Table~A1)
depend
mostly on their large-scale properties and are
relatively insensitive to these details.

%
\begin{deluxetable*}{lllll}
\tabletypesize{\scriptsize}
\tablewidth{0pc}
\tablecaption{Detached Shell Model Parameters}
\tablehead{\colhead{Parameter}  & \colhead{RX Lep} & \colhead{Y UMa}& \colhead{Y CVn}&
\colhead{V1942 Sgr}
}

\startdata
${\dot M}$ ($ M_{\odot}$ yr$^{-1}$) & 2.0$\times10^{-7}$ &
2.6$\times10^{-7}$ & $1.7\times10^{-7}$ & $1.0\times10^{-7}$\\
$\mu$ & 1.3 & 1.3 & 1.3 & 1.3\\
$t_{1}$ (yr) & 2.0$\times 10^{4}$ & 3.6$\times 10^{4}$ &
2.8$\times10^{4}$ & $2.4\times10^{4}$ \\
$t_{\rm DS}$ (yr) &  7.0$\times 10^{4}$ & 8.8$\times 10^{4}$ &
4.2$\times10^{5}$ & $7.0\times10^{5}$ \\
$r_{1}$ (pc) & 0.087 (\am{2}{0}) & 0.22 (\am{2}{0}) & 0.22
(\am{2}{8}) & 0.12 (\am{0}{8})\\
$r_{f}$ (pc) & 0.11 (\am{2}{6}) & 0.26 (\am{2}{3}) & 0.32
(\am{4}{0}) & 0.22 (\am{1}{4})\\
$r_{2}$ (pc) & 0.17 ($4'$) & 0.34 (\am{3}{0}) & 0.40 (\am{5}{1}) &
  0.44 (\am{2}{8})\\
$T_{0}(\equiv T^{-1}_{1})$,$T^{+}_{1}$ (K) & 20, 528 & 20, 1071 & 20,
1804 & 20, 746\\
$T_{f}(\equiv T_{2})$ (K) & 157 & 221 & 173 & 87 \\
$v_{0}(\equiv v^{+}_{1})$,$v^{+}_{1}$ (\kms) & 4.2, 1.07 & 6.0, 1.52 &
7.8, 1.96 & 5.0, 1.27\\
$v_{f}$ (\kms) & 0.16 & 0.18 & 0.07 & 0.04\\
$v_{2}$ (\kms) & 1.35 & 1.2 & 1.0 & 1.26\\
$n^{-}_{1}$,$n^{+}_{1}$ (H cm$^{-3}$) & 14.0, 54.8 & 1.9, 7.55 & 1.0,
3.8 & 2.8, 11.2\\
$n^{-1}_{f}$,$n^{+}_{f}$ (H cm$^{-3}$) & 228, 2.3 & 45.0, 0.5 & 50.0,
0.6 & 119, 1.6\\
$n_{2}$ (H cm$^{-3}$) & 1.0 & 0.2 & 0.4 & 0.4\\
$M_{r<r_{1}}$ ($M_{\odot}$) & $4.0\times10^{-3}$ & $9.5\times10^{-3}$
& $4.8\times10^{-3}$ & $2.4\times10^{-3}$\\
$M_{\rm DT,CS}$ ($M_{\odot}$) & $1.4\times10^{-2}$ &
$2.3\times10^{-2}$ & $7.17\times10^{-2}$ & $7.0\times10^{-2}$\\
$M_{\rm DT,EX}$ ($M_{\odot}$) & $7.7\times10^{-4}$ &
$1.1\times10^{-3}$ & $2.4\times10^{-3}$ & $7.3\times10^{-3}$\\
\enddata

\tablecomments{Notations follow those adopted by Libert et al. 2007. 
Matter lost is assumed to be 90\% neutral hydrogen, 10\% helium by
number density. Distances are
  adopted from Table~1.}

\end{deluxetable*}


\begin{references}

Begum, A. et al. 2010, ApJ, 722, 395

Bergeat, J., Knapik, A., \& Rutily, B. 2001, A\&A, 369, 178

Bergeat, J., Knapik, A., \& Rutily, B. 2002, A\&A, 390, 967

Bergman, P., Kerschbaum, F., \& Olofsson, H. 2000, A\&A, 353, 257

Bowers, P. F. \& Knapp, G. R. 1987, ApJ, 315, 305

Bowers, P. F. \& Knapp, G. R. 1988, ApJ, 332, 299

Castro-Carrizo, A. et al. 2010, A\&A, 523, A59 

Chiotellis, A., Schure, K. M., \& Vink, J. 2012, A\&A, 537, A139

Claussen, M. J., Sjouwerman, L. O., Rupen, M. P., Olofsson, H.,
Sch\"oier, F. L., Bergman, P., \& Knapp, G. R. 2011, ApJ, 739, L5

Clayton, G. C. 2012, JAVSO, 40, 539

Comer\'on, F. \& Kaper, L. 1998, A\&A, 338, 273

Condon, J. J., Cotton, W. D., Greisen, E. W., Yin, Q. F.,
Perley, R. A., Taylor, G. B., \& Broderick, J. J. 1998, AJ, 115, 1693

Cox, N. et al. 2012a, A\&A, 537, A35

Cox, N. et al. 2012b, A\&A, 543, C1

Cristallo, S., Straniero, O., Gallino, R., Piersanti, L., 
Dom\'\i nguez, I., \& Lederer, M. T. 2009, ApJ, 696, 797

Decin, L. et al. 2010, A\&A, 521, L4

Dickinson, D. F. \& Dinger, A. S. C. 1982, ApJ, 254, 136

Dominy, J. F. 1984, ApJS, 55, 27

Dumm, T. \& Schild, H. 1998, New Astron., 3, 137

Dyck, H. M., Lockwood, G. W., \& Capps, R. W. 1974, ApJ, 189, 89

Eriksson, K., Gustafsson, B., Johnson, H. R., Querci, F., Querci, M.,
Baumert, J. H., Carlsson, M., \& Olofsson, H. 1986, A\&A, 161, 305

Feast, M. W. 1996, MNRAS, 278, 11

Feast, M. W., Whitelock, P. A., Catchpole, R. M., Roberts, G., \&
Carter, B. S. 1985, MNRAS, 215, 63P

Ferguson, B. A \& Ueta, T. 2010, ApJ, 711, 613

Gardan, E., G\'erard, E., \& Le Bertre, T. 2006, MNRAS, 365, 245

Geise, K. M. 2011, Masters Thesis, University of Denver

Gehrz, R. D., Hackwell, J. A., \& Briotta, D. 1978, ApJ, 221, L23

G\'erard, E. \& Le Bertre, T. 2003, A\&A, 397, L17

G\'erard, E. \& Le Bertre, T. 2006, AJ, 132, 2566

G\'erard, E., Le Bertre, T., \& Libert, Y. 2011a, in Why Galaxies Care
about AGB Stars, ASP Conf. Series Vol. 445, edited by 
F. Kerschbaum, T. Lebzelter, and R.F. Wing, 329

G\'erard, E., Le Bertre, T., \& Libert, Y. 2011b, in SF2A-2011: Proceedings of
the Annual Meeting of the French Society of Astronomy and
Astrophysics, edited by G. Alecian, K. Belkacem, R. Samadi, and
D. Valls-Gabaud, 419

Glassgold, A. E. \& Huggins, P. J. 1983, MNRAS, 203, 517

Gonz\'alez Delgado, D., Olofsson, H., Kerschbaum, F., Sch\"oier, F. L.,
Lindqvist, M., \& Groenewegen, M. A. T. 2003, A\&A, 411, 123

Groenewegen, M. A. T. \& Whitelock, P. A. 1996, MNRAS, 281, 1347

Hawkins, G. \& Proctor, D. 1993, in Mass Loss on the AGB and Beyond,
ed. H. E. Schwarz (ESO: Munich), 461

Heske, A. 1990, A\&A, 229, 494

Heske, A., te Lintel Hekkert, P, \& Maloney, P. R. 1989, A\&A, 218, L5

Imai, H., Obara, K., Diamond, P. J., Omodaka, T., \& Sasao,
T. 2002, Nature, 417, 829

Inomata, N., Imai, H., \& Omodaka, T. 2007, PASJ, 59, 799

Isaacman, R. 1979, A\&A, 77, 327

Izumiura, H., Hashimoto, O., Kawara, K., Yamamura, I., \& Waters,
L. B. F. M. 1996, A\&A, 315, L221

Johnson, D. R. H. \& Soderblom, D. R. 1987, AJ, 93, 864

Jorissen, A. et al. 2011, A\&A, 532, 135

Josselin, E., Mauron, N., Planesas, P., \& Bachiller, R. 2000, A\&A,
362, 255

Judge, P. G. \& Stencel, R. E. 1991, ApJ, 371, 357

Jura, M. 1986, ApJ, 303, 327

Jura, M., Kahane, C., \& Omont, A. 1988, A\&A, 201, 80

Jura, M. \& Kleinmann, S. G. 1992, ApJS, 79, 105

Kahane, C. \& Jura, M. 1996, A\&A, 310, 952

Kalberla, P. M. W., Burton, W. B., Hartmann, D., Arnal, E. M., Bajaja,
E., Morras, R., \& P\"oppel, W. G. L. 2005, A\&A, 440, 775

Kiss, L. L., Szatm\'ary, K., Cadmus, R. R. Jr., \& Mattei, J. A. 1999,
A\&A, 346, 542

Klotz, D., Paladini, C., Hron, J., Aringer, B., Sacuto, S., Marigo,
P., \& Verhoelst, T. 2013, submitted to A\&A (arXiv:1301.0404)

Knapp, G. R. \& Bowers, P. F. 1983, ApJ, 266, 701

Knapp, G. R. \& Morris, M. 1985, ApJ, 292, 640

Knapp, G. R., Pourbaix, D., Platais, I., \& Jorissen, A. 2003, A\&A,
403, 993

Knapp, G. R., Young, K., Lee, E., \& Jorissen, A. 1998, ApJS, 117, 209

Lambert, D. L., Gustafsson, B., Eriksson, K., \& Hinkle, K. H. 1986,
62, 373

Lamers, H. J. G. L. M. \& Cassinelli, J. P. 1999, Introduction to Stellar
Winds (Cambridge University Press: Cambridge)

Le Bertre, T. 1992, A\&AS, 94, 377

Le Bertre, T. 1993, A\&AS, 97, 729

Le~Bertre, T. \& G\'erard, E. 2001, A\&A, 378, 29

Le~Bertre, T. \& G\'erard, E. 2004, A\&A, 419, 549

Le~Bertre, T., Matsuura, M., Winters, J. M., Murakami, H., Yamamura,
I., Freund, M., \& Tanaka, M. 2001, A\&A, 376, 997

Le~Bertre, T., Matthews, L. D., G\'erard, E., \& Libert, 
Y. 2012, MNRAS, 422, 3433

Le~Bertre, T. \& Winters, J. M. 1998, A\&A, 334, 173

Lebzelter, T. \& Hinkle, K. H. 2002, A\&A, 393, 563

Lebzelter, T. \& Hron, J. 1999, A\&A, 351, 533

Le Sidaner, P. \& Le Bertre, T. 1996, A\&A, 314, 896

Leitner, S. N. \& Kravtsov, A. V. 2011, ApJ,
734, 48

Libert, Y., G\'erard, E., \& Le~Bertre, T. 2007, MNRAS, 380, 1161

Libert, Y., Le~Bertre, T., G\'erard, E., \& Winters, J. M. 2008, A\&A,
491, 789

Libert, Y., G\'erard, E., Thum, C., Winters, J. M., Matthews, L. D.,
\& Le~Bertre, T. 2010a, A\&A, 510, 14

Libert, Y., Winters, J. M., Le~Bertre, T., G\'erard, E., \& Matthews,
L. D. 2010b, A\&A, 515, 112

Little, S. J., Little-Marenin, I. R., \& Hagen Bauer, W. 1987, AJ, 94, 981

Lorenz-Martins, S. 1996, A\&A, 314, 209

Loup, C., Forveille, T., Omont, A., \& Paul, J. F. 1993, A\&AS, 99, 291

Martin, D. C. et al. 2007, Nature, 448, 780

Matthews, L. D. \& Reid, M. J. 2007, AJ, 133, 2291

Matthews, L. D., Libert, Y., G\'erard, E., Le~Bertre, T., \& Reid,
M. J. 2008, ApJ, 684, 603

Matthews, L. D., G\'erard, E., Johnson, M. C., Le~Bertre, T., Libert,
Y., \& Reid, M. J. 2011a, in Why Galaxies Care about AGB Stars II:
Shining Examples and Common Inhabitants, ed. F. Kerschbaum,
T. Lebzelter, \& R. F. Wing, (ASP: San Francisco), 445, 305

Matthews, L. D., Libert, Y., G\'erard, E., Le~Bertre, T.,
Johnson, M. C., \& Dame, T. M. 2011b, AJ, 141, 60

Matthews, L. D., Marengo, M., Evans, N. R., Bono, G. 2012, ApJ, 744, 53

Mennessier, M. O., Mowlavi, N., Alvarez, R., \& Luri, X. 2001, A\&A,
374, 968

Menten, K. M., Reid, M. J., Kami\'nski, T., \& Claussen, M. J. 2012,
A\&A, 543, 73

Merrill, P. W. 1956, PASP, 68, 70

Milam, S. N., Apponi, A. J., Woolf, N. J., \& Ziurys, L. M. 2007, ApJ,
668, L131

Miller, A. A., Richards, J. W., Bloom, J. S., Cenko, S. B., Silverman,
J. M., Starr, D. L., \& Staussun, K. G. 2012, ApJ, 755, 98

Mohamed, S., Mackey, J., \& Langer, N. 2012, A\&A, 541, 1

Mohamed, S. \& Podsiadlowski, P. 2007, in 15th European Workshop
on White Dwarfs, ASP Conf. Series, Vol. 372, ed. R. Napiwotzki and
M. R. Burleigh, (ASP: San Francisco), 397

Neri, R., Kahane, C., Lucas, R., Bujarrabal, V., \& Loup, C. 1998,
A\&AS, 130, 1

Olnon, F. M. \& Raimond, E. 1986, A\&AS, 65, 607

Olofsson, H., Eriksson, K., Gustafsson, B., \& Carlstr\"om, U. 1993,
ApJS, 87, 267

Olofsson, H., Gonz\'alez Delgado, D., Kerschbaum, F., \& Sch\"oier,
F. L. 2002, A\&A, 391, 1053

Olofsson, H., Maercker, M., Eriksson, K., Gustafsson, B., \&
Sch\"oier, F. 2010, A\&A, 515, 27

Perley, R. A. \& Taylor, G. B. 2003, VLA Calibration Manual \\
(http://www.vla.nrao.edu/astro/calib/manual/index.shtml)

Perrin, G., Ridgway, S. T., Coud\'e du Foresto, V., Mennesson, B.,
Traub, W. A., \& Lacasse, M. G. 2004, A\&A, 418, 675

Raga, A. C. \& Cant\'o, J. 2008, ApJ, 685, L141

Roy, N., Kantharia, N. G., Eyres, S. P. S., Anupama, G. C., Bode,
M. F., Prabhu, T. P., \& O'Brien, T. J. 2012, MNRAS, in press (arXiv:1209.2431)

Sahai, R. 1990, ApJ, 362, 652

Sahai, R. \& Chronopoulos, C. K. 2010, ApJ, 711, L53

Sahai, R., Morris, M., Knapp, G. R., Young, K., \& Barmbaum, C. 2003,
Nature, 426, 261

Samus, N. N. et al. 2004, Combined General Catalogue of Variable Stars

Sch\"oier, F. L., Lindqvist, M., \& Olofsson, H. 2005, A\&A, 391, 577

Sch\"onrich, R., Binney, J., \& Dehnen, W. 2010, MNRAS, 403, 1829

Schr\"oder, K.-P. \& Sedlmayr, E. 2001, A\&A, 366, 913

Smith, H. 1976, MNRAS, 175, 419

Soszy\'nski, I., Udalski, A., Szyma\'nski, M. K., Kubiak, M.,
Pietrzyn\'nski, G., Wyrzykowski, \L., Szewczyk, O., Ulaczyk, K., \&
Poleski, R. 2009, Acta Astron., 59, 335

Spergel, D. N., Giuliani, J. L. Jr., \& Knapp, G. R. 1983, 275, 330

Stencel, R. E. 2009, in The Biggest, Baddest, Coolest Stars,
ASP Conf. Series, Vol. 412, ed. D. G. Luttermoser, B. J. Smith, \&
R. E. Stencel, 197

Straniero, O., Chieffi, A., Limongi, M., Busso, M., Gallino, R., \&
Arlandini, C. 1997, ApJ, 478, 332

Tisserand, P. et al. 2009, A\&A, 501, 985

Ueta, T. 2008, ApJ, 687, L33

Ueta, T. 2011, in Why Galaxies Care about AGB Stars II: Shining
Examples and common Inhabitants, ed. F. Kerschbaum, T. Lebzelter, \&
R. F. Wing, (ASP: San Francisco), 445, 295

Ueta, T. et al. 2006, ApJ, 648, L39

Ueta, T. et al. 2010, A\&A, 514, 16

Ueta, T., Izumiura, H., Yamamura, I., Nakada, Y., Matsuura, M., Ita,
Y., Tanab\'e, T., Fukushi, H., Matsunaga, N., \& Mito, H. 2008, PASJ,
60, 407

Utsumi, L. 1985, in Cool Stars with Excesses of Heavy Elements,
(Reidel: Dordrecht), 243

van Belle, G., Dyck, H. M., Benson, J. A., \& Lacasse, M. G. 1996, AJ,
112, 2147

van Leeuwen, F. 2007, A\&A, 474, 653

van Marle, A. J., Maliani, Z., Keppens, R., \& Decin, L. 2011, ApJ,
734, L26

Vassiliadis, E. \& Wood, P. R. 1993, ApJ, 413, 641

Villaver, E., Garc\'\i a-Segura, G., \& Manchado, A. 2002, ApJ, 571,
880

Villaver, E., Manchado, A., \&  Garc\'\i a-Segura, G. 2012, ApJ, 748,
94

Wang, L., Baade, D., H\"oflick, P., Wheeler, J. C., Kawabata, K., \&
Nomoto, K. 2004, ApJ, 604, L53

Wareing, C. J. 2012, ApJ, 748, L19

Wareing, C. J. et al. 2006, MNRAS, 372, L63

Wareing, C. J., Zijlstra, A. A., \& O'Brien, T. J. 2007a, ApJ, 660,
L129

Wareing, C. J., Zijlstra, A. A., \& O'Brien, T. J. 2007b, MNRAS, 382,
1233

Wareing, C. J., Zijlstra, A. A., O'Brien, T. J., Seibert, M. 2007c, ApJ 670,
L125

Wareing, C. J., et al. 2006b, MNRAS, 372, L63

Wasatonic, R. 1995, IBVS, 4159, 1

Wasatonic, R. P. 1997, JAVSO, 26, 1

Whitelock, P. A., Feast, M. W., \& van Leeuwen, F. 2008, MNRAS, 386, 313

Wilkin, F. P. 1996, ApJ, 459, L31

Winters, J. M., Le Bertre, T., Jeong, K. S., Nyman, L.-\ang., \&
Epchtein, N. 2003, A\&A, 409, 715 

Winters, J. M., Le Bertre, T., Pety, J., \& Neri, R. 2007, A\&A, 475, 559

Young, K., Phillips, T. G., \& Knapp, G. R. 1993a,
ApJS, 86, 517

Young, K., Phillips, T. G., \& Knapp, G. R. 1993b, ApJ, 409, 725

Zijlstra, A. A. \& Weinberger, R. 2002, ApJ, 572, 1006

Zucker, D. B. \& Soker, N. 1993, ApJ, 408, 579

Zuckerman, B., Terzian, Y., \& Silverglate, P. 1980, ApJ, 241, 1014
\end{references}
\end{document}